\documentclass[aps,
prd,
a4paper,
notitlepage,
12pt,
onecolumn,
nofootinbib,
showpacs,
eqsecnum,
superscriptaddress,
longbibliography,
preprintnumbers]{revtex4-1}

\usepackage{amsmath, dsfont}
\allowdisplaybreaks %allows the equations to break
\usepackage{amssymb}
\usepackage{epsfig}
\usepackage{calc}
\usepackage{amsfonts}
\usepackage{mathtools}
\usepackage{enumerate}
\usepackage{graphicx}
\usepackage{caption}
\usepackage{subcaption}
\usepackage{ragged2e}

\DeclareCaptionJustification{bhjustified}{\justifying}

\usepackage[ansinew]{inputenc}
\usepackage{multirow}
\usepackage{tensor}

\usepackage[dvipsnames]{xcolor}
\usepackage{natbib}
\usepackage{float,booktabs}
\usepackage{soul}
\usepackage[T1]{fontenc}
\usepackage[normalem]{ulem}
\usepackage{etoolbox}
\makeatletter
\patchcmd{\@bibsetup}{\raggedright}{}{}{}
\usepackage{array}
\usepackage{makecell}
\usepackage{setspace}
\usepackage[top=1in, bottom=1in, left=0.8in, right=0.8in]{geometry}
\usepackage{mathrsfs}
\usepackage{cancel}

\usepackage{fancyhdr}
\usepackage{empheq} %emphasize equations
\usepackage[skins]{tcolorbox}

\graphicspath{{Figs/}}

\usepackage{libertine}

\definecolor{azure(colorwheel)}{rgb}{0.0, 0.5, 1.0}
\definecolor{PRDblue}{RGB}{48,46,146}
\definecolor{nicered}{rgb}{0.7,0.1,0.1}
\definecolor{DarkViolet}{RGB}{148,0,211}
\definecolor{myDarkBlue}{rgb}{0,0.1,0.7}
\definecolor{tealgreen}{rgb}{0.0, 0.51, 0.5}

\usepackage{hyperref}
\hypersetup{colorlinks,
    bookmarksopen,
	bookmarksnumbered,
	citecolor={nicered},
	linkcolor={PRDblue},
	urlcolor={tealgreen},
	pdfstartview=FitH}

\newcommand{\dd}{\mathrm{d}}

\newcommand\Veff{V_{\mathrm{eff}}}
\newcommand\rmax{r_*^{\max}}
\newcommand\Det{\Delta}
\newcommand\Schroop{{\cal{L}}}
\newcommand\SchroopE{\Schroop_{{E}}}

\newcommand\pmin{p_{\rm min}}
\newcommand\Bterm{\mathscr{B}}

\usepackage{orcidlink}
\def\idbahamonde{\orcidlink{0000-0001-6235-120X}}
\def\idnakas{\orcidlink{0000-0002-3522-5803}}

\usepackage{titletoc}
\titlecontents{section}[1.5em]{\addvspace{0.5em}\bfseries}
    {\contentslabel{2.5em}}{}
    {\titlerule*[0.5pc]{.}\contentspage}
\titlecontents{subsection}[4em]{\normalfont}
    {\contentslabel{1.5em}}{}
    {\titlerule*[0.5pc]{.}\contentspage}
\titlecontents{subsubsection}[6.2em]{\normalfont}
    {\contentslabel{1.5em}}{}
    {\titlerule*[0.5pc]{.}\contentspage}

\usepackage[splitrule]{footmisc}
\begin{document}

\title{\Large Nonreciprocal black holes with primary hair in beyond Horndeski gravity: Exact solutions and stability}

\author{Sebastian Bahamonde\,\idbahamonde}
\email{sbahamondebeltran@gmail.com}
\affiliation{Cosmology, Gravity, and Astroparticle Physics Group, Center for Theoretical Physics of the Universe, Institute for Basic Science (IBS), Daejeon 34126, Korea}

\author{Theodoros Nakas\,\idnakas}
\email{theodoros.nakas@gmail.com}
\affiliation{Cosmology, Gravity, and Astroparticle Physics Group, Center for Theoretical Physics of the Universe, Institute for Basic Science (IBS), Daejeon 34126, Korea}

\author{Simon Iteanu\,}
\email{simon.iteanu@ijclab.in2p3.fr}
\affiliation{Universit\'e Paris-Saclay, CNRS/IN2P3, IJCLab, 91405 Orsay, France}

 \begin{abstract}
    \noindent\textbf{Abstract.} We develop a novel systematic method for constructing exact, static, and asymptotically flat black-hole solutions in shift-symmetric and $Z_2$-symmetric beyond Horndeski theories. In contrast to previous studies, our approach does not rely on the restrictive reciprocal condition $g_{tt}g_{rr}=-1$, allowing for more general geometries characterized by two independent metric functions.  
    The resulting solution space contains two qualitatively distinct types of hair originating from different branches of the theory. One branch is characterized by khronometric hair, while the other by genuine scalar hair. In both, the scalar profile defines a nontrivial preferred foliation, but the shift-symmetry Noether current and the associated propagating degree of freedom are trivial for the former and nonvanishing for the latter. Focusing mainly on the second branch, we derive several distinct families of exact black-hole solutions. In general, these solutions are singular and exhibit primary scalar hair associated with an independent scalar charge. Nevertheless, we demonstrate that they can be regularized to yield regular black-hole or solitonic configurations, for which the scalar hair becomes secondary. We further investigate the stability properties of these solutions under linear perturbations by analyzing the axial sector and the monopole $(\ell=0)$ mode of the even-parity sector. For representative solutions with scalar hair, the perturbative analysis provides evidence for linear stability under the modes considered.
\end{abstract}

\maketitle

\tableofcontents

\flushbottom

\vspace{2mm}

%==========================================
%Introduction 
%==========================================
\section{Introduction}

%box-environment
\tcbset{frame style={top color=gray!20,
                     bottom color=gray!20},
        fonttitle=\bfseries,coltitle=black, sharp corners,boxrule=0pt}

The simplicity of black holes (BH), as characterized by the uniqueness and no-hair theorems, stands as one of the most substantial results in General Relativity (GR).
In the 1970s, a sequence of uniqueness theorems~\cite{Israel:1967wq, Carter:1971zc, Robinson:1975bv}, established that stationary, asymptotically flat vacuum black holes are completely characterized by a small set of global charges, namely their \emph{mass} and \emph{angular momentum}, leading uniquely to the Kerr family of solutions \cite{Kerr:1963ud}.
In the presence of electromagnetism, this result extends to the Kerr-Newman spacetime \cite{Newman:1965my}, such that stationary black holes in Einstein-Maxwell theory are fully specified by their \emph{mass}, \emph{angular momentum}, and \emph{electric charge}.
These results are closely related to the no-hair paradigm \cite{Ruffini:1971bza, Bekenstein:1995un}, according to which black holes cannot support additional independent fields beyond the conserved charges measurable at infinity.
Consequently, astrophysical black holes in GR are universal and remarkably simple objects, irrespective of the details of their formation history.

Nevertheless, despite their remarkable success, the uniqueness and no-hair theorems rely on a number of assumptions that can be violated in theories beyond GR.
Specifically, for scalar hair, Bekenstein's formulation \cite{Bekenstein:1995un} assumes that the action depends on the derivatives of the scalar field solely via the $(\partial \phi)^2$ term.
Higher-derivative scalar interactions, such as Galileon-type terms~\cite{Nicolis:2008in}, and nonminimal scalar-curvature couplings can evade the assumptions of Bekenstein's theorem. An early example is provided by black holes with dilatonic Gauss--Bonnet hair~\cite{Kanti:1995vq}.
Later, a novel no-scalar-hair theorem for Galileon theories was formulated \cite{Hui:2012qt}, but it was soon realized that some of the assumptions underlying the theorem can be relaxed, thereby allowing it to be evaded. For a non-exhaustive list of works studying black holes with nontrivial scalar hair, the reader is referred to \cite{Sotiriou:2013qea, Babichev:2013cya, Sotiriou:2015pka, Herdeiro:2015waa, Antoniou:2017acq, Babichev:2017guv, Doneva:2017bvd, Silva:2017uqg, Barack:2018yly, BenAchour:2018dap, Creminelli:2020lxn, BenAchour:2020fgy, Baake:2021jzv, Bakopoulos:2023fmv, Baake:2023zsq, Bakopoulos:2023sdm, Lara:2025hqh, Charmousis:2025xug, Smulders:2026qwc} and references therein.
The works of Refs.~\cite{Bakopoulos:2023fmv, Baake:2023zsq,
Bakopoulos:2023sdm} were the first to establish black-hole solutions endowed with primary scalar hair, a construction subsequently extended along several directions: to hedgehog scalar hair~\cite{Bahamonde:2026bvh}, torsion hair in cubic Holst-type Poincar\'e gauge gravity \cite{Bahamonde:2026yjl}, Proca hair in regularized Gauss-Bonnet gravity \cite{Charmousis:2025jpx}, an infinite tower of regularized Lovelock-Proca corrections \cite{Fernandes:2025mic}, and rotating configurations in beyond Horndeski gravity \cite{Nozari:2026wjo}, alongside systematic theoretical constraints on the viable space of shift-symmetric Horndeski constructions~\cite{Bakopoulos:2025byy}.
Another important aspect of modified theories of gravity concerns the existence of regular (non-singular) black holes, i.e. configurations in which spacetime singularities are absent~\cite{Bardeen1968, Dymnikova:1992ux, Hayward:2005gi,Lan:2023cvz}.\\
\indent In GR, the singularity theorems imply causal geodesic incompleteness under
their respective assumptions~\cite{Penrose:1964wq,Hawking:1970zqf}.
Black-hole solutions such as Schwarzschild and Kerr additionally exhibit
curvature singularities hidden behind an event horizon.
The presence of spacetime singularities is commonly regarded as signaling the breakdown of the classical description of gravity and suggests that GR should be modified in the ultraviolet regime.
In this context, modified gravity theories provide a framework in which regular black holes may arise through additional dynamical degrees of freedom or higher-order interactions.
Scalar-tensor theories and specifically Horndeski \cite{Horndeski:1974wa} and beyond Horndeski \cite{Langlois:2015cwa} can support nontrivial scalar configurations capable of modifying the near-core geometry and giving rise to regular BH \cite{Baake:2021jzv, Bakopoulos:2023fmv, Baake:2023zsq, Bakopoulos:2023sdm}.
Besides their mathematical interest, such solutions provide a useful setting for exploring strong-field gravitational physics beyond the Schwarzschild \cite{Schwarzschild:1916uq} or Kerr \cite{Kerr:1963ud} paradigm and for investigating possible observational signatures of modified gravity.
This has motivated extensive investigations of modified theories of gravity, both from a theoretical perspective and in light of the growing observational access to the strong-field regime through gravitational-wave astronomy (see e.g. \cite{LIGOScientific:2016aoc, LIGOScientific:2017vwq, LIGOScientific:2021qlt, Barausse:2020rsu}) and horizon-scale imaging experiments, such as \cite{EventHorizonTelescope:2019dse, EventHorizonTelescope:2020qrl}.
\enlargethispage{-3\baselineskip}
In this work, we generalize the results presented in~\cite{Bakopoulos:2023fmv, Bakopoulos:2023sdm}.
In particular, we develop a novel systematic and analytic method for solving the equations of motion for static, spherically symmetric, and asymptotically flat black holes in shift-symmetric ($\phi \to \phi + {\rm const.}$) and $Z_2$-symmetric ($\phi \to -\phi$) beyond Horndeski theories \cite{Gleyzes:2014dya}, which constitute a subclass of the Quadratic Degenerate Higher-Order Scalar-Tensor (DHOST) theories \cite{Langlois:2015cwa, BenAchour:2016cay}.
In contrast to the analysis performed in~\cite{Bakopoulos:2023sdm}, our construction does not rely on the restrictive condition $g_{tt}g_{rr}=-1$, thereby allowing for new and more general classes of geometries with two independent metric functions.
In the literature, black-hole solutions satisfying this condition are sometimes referred to as homogeneous solutions, even though the corresponding spacetimes are not homogeneous in the standard geometrical sense.
To avoid possible ambiguity associated with this terminology, throughout the present article we refrain from using the term homogeneous to describe black holes obeying $g_{tt}g_{rr}=-1$. Instead, we refer to such configurations as reciprocal black holes. Consequently, black-hole solutions for which $g_{tt}g_{rr}\neq -1$ will be termed \textbf{non-reciprocal black holes}.
Subsequently, by applying our method, we construct several distinct families of exact non-reciprocal black-hole solutions, with $g_{tt}g_{rr}\neq -1$.\\
\indent As we will show below, the field equations naturally separate the nontrivial
solution space into two main branches, which we denote as Branch~II and
Branch~III. In Branch~II, the shift-symmetry Noether current vanishes
identically and no independent scalar degree of freedom propagates. The scalar
field nevertheless defines a preferred foliation, and we refer to this
configuration as \emph{khronometric hair}. By contrast, Branch~III possesses a
nonvanishing Noether current and supports an independent propagating scalar
mode, and we refer to this as \emph{scalar hair}. The distinction between
primary and secondary hair will be specified in Sec.~\ref{Sec:General-Approach}. \\
\indent Another important question regarding black-hole solutions in modified theories of gravity is whether they are stable under linear perturbations.
Indeed, from both theoretical and phenomenological perspectives, physically relevant black-hole configurations are expected to remain stable against small fluctuations of the metric and scalar field.
Linear perturbations of black holes in Horndeski \cite{Horndeski:1974wa} and DHOST \cite{Langlois:2015cwa} theories have been extensively studied in recent years~\cite{Kobayashi:2012kh, Cisterna:2015uya, Takahashi:2016dnv, Takahashi:2019oxz, deRham:2019gha, Charmousis:2019fre, Khoury:2020aya, Tomikawa:2021pca, Langlois:2021xzq, Langlois:2021aji, Takahashi:2021bml, Minamitsuji:2022mlv, Minamitsuji:2022vbi, Noui:2023ksf, Antoniou:2024gdf}, and, very recently, for black holes with static scalar hair in general GLPV theories \cite{Boonserm:2026qwz}.
For static and spherically symmetric black holes, perturbations can be decomposed into polar (even-parity) and axial (odd-parity) sectors \cite{Regge:1957td, Zerilli:1970se, Vishveshwara:1970cc, Vishveshwara:1970zz}.
In scalar-tensor theories, the axial sector is considerably simpler than the polar one, since scalar perturbations belong exclusively to the even-parity sector under the spherical harmonic decomposition of linear perturbations.
We investigate the axial perturbations of the families constructed below, as well as the monopole $\ell=0$ sector of even-parity perturbations.
The structure of this paper is as follows. \\
\indent In Sec.~\ref{Sec:General-Approach}, we introduce our systematic method for constructing non-reciprocal, exact, static, spherically symmetric, and asymptotically flat black-hole solutions in two distinct classes of theories originating from shift- and $Z_2$-symmetric beyond Horndeski theory. 
The first, and simplest, is a subclass of khronometric theories, whose solutions are characterized by khronometric hair.
The second is a subclass of quadratic DHOST theory, whose solutions are characterized, in general, by primary scalar hair. 
In the latter class, a unique relation between the mass and the scalar
charge removes the central curvature singularity altogether; hence, the resulting solutions describe either regular black holes or horizonless solitons. 
Because this relation ties the scalar charge to the mass, the hair carried by the regular solutions is secondary.
In Sec.~\ref{Sec:Disformal-Frame}, we show that every non-reciprocal solution obtained in this way is related, through a \emph{unique} purely disformal transformation, to a solution satisfying the reciprocal condition $g_{tt}g_{rr}=-1$. 
We then determine the reciprocal-frame theories associated with the theories under examination. In Sec.~\ref{Sec:BranchII}, we present an explicit solution describing regular black holes in khronometric theory, and we discuss its characteristics.
In Sec.~\ref{sec:branch3}, we specialize to the scalar-hair Branch~III focusing on the beyond Horndeski theory exhibiting primary scalar hair and construct four explicit, distinct families of non-reciprocal black-hole solutions, generated from a specific choice of the auxiliary function $Z(X)$.
These include a power-law family for the auxiliary function $\mathcal F(\Xi)$, containing as special cases a stealth extension of Schwarzschild ($p=1/2$) and a non-reciprocal extension of the Bardeen solution ($p=5/2$), a superposition of power-law contributions, and two Pad\'e-type ans\"atze. In Sec.~\ref{sec:IVB}, we then show that this construction is not tied to a single choice of $Z(X)$. We show that a different $Z(X)$---while keeping the same auxiliary function $\mathcal F(\Xi)$---produces a genuinely different, inequivalent geometry in the non-reciprocal frame, while both remain disformally connected to one and the same reciprocal-frame solution. In Sec.~\ref{Sec:Axial}, we study the axial (odd-parity) perturbations of every solution constructed above, deriving a master Schr\"odinger-like equation and establishing the conditions under which axial mode stability follows. In Sec.~\ref{Sec:Polar}, we study the polar monopole ($\ell=0$) sector, explaining the absence of an independent propagating scalar in khronometric theories provided no additional matter is included. More importantly, we analyze the stability of solutions with primary scalar hair originating from the other class of theories. We conclude and discuss future directions in Sec.~\ref{Sec:Conclusions}.

%==========================================
%Main Part
%==========================================

\section{A systematic approach for non-reciprocal black holes with primary hair and regular counterparts}
\label{Sec:General-Approach}

\raggedbottom

In this section, we develop a method for constructing exact, static, and asymptotically flat black holes with primary hair, together with the conditions under which regular counterparts can be obtained.
Since our approach does not impose the restrictive reciprocal condition $g_{tt}g_{rr}=-1$, the line element can be written in the more general form
\begin{equation}
    \dd s^2 = -A(r) \dd t^2 + \frac{[B(r)]^2}{A(r)}\, \dd r^2 + r^2 \dd \Omega^2\,,\qquad \dd\Omega^2 = \dd \vartheta^2 + \sin^2\vartheta\, \dd \varphi^2\,,
    \label{eq:metric}
\end{equation}
which contains two independent metric functions.

Our starting point is the shift-symmetric ($\phi\to \phi+{\rm const.}$) and $Z_2$-symmetric ($\phi\to -\phi$) beyond Horndeski action \cite{Langlois:2015cwa}, which in geometrized units ($c=G=1$) takes the form 
\begin{equation}
    S[g_{\mu\nu},\phi] = \frac{1}{16\pi}\int \dd^4x\,\sqrt{-g}\big( \mathcal{L}^H_2 + \mathcal{L}^H_4 + \mathcal{L}^{bH}_4\big)\,,
    \label{eq:act}
\end{equation}
with
\begin{alignat}{2}
& \mathcal{L}^H_2 && = G_2(X)\,, 
    \label{eq:L2H}\\[2mm]
& \mathcal{L}^H_4 && = G_4(X)R+G_{4X}\bigl[(\Box\phi)^2-\phi_{;\mu\nu}\phi^{;\mu\nu}\bigr]\,,
    \label{eq:L4H}\\[2mm]
& \mathcal{L}^{bH}_4 && = F_4(X)\epsilon^{\mu\nu\rho\sigma}\epsilon^{\alpha\beta\gamma}{}_{\sigma}\phi_{;\mu}\phi_{;\alpha}\phi_{;\nu\beta}\phi_{;\rho\gamma}\,.
    \label{eq:L4bH}
\end{alignat}
The Lagrangian densities $\mathcal{L}^H_2$ and $\mathcal{L}^H_4$ belong to Horndeski theory, while $\mathcal{L}^{bH}_4$ is associated with the beyond Horndeski sector.
In the above, $\phi_{;\mu}\equiv\nabla_\mu\phi$, $\phi_{;\mu\nu}\equiv\nabla_\mu \nabla_\nu \phi$, $G_{iX}\equiv \dd G_i/\dd X$, and $\displaystyle X\equiv-(\partial\phi)^2/2$ is the kinetic term of the scalar field.
Note also that for a dimensionless scalar field, the model functions $G_2$ and $F_4$ have dimensions of $(\text{length})^{-2}$ and $(\text{length})^4$, respectively, while $G_4$ is dimensionless.
The shift-symmetric nature of the theory \eqref{eq:act} allows the scalar field to exhibit, in addition to the radial dependence, a linear dependence on time, namely
\begin{equation}
    \phi(t,r)=qt+\psi(r)\,,
    \label{eq:scalar}
\end{equation}
without spoiling the staticity of the resulting solutions.
Moreover, as pointed out in \cite{Bakopoulos:2023sdm}, whenever the Noether current associated with the shift symmetry is nonvanishing, the parameter $q$ is directly related to the corresponding scalar charge.
Additionally, due to this symmetry, the scalar-field equation is equivalent to $\nabla_\mu J^\mu=0$, where
\begin{align}
J^\mu = &-\big\{G_{2X}+G_{4X}R+G_{4XX}\big[(\square \phi)^2-\phi_{;\alpha\beta}\phi^{;\alpha\beta}\big]+F_{4X}\epsilon^{\lambda\nu\rho\sigma}\epsilon^{\alpha\beta\gamma}{}_{\sigma}\phi_{;\lambda}\phi_{;\alpha}\phi_{;\nu\beta}\phi_{;\rho\gamma}\big\}\phi^{;\mu}\nonumber\\
&+2F_4 \epsilon^{\mu\nu\rho}{}_\sigma\,\epsilon^{\alpha\beta\gamma\sigma}\phi_{;\alpha}\phi_{;\nu\beta}\phi_{;\rho\gamma}-2\nabla_\nu\big[G_{4X}(g^{\mu\nu}\square \phi-\phi^{;\mu\nu})+F_4 \epsilon^{\alpha\mu\rho}{}_{\sigma}\,\epsilon^{\beta\nu\gamma\sigma}\phi_{;\alpha}\phi_{;\beta}\phi_{;\rho\gamma}\big]\,.
\label{eq:J_total}
\end{align}
Interestingly, due to the connection between the $(tr)$-component of the metric field equations and the Noether current (see \cite{Babichev:2015rva}), the only potentially nonvanishing (on-shell) component of the Noether current is the $t$-component.
Whenever $J^t\neq0$ and the corresponding integral converges, one may define
the scalar Noether charge
\begin{equation}
    Q_s\propto \int_{\Sigma_t}\star\mathbf J
    =4\pi\int \dd r\,r^2B(r)J^t(r).
\end{equation}
Thus, $q$ parametrizes the nonvanishing Noether-current sector. For the
explicit models for which the above integral is finite, the resulting scalar
charge is a model-dependent function of $q$.
The Lagrangian of Eq. \eqref{eq:act} is part of the Quadratic DHOST theory \cite{Langlois:2015cwa, BenAchour:2016cay}, which according to \cite{Langlois:2021aji} can be expressed as
\begin{equation}
S[g_{\mu\nu},\phi]=\int \dd^{4}x\,\sqrt{-g}\left[
P(\phi,X)+Q(\phi,X)\,\Box\phi+F(\phi,X)\,R+\sum_{i=1}^{5}A_i(\phi,X)\,L_i
\right],
\label{eq:Quad-DHOST}
\end{equation}
where
\begin{equation}
\begin{gathered}
\mathcal{L}_1 \equiv \phi_{;\mu\nu}\phi^{;\mu\nu}, \qquad
\mathcal{L}_2 \equiv (\Box\phi)^2, \qquad
\mathcal{L}_3 \equiv \phi^{;\mu}\phi_{;\mu\nu}\phi^{;\nu}\,\Box\phi,\\
\mathcal{L}_4 \equiv \phi^{;\mu}\phi_{;\mu\nu}\phi^{;\nu\rho}\phi_{;\rho}, \qquad
\mathcal{L}_5 \equiv \big(\phi^{;\mu}\phi_{;\mu\nu}\phi^{;\nu}\big)^2.
\end{gathered}
\end{equation}
One can readily show that the general Quadratic DHOST Lagrangian \eqref{eq:Quad-DHOST} matches the Lagrangian \eqref{eq:act} for the following model functions:
\begin{gather}
    P(\phi,X)=G_2(X)\,,\qquad  Q(\phi,X)=0\,,\qquad   F(\phi,X)=G_4(X)\,,
    \label{eq:quad-dhost1}\\[2mm]
    A_1(\phi,X)=-A_2(\phi,X)=-G_{4X}(X)-2XF_4(X)\,,
    \label{eq:quad-dhost2}\\[2mm] 
    A_3(\phi,X)=-A_4(\phi,X)=2F_4(X)\,,\qquad A_5(X)=0\,.  
    \label{eq:quad-dhost3}
\end{gather}
It is important to note that our convention for $X$ differs from that of \cite{Langlois:2015cwa, BenAchour:2016cay, Langlois:2021aji}. 
This difference affects every relation that depends explicitly on $X$ or on derivatives with respect to $X$.

The independent field equations of the theory \eqref{eq:act}, for the metric ansatz \eqref{eq:metric} and the scalar field \eqref{eq:scalar}, can be brought into the following form
\begin{gather}
    \frac{1}{B^2}=\frac{1}{Z^2}\,,\label{eq1-new}\\[2mm]
    r^2(G_2 Z)_{X}+2(G_4 Z)_{X}
    \left(1-\frac{q^2 }{2 Z^2 X}\right)=0\,,\label{eq2-new}\\[2mm]
    2 \left(A r-\frac{q^2 r}{2 X}\right)'=r^2 G_2 Z + 2G_4 Z\left(1-\frac{q^2 }{2 Z^2 X}\right) - \frac{2 q^2 r}{Z \sqrt{X}} \left(\frac{G_4}{\sqrt{X}}\right)' \,,
    \label{eq3-new}
\end{gather}
where prime denotes differentiation with respect to the radial coordinate $r$, while $Z(X)$ is an auxiliary function defined as
\begin{equation}
    Z(X)\equiv G_4(X) - 2XG_{4X} - 4X^2 F_4(X)\,.
    \label{eq:def-Z}
\end{equation}
Notice that the above equations differ slightly from the equations presented in  \cite{Bakopoulos:2023fmv} despite being derived within the same theoretical framework.
This discrepancy originates from the different definition of the auxiliary function, which in our convention differs by an overall minus sign, as well as from rewriting the last term in \eqref{eq3-new} as a derivative with respect to $r$, a form that will prove useful in the subsequent analysis. 
Finally, to further simplify the equations, we absorb the dimensionless parameter $\gamma$ appearing in \cite{Bakopoulos:2023fmv} into the auxiliary function $Z$, without loss of generality.

Eq.~\eqref{eq1-new} fixes the metric function $B(r)$ entirely in terms of the model functions $G_4(X)$ and $F_4(X)$ through $Z(X)$.
To solve Eq.~\eqref{eq2-new} it is convenient to introduce the auxiliary functions $\mathcal{G}_2$ and $\mathcal{G}_4$, defined through
\begin{equation}
(G_2Z)_X \equiv \mathcal{G}_2(X),
\qquad
(G_4Z)_X \equiv \mathcal{G}_4(X)\,.
\label{eq:G2-G4-aux}
\end{equation}
In terms of these functions, Eq. \eqref{eq2-new} becomes
\begin{equation}
r^2\mathcal{G}_2(X)+\mathcal{G}_4(X)\left(2-\frac{q^2}{XZ^2}\right)=0,
\label{eq2-aux}
\end{equation}
which can now be solved algebraically in terms of $X$, given the expressions for $\mathcal{G}_2$ and $\mathcal{G}_4$. 
The essential point of \eqref{eq2-aux} is that it splits the solution space into branches.

\textbf{Branch I.} The simplest assumption one could make to solve~\eqref{eq2-aux} is
\begin{equation}
    \mathcal{G}_2=0\,,
    \qquad 
    Z^2 X = \frac{q^2}{2}\,,
\end{equation}
which implies that $G_2 Z = g_2$ and that $X$ is constant. 
This branch is a very special case because $X$ is constant on the background,
so that all model functions entering the background equations are evaluated
at the fixed value $X=X_0$.
As a consequence, Eq.~\eqref{eq3-new} can be straightforwardly integrated to determine $A(r)$. 
After performing the time redefinition $\tau = Z t$, with $Z=\mathrm{const.}$ the resulting spacetime is simply the Schwarzschild-de Sitter solution.
Hence, this branch offers no flexibility to support new black-hole solutions.

\textbf{Branch II.} Another possibility is obtained by assuming that $\mathcal{G}_4=0$, this, together with \eqref{eq2-aux}, directly force $\mathcal{G}_2$ to vanish as well.
Therefore, this branch of solutions is characterized by
\begin{equation}
\mathcal{G}_2=\mathcal{G}_4=0.
\end{equation}
The above relation immediately implies
\begin{equation}
G_2 Z = g_2,
\qquad
G_4 Z = g_4,
\label{eq:br2-G2-G4}
\end{equation}
where \(g_2\) and \(g_4\) are constants of the theory. The constant $g_2$ has dimensions of $({\rm length})^{-2}$, while $g_4$ is dimensionless. 
Under this choice, Eq.~\eqref{eq2-aux} is automatically satisfied, and therefore no independent equation is left to determine the radial profile of $X(r)$. 
The remaining equation \eqref{eq3-new} then determines the metric function $A(r)$. This property already distinguishes Branch~II sharply from the scalar-hair branch discussed below.
Although the scalar profile, in this case, is nontrivial and may contain an independent parameter, it is not associated with a nonzero scalar Noether charge.
Moreover, as we will show in Sec.~\ref{Sec:Rec-BranchII}, the Branch-II theory, expressed in the reciprocal frame, belongs to a special khronometric subclass that is related to GR by an invertible disformal transformation and propagates only the two tensor degrees of freedom, corresponding to the two graviton polarizations, with no independent scalar mode.
For this reason, throughout this work we refer to the nontrivial scalar configuration of Branch~II as \emph{khronometric hair}, rather than scalar hair.

Utilizing \eqref{eq:br2-G2-G4}, one can readily solve Eq.~\eqref{eq3-new} for $A(r)$:
\begin{equation}
    A(r)= g_4+\frac{C_0}{r}+\frac{g_2}{6}\,r^2+\frac{q^2}{2X}\bigg(1 - \frac{G_4^2}{g_4}\bigg)\,,
    \label{Branch2MetricA}
\end{equation}
where $C_0$ is an integration constant. 
Additionally, from  Eqs.~\eqref{eq1-new} and \eqref{eq:br2-G2-G4} we have 
\begin{equation}
    B^2(r)=Z^2(X)=\frac{g_4^2}{G_4^2(X)}.\label{Branch2MetricB}
\end{equation}
Thus, the spacetime geometry is controlled by the radial profile \(X(r)\) and the model function $G_4(X)$. 
The fact that the field equations do not determine \(X(r)\) independently constitutes a distinctive feature of this branch, reflecting the identically vanishing Noether current.
We will return to this Branch in Sec.~\ref{Sec:Disformal-Frame}, where its geometrical properties and connection to General Relativity when $g_4=1$ are further elucidated through a disformal transformation.
Additionally, in Secs. \ref{Sec:Axial} and \ref{Sec:Polar} we investigate the stability of solutions in this branch.

\textbf{Branch III.} From the above discussion, it becomes clear that, in order to obtain new black-hole solutions within the framework of \eqref{eq:act} with a nonvanishing Noether current \eqref{eq:J_total}, one must require $\mathcal{G}_2\neq0$ and $\mathcal{G}_4\neq0$.
In contrast to Branch~II, this branch therefore belongs to the nonvanishing Noether-current sector.
Furthermore, as shown explicitly by the monopole analysis in Sec.~\ref{Sec:Polar}, Branch~III supports an independent propagating scalar degree of freedom.
We shall consequently refer to Branch~III as the \emph{scalar-hair branch}. This distinction concerns the physical nature of the hair and is independent
of the usual classification into primary and secondary hair. We call the hair
\emph{primary} whenever it introduces an additional independent parameter
characterizing the black-hole solution beyond the standard gravitational
charges. Conversely, the hair is \emph{secondary} when the corresponding hair
parameter is fixed by the gravitational charges and the couplings of the
theory. Accordingly, Branch~II may carry primary or secondary khronometric
hair, whereas Branch~III may carry primary or secondary scalar hair. For the
Branch-III solutions studied below, the parameter $q$ associated with
$\phi(t,r)=qt+\psi(r)$ parametrizes the nonvanishing Noether-current sector and,
whenever the corresponding charge integral is finite, is related to the scalar
Noether charge.\\
\indent Solving the full system of equations \eqref{eq1-new}--\eqref{eq3-new} in complete generality is, however, a highly nontrivial task.
Hence, to simplify our analysis, we assume that
\begin{equation}
    \frac{\mathcal{G}_2}{\mathcal{G}_4}=\frac{(G_2 Z)_X}{(G_4 Z)_X}=\frac{2}{\lambda^2}\quad \Rightarrow\quad G_4 Z = 1 +  \frac{\lambda^2}{2}\, G_2 Z \,,
    \label{eq:G2-G4}
\end{equation}
where $\lambda$ is a coupling constant with dimensions of $\text{length}$ and the integration constant is taken to be unity.
Therefore, this ansatz imposes one functional relation among the three model functions, reducing the number of independent functions from three to two. Using now \eqref{eq:G2-G4} in \eqref{eq2-new} it is straightforward to algebraically and implicitly solve \eqref{eq2-new} for $X$ and obtain
\begin{equation}
    \Xi(X)\equiv Z^2 X = \frac{q^2}{2}\, \frac{1}{1+r^2/\lambda^2}\,.
    \label{eq:Xi}    
\end{equation}
Notice that the combination $Z^2 X$ is an invariant quantity in theories satisfying \eqref{eq:G2-G4}.
Motivated by this property, in the preceding equation, we have introduced the auxiliary function $\Xi(X)$.
It is worth noting that for $Z=1$, or equivalently $g_{tt}g_{rr}=-1$ (corresponding to reciprocal black holes), the above expression for the kinetic term of the scalar field reduces to that obtained in \cite{Bakopoulos:2023fmv, Bakopoulos:2023sdm}.
Here, however, we focus on non-reciprocal solutions, for which $g_{tt}g_{rr}\neq-1$ and consequently $Z\neq 1$.
Hence, in this case, $\Xi$ plays the role that $X$ played in \cite{Bakopoulos:2023fmv, Bakopoulos:2023sdm}.
In order to solve Eq.~\eqref{eq3-new}, we reparametrize the product $G_2 Z$ as a function of $\Xi$ and we write
\begin{equation}
    G_2 Z = \mathcal{F}(\Xi)\,.
    \label{eq:G2-F}
\end{equation}
Combining now Eqs.~\eqref{eq:G2-G4}-\eqref{eq:G2-F}, one can readily show that \eqref{eq3-new} can be solved for $A(r)$ and brought into the form
\begin{equation}
    A(r) = 1 -\frac{C_0}{r}+\bigg(1+\frac{r^2}{\lambda^2}\bigg)\big[B^2(r)-1\big] + \frac{1}{r}\int \dd r\, r^2 \big[2\,\Xi\, \mathcal{F}_\Xi -\mathcal{F}(\Xi)\big]\bigg|_{\text{\fontsize{10}{10}\selectfont $\Xi=\frac{q^2}{2}\frac{1}{1+r^2/\lambda^2}$}}\,.
    \label{eq:A-sol}
\end{equation}

Since we are interested in asymptotically flat spacetimes, we define
the ADM mass $M$ through the asymptotic behaviour
\begin{equation}
   A(r)=1-\frac{2M}{r}+o(r^{-1}),
\qquad r\to+\infty.
\end{equation}
In general, the coefficient of the $1/r$ term in
Eq.~\eqref{eq:A-sol} receives contributions from the integration
constant $C_0$, from the asymptotic behaviour of
$(1+r^2/\lambda^2)[B^2(r)-1]$, and from the radial integral.
Therefore, $C_0$ should not in general be identified directly with the
ADM mass.
Note that a generic compact-object solution obtained from Eq.~\eqref{eq:A-sol} contains two independent parameters: the ADM mass and $q$. 
Since in Branch~III the parameter $q$ parametrizes the nonvanishing
shift-symmetry Noether-current sector and the propagating scalar sector, it
constitutes primary scalar hair whenever it is independent of the ADM mass.
For regular solutions, such as regular black holes or gravitational solitons, regularity at the origin demands that the $1/r$ term vanish as $r \to 0$. 
This imposes a relation between the two parameters, eliminating one degree of freedom and leaving only the ADM mass free, thereby yielding solutions with secondary scalar hair. 

Before turning to explicit examples, it is useful to work out the conditions on the metric functions $A(r)$ and $B(r)$ in the line element \eqref{eq:metric} that ensure asymptotic flatness, along with the additional conditions needed for the resulting black hole solutions to be regular, i.e.\ non-singular at the center.
Hence, as shown in Appendix~\ref{App:Asym-III}, for the class of model functions considered here, admitting asymptotic expansions in powers of $\Xi$, sufficient conditions for asymptotic flatness are
\begin{gather}
    B^2(r) = 1 + \mathcal{O}(r^{-3})\,,\qquad \text{as\ \ $r\to+\infty$},
    \label{eq:B-asym-con}
\end{gather}
and
\begin{equation}
    \mathcal{F}(\Xi) = \left\{ \begin{array}{cl}
        o(\Xi^{3/2})\,, & \text{falls faster than $\Xi^{3/2}$} \\[2mm]
        \multicolumn{2}{c}{\text{or}} \\[2mm]
        \propto \Xi^{1/2}\,,  & \text{proportional to $\Xi^{1/2}$}
    \end{array} \right.\,,\qquad \text{as\ \ } \Xi \xrightarrow{r\to+\infty} 0\,.
    \label{eq:Fcal-asym-con}
\end{equation}
The case $\mathcal F(\Xi)\propto \Xi^{1/2}$ is special because the last term on the r.h.s. of Eq.~\eqref{eq:A-sol} vanishes altogether.
Regarding the regularity conditions, as $r\to0$ the metric functions must behave as
\begin{equation}
    A(r)=A_0 + \sum_{n=1}^{+\infty} A_{2n} r^{2n} ,
    \qquad
    B(r)=B_0 + \sum_{n=1}^{+\infty} B_{2n} r^{2n},
    \qquad
    A_0 = B_0^2 .
    \label{eq:regular-AB}
\end{equation}
As shown in Appendix \ref{App:Reg-Cond}, these conditions guarantee the finiteness of the Riemann tensor components, and hence of all curvature invariants. 
Notice that, for reciprocal solutions, where $B(r)=1$, $\forall r\geq 0$, the above relation indicates that $A(0)=1$.
A regular asymptotically flat solution must satisfy both sets of requirements. 
The asymptotic conditions \eqref{eq:B-asym-con} and \eqref{eq:Fcal-asym-con} fix the large-distance behavior, while the regularity conditions \eqref{eq:regular-AB}, together with the removal of the $1/r$ term in the expression of the metric function $A(r)$, fix the relation between the ADM mass and the scalar hair of the solution. 
The forthcoming analysis and examples in Sec.~\ref{sec:branch3} illustrate this mechanism explicitly.
In Secs. \ref{Sec:Axial} and \ref{Sec:Polar} we investigate the stability of solutions in this branch.

Before discussing particular solutions within this family of theories, it is useful to first understand how the non-reciprocal metric can be mapped to its reciprocal counterpart through a disformal transformation. 
Disformal transformations provide a particularly useful redefinition of the metric tensor and can also be used to obtain new compact-object solutions.
On top of that, as shown in \cite{Langlois:2022ulw, Tomikawa:2021pca, Charmousis:2025xug, Charmousis:2026bhe}, disformal mappings offer valuable insight into the perturbative structure and stability properties of black-hole solutions. 
In particular, as demonstrated in \cite{Langlois:2022ulw}, there exists a remarkable correspondence between the dynamics of axial perturbations in quadratic DHOST theories and those of axial modes in General Relativity, albeit propagating in a different geometry.
A similar behavior has also been found in the monopole case of the polar sector in \cite{Charmousis:2026bhe}, and will be used later, in Sec.~\ref{Sec:Polar}, to study the monopole stability of the solutions.

%==========================================
%Unique disformal map to reciprocal black holes
%==========================================

\section{Unique purely disformal transformation to reciprocal black holes}
\label{Sec:Disformal-Frame}

In this section, we demonstrate that the reciprocal counterparts of the preceding solutions can be uniquely determined through a purely disformal transformation~\cite{Zumalacarregui:2013pma, Gleyzes:2014qga, BenAchour:2016cay, Langlois:2020xbc}. 
In particular, every well-behaved non-reciprocal black-hole solution obtained from the preceding construction can be uniquely mapped, via a purely disformal transformation, to a reciprocal solution belonging to a different DHOST theory.
Throughout this section, untilded quantities refer to the seed DHOST theory and its solutions, while tilded quantities refer to the target DHOST theory and its corresponding solutions. 

At first, starting from a metric of the form \eqref{eq:metric}, with $B^2(r)=Z^2(X(r))$, as dictated by \eqref{eq1-new}, and the scalar field ansatz \eqref{eq:scalar}, one can readily verify that there is only a unique purely disformal transformation that renders the original metric \eqref{eq:metric} into the reciprocal form $\widetilde{g}_{tt}\,\widetilde{g}_{rr}=-1$. 
This transformation is
\begin{equation}
    \widetilde{g}_{\mu\nu} =  g_{\mu\nu} + D(X)\,\partial_\mu\phi\, \partial_\nu\phi,
    \qquad
    D(X) = \frac{1}{2X} \left(1-\frac{1}{Z^2}\right)\,,
    \label{eq:dis}
\end{equation}
where the kinetic term $X$ corresponds to the original/untilded frame.
A legitimate disformal transformation is characterized by $1-2XD(X)>0$, which is automatically satisfied in this case since
\begin{equation}
    1-2XD(X)=\frac{1}{Z^2(X)}\,.
    \label{eq:legit-dis}
\end{equation}
By applying the above transformation, and after performing the time redefinition
\begin{equation}
    t = \tau + \int \dd r\, \frac{D(X)\,q\,\psi'(r)}{A(r)-q^2D(X)}
\end{equation}
we obtain the line element
\begin{equation}
    \dd \widetilde{s}^2 = -\widetilde{A}(r)\dd \tau^2 + \frac{\dd r^2}{\widetilde{A}(r)} + r^2\dd \Omega^2,
    \qquad
    \widetilde{A}(r)=A(r)-q^2D(X)\,.
    \label{eq:metric-rec}
\end{equation}

The scalar field and its kinetic term, after the transformation and in the new coordinate system, are given by
\begin{equation}
    \phi(\tau,r) = q\tau + \int \dd r\,\frac{A(r)}{\widetilde{A}(r)}\psi'(r),
    \qquad 
    \widetilde{X} = \frac{X}{1-2XD(X)} = XZ^2\,,
    \label{eq:Xnew}
\end{equation}
while the model functions of the target DHOST theory are determined through the following expressions:
\begin{equation}
    \begin{gathered}
     \widetilde{G}_2(X) = \frac{G_2(X)}{\sqrt{1-2XD(X)}},
     \qquad 
     \widetilde{G}_4(X) = \frac{G_4(X)}{\sqrt{1-2XD(X)}}\,,
     \\[5mm]
     \widetilde{F}_4(X) = \frac{\big[1-2XD(X)\big]^{5/2}}{2}\frac{2F_4(X)-Z(X)D_X(X)}{1+2X^2D_X(X)}\,.
     \end{gathered}
     \label{eq:G2G4F4new-Xold}
\end{equation}
Notice that the model functions in the target (tilded) DHOST frame are given in terms of the seed kinetic term $X$, instead of $\widetilde{X}$, since we have not yet specified a particular functional form for $D(X)$ that would allow us to invert the function $\widetilde{X}=\widetilde{X}(X)$ and express the transformed model functions in terms of $\widetilde{X}$.
Hence, whenever the disformal factor $D(X)$ is such that the function $\frac{X}{1-2XD(X)}$ is invertible, the preceding model functions can always be expressed in terms of the new kinetic term $\widetilde{X}$.
Combining the above relations, one can verify that
\begin{equation}
    \widetilde{Z}(X) = \sqrt{1-2XD(X)}\,Z(X)\,.
    \label{eq:Z-dis-Xold}
\end{equation}
Notice now that for $Z(X(r))>0$, the combination of Eqs.~\eqref{eq:legit-dis} and \eqref{eq:Z-dis-Xold} indicates that
\begin{equation}
    \widetilde{Z}(X) = \widetilde{Z}(\widetilde{X}) = 1\,,
    \label{eq:Znew}
\end{equation}
which is expected and related to the results in \cite{Bakopoulos:2023sdm}.

%==========================================
%Reciprocal-Branch II
%==========================================

\subsection{Reciprocal frame of Branch II}
\label{Sec:Rec-BranchII}
It is now useful to understand how Branch II transforms under \eqref{eq:dis}. 
Making use of Eqs.~\eqref{eq:br2-G2-G4}, \eqref{eq:legit-dis}, \eqref{eq:Xnew}, and \eqref{eq:G2G4F4new-Xold}, one immediately finds
\begin{equation}
    \widetilde G_2(\widetilde X)=g_2,
    \qquad
    \widetilde G_4(\widetilde X)=g_4\,.
    \label{eq:b2-dis-G2-G4}
\end{equation}
Moreover, since the transformation gives $\widetilde Z=1$, the defining relation for $\widetilde Z$, i.e. \eqref{eq:def-Z}, implies that
\begin{equation}
    \widetilde F_4(\widetilde X) = \frac{g_4-1}{4\widetilde X^2}\,.
    \label{eq:b2-dis-F4}
\end{equation}
It is important to distinguish two separate properties of the reciprocal frame. 
First, at the level of the theory, Eq.~\eqref{eq:b2-dis-F4} shows that the beyond-Horndeski interaction vanishes only for $g_4=1$. 
Hence, for $g_4=1$ the reciprocal-frame theory reduces to GR (up to the contribution of $g_2$), whereas for $g_4\neq1$ it remains a nontrivial DHOST theory. 
Second, this condition should not be confused with the asymptotic normalization of the reciprocal geometry. 
Whether the metric approaches the standard asymptotically flat form, without a solid-angle deficit or excess, depends on the large-distance behaviour of the reciprocal lapse and therefore on the combined asymptotic behaviour of $g_4$ and $\widetilde X(r)$.

Indeed, from Eqs.~\eqref{Branch2MetricA} and \eqref{eq:metric-rec}, the Branch-II reciprocal metric takes the form
\begin{equation}
    \dd \widetilde{s}^2 = -\widetilde{A}_{\mathrm{II}}(r)\dd \tau^2 + \frac{\dd r^2}{\widetilde{A}_{\mathrm{II}}(r)} + r^2\dd\Omega^2\,,
\end{equation}
with
\begin{equation}
    \widetilde{A}_{\mathrm{II}}(r) = g_4 + \frac{C_0}{r} + \frac{g_2}{6}r^2 +
    (1-g_4)\frac{q^2}{2\widetilde X(r)}\,,
    \label{eq:b2-reciprocal-A}
\end{equation}
where we used $\widetilde X=XZ^2=Xg_4^2/G_4^2$. 
Thus, $g_2$ contributes as an effective cosmological constant, while $C_0$ is an integration constant. 
In the asymptotically flat case, one removes the $r^2$ term by setting $g_2=0$. 
The absence of a conical deficit or excess then requires the standard normalization $\widetilde{A}_{\mathrm{II}}\to 1$. 
This can be achieved with $g_4=1$, but it can also be achieved for $g_4\neq1$ by an appropriate large-distance behavior of $\widetilde X(r)$. 
Therefore, $g_4=1$ should be understood as the condition that the reciprocal-frame theory becomes GR, while the absence of an asymptotic solid-angle defect is a condition on the combined asymptotic behavior of $g_4$ and $\widetilde X(r)$.

More generally, the reciprocal Branch-II theory belongs to the special
khronometric family identified in Ref.~\cite{BenAchour:2016cay}.
Using our convention $X\equiv-(\partial\phi)^2/2$ and the quadratic
DHOST functions $A_i$ introduced above, after normalizing the coefficient
of the Ricci scalar to unity, the quadratic DHOST functions take the form
\begin{equation}
    \widetilde{A}_1=-\widetilde{A}_2=-\frac{c_2}{2\widetilde{X}},
    \qquad
    \widetilde{A}_3=-\widetilde{A}_4=\frac{c_2}{2\widetilde{X}^2},
    \qquad
    \widetilde{A}_5=0,
    \qquad
    c_2=\frac{g_4-1}{g_4}\,.
\end{equation}
These are precisely the functions characterizing the khronometric
family given in Eq.~(7.9) of Ref.~\cite{BenAchour:2016cay}, rewritten
in our conventions.
Importantly, this is a special degenerate subclass of khronometric
gravity rather than a generic khronometric theory.

As shown in Ref.~\cite{BenAchour:2016cay}, this family is closed under
the one-parameter conformal--disformal transformation
\begin{equation}
  \hat{g}_{\mu\nu}
    =
    a\,\widetilde{g}_{\mu\nu}
    +\frac{a-a^{-1}}{2\widetilde X}\,
    \partial_\mu\phi\,\partial_\nu\phi,
    \qquad
   0<|a|<+\infty\,,
\end{equation}
under which the parameter $c_2$ transforms according to
\begin{equation}
    c_2\longrightarrow (c_2-1)a^2+1\,.
\end{equation}
For the Branch-II value $c_2=(g_4-1)/g_4$, with $g_4>0$, choosing $a^2=g_4$ maps the theory to $c_2=0$, namely to GR. The constant $g_2$
provides only a cosmological-constant contribution and does not
introduce an additional scalar degree of freedom. Therefore, whenever
this disformal transformation is regular and invertible,
Branch~II propagates only the two tensor degrees of freedom corresponding
to the two graviton polarizations, with no independent scalar mode. This two-degree-of-freedom structure provides the dynamical justification for the terminology introduced above: the nontrivial Branch-II scalar profile represents khronometric hair associated with the preferred foliation, rather than genuine scalar hair associated with an independent propagating scalar degree of freedom.

The special case $g_4=1$ is particularly simple: the reciprocal-frame
theory is already General Relativity, up to the cosmological-constant
contribution $g_2$, without requiring any further disformal
transformation. In this case,
\begin{equation}
    \widetilde G_2(\widetilde X)=g_2\,,
    \qquad
    \widetilde G_4(\widetilde X)=1\,,
    \qquad
    \widetilde F_4(\widetilde X)=0\,,
\end{equation}
and the reciprocal-frame action reduces to General Relativity with an effective cosmological constant $\Lambda_{\rm eff}=-g_2/2$.
Accordingly, in this special case, the reciprocal vacuum solution is the Schwarzschild--(anti-)de Sitter geometry.

%==========================================
%Reciprocal-Branch III
%==========================================

\subsection{Reciprocal frame of Branch III}

Turning now to Branch III, one readily notices that from \eqref{eq:legit-dis}, \eqref{eq:Xi}, and \eqref{eq:Xnew}, the kinetic term in the reciprocal frame is fixed by
\begin{equation}
    \widetilde{X} = \Xi = \frac{q^2/2}{1+r^2/\lambda^2}\,.
    \label{eq:Xtilde-branchIII}
\end{equation}
This particular kinetic profile for black-hole solutions in beyond Horndeski theories was first found in \cite{Bakopoulos:2023fmv} and subsequently employed in later works, such as \cite{Baake:2023zsq, Bakopoulos:2023sdm, Bakopoulos:2024ogt, Fathi:2025uwa, Charmousis:2025xug, Myung:2025afs, Candan:2025fbl, Konoplya:2026cus, Huang:2026ubc, Charmousis:2026bhe, Boumaza:2026dix, Antoniou:2026vrm}.
Using Eq.~\eqref{eq:Xtilde-branchIII} together with \eqref{eq:G2-F}, \eqref{eq:G2-G4}, \eqref{eq:G2G4F4new-Xold}, \eqref{eq:Znew}, and the defining expression for $\widetilde Z$, one obtains
\begin{equation}
    \begin{gathered}
        \widetilde{G}_2(\widetilde{X}) = \mathcal{F}(\widetilde{X})\,,
        \\[2mm]
        \widetilde{G}_4(\widetilde{X}) = 1+\frac{\lambda^2}{2}\mathcal{F}(\widetilde{X})\,,
        \\[2mm]
        \widetilde{F}_4(\widetilde{X}) = \frac{\widetilde{G}_4(\widetilde{X}) - 2\widetilde{X}\widetilde{G}_{4\widetilde{X}}(\widetilde{X}) -1}{4\widetilde{X}^2}\,.
        \label{eq:G2G4F4new-Xnew}
    \end{gathered}
\end{equation}

Interestingly, by using Eqs.~\eqref{eq:Xi}, \eqref{eq:dis}, and the fact that $B^2(r)=Z^2(X(r))$, one can easily verify that 
\begin{equation}
    \bigg(1+\frac{r^2}{\lambda^2}\bigg)\big[B^2(r)-1\big] = q^2 D(X)\,,
\end{equation}
which by its turn means that
\begin{equation}
    \widetilde{A}(r) = 1 -\frac{C_0}{r} + \frac{1}{r}\int \dd r\, r^2 \big[2\,\widetilde{X}\, \mathcal{F}_{\widetilde{X}} -\mathcal{F}(\widetilde{X})\big]\,.
    \label{eq:A-rec-frame}
\end{equation}
We therefore realize that, in the reciprocal frame, the initial choice of $Z(X)$ does not make any difference.
The choice of $Z(X)$ affects the theory and the solution only in the non-reciprocal frame. 
Indeed, as we have shown above, in the reciprocal frame $\widetilde{Z}(\widetilde{X})=1$.
The solution in the reciprocal frame depends only on the functional form of $\mathcal{F}$.

Hence, the non-reciprocal Branch-III frame is mapped to the reciprocal DHOST theory \eqref{eq:G2G4F4new-Xnew}, with the line element \eqref{eq:metric-rec} characterized by the metric function \eqref{eq:A-rec-frame}, and the fixed kinetic profile \eqref{eq:Xtilde-branchIII}. 
This map is useful for understanding the geometry and the perturbations, but it should not be interpreted as a physical equivalence once matter is included. Matter minimally coupled to the original metric becomes disformally coupled in the reciprocal frame, and the two frames may therefore lead to different phenomenology.

%==========================================
%Branch II-solutions
%==========================================

\section{Regular black holes with khronometric hair}
\label{Sec:BranchII}
In this section, we construct an exact family of asymptotically flat
Branch-II black-hole solutions for which both the spacetime geometry and the
scalar field are smooth at the center.

As established in Sec.~\ref{Sec:Rec-BranchII}, Branch~II belongs to a
special two-degree-of-freedom khronometric subclass related to GR by an
invertible field redefinition. 
Black-hole solutions in Einstein--aether theory and in the infrared limit of
Ho\v{r}ava gravity have been studied previously~\cite{Barausse:2011pu}. Both
theories are closely related to khronometric theories.
The infrared limit of non-projectable Ho\v{r}ava gravity is a khronometric theory, while
Einstein--aether theory reduces to it when the aether is hypersurface-orthogonal.
More recently, regular black-hole and
horizonless geometries have also been explored in khronometric gravity
\cite{Mazza:2023iwv}. The solutions of~\cite{Mazza:2023iwv}, however, were introduced as possible
regularizations of a singular khronometric black hole rather than exact
vacuum solutions of the low-energy theory. Here we instead construct
an explicit exact Branch-II family for which both the metric and scalar
field are smooth at the center.

We restrict ourselves to
\begin{equation}
    g_4=1,\qquad g_2=C_0=0.
\end{equation}
Equations~\eqref{eq:br2-G2-G4}, \eqref{eq:def-Z},
\eqref{Branch2MetricA}, and \eqref{Branch2MetricB} then give
\begin{equation}
    G_2=0,\qquad
    Z=\frac{1}{G_4},\qquad
    F_4=\frac{G_4-2XG_{4X}-G_4^{-1}}{4X^2},
    \label{eq:BII-smooth-general-theory}
\end{equation}
together with
\begin{equation}
    A(r)=1+\frac{q^2}{2X(r)}
    \left[1-G_4^2(X(r))\right],
    \qquad
    B^2(r)=\frac{1}{G_4^2(X(r))}.
    \label{eq:BII-smooth-general-metric}
\end{equation}

Before specifying the solution, let us notice an important consequence
of regularity of the scalar field. From the definition of $X$,
\begin{equation}
    \psi'^2=
    \frac{B^2}{A^2}\left(q^2-2AX\right).
    \label{eq:BII-smooth-scalar-general}
\end{equation}
A spherically symmetric scalar field is smooth in the center only if
$\psi'(0)=0$. Hence, denoting $X(0)=X_0$, smoothness requires
\begin{equation}
    q^2=2A(0)X_0.
    \label{eq:BII-smooth-center-scalar}
\end{equation}
The family below is chosen such that this condition coincides with
the metric regularity condition $A(0)=B^2(0)$.

%=========================================================
\subsection{Explicit regular black-hole solution}
%=========================================================

As an explicit example, consider the theory
\begin{equation}
    G_4(X)=
    \sqrt{\frac{1+2\eta^2 X^2}{1+\eta^2 X^2}},
    \qquad
    F_4(X)=
    \frac{\eta^2(\eta^2 X^2-1)}
    {4(1+\eta^2 X^2)^{3/2}\sqrt{1+2\eta^2 X^2}},
    \label{eq:BII-smooth-model-I}
\end{equation}
where $\eta$ has dimensions of $(\mathrm{length})^2$. Since the coupling
functions depend only on $\eta^2$, we take $\eta>0$ without loss of
generality.  
We now assume the kinetic term of the form
\begin{equation}
    X(r)=X_0\left[
    1+\left(\frac{\eta^2 q^2X_0\,r}{4M}\right)^2
    \right]^{-1/2},
    \label{eq:BII-smooth-X-I}
\end{equation}
where $M>0$ and $q>0$ are the independent mass and khronometric-hair
parameters. The central value $X_0\equiv X(0)$ is not an additional
free parameter, but is determined by regularity as the unique positive
root of
\begin{equation}
    2X_0\left(1+\eta^2 X_0^2\right)
    -q^2\left(1+2\eta^2 X_0^2\right)=0.
    \label{eq:BII-smooth-X0-I}
\end{equation}
The normalisation of the profile in Eq.~\eqref{eq:BII-smooth-X-I}
is chosen so that $M$ coincides with the ADM mass, as verified below.

Substitution into Eq.~\eqref{eq:BII-smooth-general-metric} gives
\begin{equation}
    A(r)=
    1-\frac{\eta^2 q^2X(r)}
    {2\left[1+\eta^2 X^2(r)\right]},
    \qquad
    B^2(r)=
    \frac{1+\eta^2 X^2(r)}
    {1+2\eta^2 X^2(r)}.
    \label{eq:BII-smooth-metric-I}
\end{equation}
The geometry is regular at the center. Indeed,
$X(r)=X_0-\eta^4q^4X_0^3r^2/(32M^2)+\mathcal O(r^4)$, while
Eq.~\eqref{eq:BII-smooth-X0-I} implies
\[
A(0)=B^2(0)
=\frac{1+\eta^2 X_0^2}{1+2\eta^2 X_0^2}.
\]
Both metric functions admit expansions in even powers of $r$ and
satisfy the regularity conditions of Appendix~\ref{App:Reg-Cond}.

The scalar field is regular as well. Using
Eq.~\eqref{eq:BII-smooth-X0-I}, the combination entering its radial
profile can be written as
\begin{equation}
\begin{aligned}
q^2-2AX
={}&\frac{2(X_0-X)}
{(1+\eta^2 X^2)(1+2\eta^2 X_0^2)}
\left[
1+\eta^2(X^2-XX_0+X_0^2)
+2\eta^4X^2X_0^2
\right].
\end{aligned}
\label{eq:BII-smooth-reality-I}
\end{equation}
Since $0<X(r)\leq X_0$, this expression is non-negative everywhere
and vanishes only at the center. Moreover,
$X_0-X=\mathcal O(r^2)$ as $r\to0$, and therefore
$q^2-2AX=\mathcal O(r^2)$. It follows from
Eq.~\eqref{eq:BII-smooth-scalar-general} that
$\psi'(r)=\mathcal O(r)$ and hence
$\psi(r)=\psi_0+\mathcal O(r^2)$. More precisely, near the center
$\psi'(r)$ is $r$ times a function analytic in $r^2$, so the scalar
field admits a smooth extension through the center.

The solution is also asymptotically flat. From
Eq.~\eqref{eq:BII-smooth-X-I},
$X(r)=4M/(\eta^2q^2r)+\mathcal O(r^{-3})$, which gives
\begin{equation}
    A(r)
    =1-\frac{2M}{r}+\mathcal O(r^{-2}),
    \qquad
    B^2(r)
    =1+\mathcal O(r^{-2}).
    \label{eq:BII-smooth-asym-I}
\end{equation}
Thus the asymptotic geometry is Minkowski and $M$ is the ADM mass.

Finally, $A(r)$ is positive both at the center and at spatial
infinity. Since $X(r)$ decreases monotonically from $X_0$ to zero,
$A(r)$ develops an interior minimum at $X(r)=1/\eta$
provided that $\eta X_0>1$. Its value is
$A^{\rm min}=1-\eta q^2/4$. Therefore, if
\begin{equation}
    \eta q^2>4,
    \label{eq:BII-smooth-BH-I}
\end{equation}
which also ensures $\eta X_0>1$, the minimum is negative and the
geometry possesses two nonextremal horizons. The solution therefore
describes a regular black hole, with $\eta q^2=4$ corresponding to
the extremal configuration.

%=========================================================
\subsection{Validity of the disformal map and khronometric hair}
%=========================================================
For completeness, we now verify explicitly that the disformal map
\eqref{eq:dis} remains valid over the full radial domain of this solution.
For this family,
\begin{equation}
    \widetilde X
    =X\,\frac{1+\eta^2 X^2}{1+2\eta^2 X^2},
    \qquad
    \frac{\dd\widetilde X}{\dd X}
    =
    \frac{1+\eta^2 X^2+2\eta^4X^4}
    {(1+2\eta^2 X^2)^2}>0,
\end{equation}
Together with $1-2XD=G_4^2>0$, this confirms explicitly that the map remains
regular throughout the solution.

The apparent divergence of the static scalar profile at a metric
horizon is only a coordinate effect. Introducing the ingoing
coordinate $v=t+\int(B/A)\,\dd r$, the scalar can be written as
\begin{equation}
    \phi=qv-
    \int\frac{2BX}{q+\sqrt{q^2-2AX}}\,\dd r,
    \label{eq:BII-smooth-ingoing}
\end{equation}
whose integrand remains finite at every simple horizon.

The solution therefore possesses a regular metric, a smooth scalar
field at the center and across the future horizons, and an
asymptotically flat geometry. Importantly, regularity does not relate
$q$ to the ADM mass. For a fixed theory, $X_0$ is determined by $q$
through Eq.~\eqref{eq:BII-smooth-X0-I}, while the profile
\eqref{eq:BII-smooth-X-I} depends directly on the independent
parameters $M$ and $q$. Hence $q$ constitutes
\emph{primary khronometric hair}.

To give this hair a physical interpretation, we take the original
metric $g_{\mu\nu}$ to be the metric minimally coupled to test matter.
The parameter $q$ then changes the geometry probed by matter while
remaining unrelated to a nonzero conserved shift charge, since
$J^\mu=0$ throughout Branch~II. For the present $g_4=1$ and
$g_2=C_0=0$ configurations,
\begin{equation}
    D(X)=\frac{1-G_4^2(X)}{2X},
    \qquad
    \widetilde A=A-q^2D=1,
    \qquad
    \widetilde B=1.
    \label{eq:BII-smooth-flat-frame}
\end{equation}
Thus the reciprocal geometry is Minkowski even though the
original-frame geometry contains nontrivial horizons and an
independent hair parameter.

There is no contradiction between these two descriptions. An
invertible field redefinition maps the gravity sector to its
reciprocal representation, but matter minimally coupled to the
original metric becomes disformally coupled in the transformed
variables. Consequently, its trajectories are not ordinary Minkowski
geodesics. The same branch can therefore have a nontrivial geometry
for matter despite its simple gravity-only description. Here we use
matter only as a probe; dynamical matter can modify the constraint
structure~\cite{Deffayet:2020qpd}, and a complete analysis of such
matter couplings lies beyond the scope of the present work.

As we shall discuss in Sec.~\ref{Sec:Axial}, the odd-parity linear perturbations
of Branch~II are governed by the corresponding GR problem on its effective
metric. For the present $g_2=C_0=0$ solutions, the effective metric is
Minkowski, so the analysis of the axial stability becomes immediate.

%==========================================
%Singular and regular hairy black holes in Branch III
%==========================================

\section{Singular and regular black holes with scalar hair in Branch III}
\label{sec:branch3}

In this section, we construct exact black-hole solutions for Branch III. 
Under the simple assumption $\mathcal{G}_2/\mathcal{G}_4=2/\lambda^2$, the equations of motion \eqref{eq1-new}--\eqref{eq3-new} can be solved implicitly. 
To obtain explicit solutions for individual theories, we proceed as follows:
\begin{tcolorbox}[enhanced,%title=My title,
top=5pt, bottom=5pt, left=10pt, right=10pt,
interior hidden]
    \begin{enumerate}[i)]
        \item Choose explicit forms for the functions $Z(X)$ and $\mathcal{F}(\Xi)$.
        
        \item Given $Z(X)$, $\mathcal{F}(\Xi)$, and the definition of $\Xi(X)$, determine $G_2(X)$ and $G_4(X)$ from Eqs.~\eqref{eq:G2-F} and \eqref{eq:G2-G4}, respectively. After that, by using the defining equation \eqref{eq:def-Z} for the auxiliary function $Z$, find the model function $F_4(X)$.
        
        \item Solve Eq.~\eqref{eq:Xi} algebraically for $X(r)$, thereby obtaining the radial profile of the kinetic term.
        
        \item Use Eq.~\eqref{eq1-new} to determine the metric function $B(r)$ through $Z(X(r))$.
        
        \item Finally, with all previous results at hand, solve \eqref{eq:A-sol} for the metric function $A(r)$.
\end{enumerate}
\end{tcolorbox}

%==========================================
%Family of solutions generated from a fixed Z profile
%==========================================

\subsection{Family of solutions generated from a specific nontrivial $Z(X)$ profile}

Let us now turn to explicit examples that yield singular black-hole solutions endowed with primary scalar hair as well as regular solutions characterized by secondary scalar hair.
Throughout this section, we will focus mainly on solutions originating from the same ansatz for the auxiliary function $Z(X)$, namely
\begin{equation}
    Z(X)=\frac{1}{\sqrt{1-\eta^2 X^2}}\,,
    \label{eq:Z-X}
\end{equation}
where $\eta$ is a coupling constant with dimensions of $({\rm length})^2$. 
Substituting \eqref{eq:Z-X} into \eqref{eq:Xi}, one can easily obtain the radial profile of the kinetic term by solving the resulting quadratic polynomial for $X$.
By doing so, we find
\begin{equation}
    X(r)=\frac{q^2}{(1+r^2/\lambda^2)[1+\Delta(r)]}\,,\qquad \Delta(r)\equiv \sqrt{1+\frac{\eta^2 q^4}{(1+r^2/\lambda^2)^2}}\,.
    \label{eq:X-r}
\end{equation}
In the above, we identified $X(r)$ with the positive root $X_{+}$, which yields a finite and positive-definite expression for all $r\in[0,+\infty)$. 
In contrast, the negative branch $X_{-}$ diverges as $r \to +\infty$.
Additionally, from \eqref{eq1-new}, \eqref{eq:Xi}, and \eqref{eq:X-r}, we can immediately determine the metric function $B(r)$, for which it holds
\begin{equation}
    [B(r)]^2=[Z(X(r))]^2=\frac{1+\Delta(r)}{2}\,.
    \label{eq:B-r}
\end{equation}
It is clear from its expression that $B^2(r)$ is finite for all $r\in[0,+\infty)$, and as one can easily verify, it is also a monotonically decreasing function in its domain.
Therefore, the highest value of $B^2(r)$ is at $r\to 0$, while its lowest value is at the asymptotic limit $r\to +\infty$.
The same holds for the function $Z(X(r))$, hence, as long as the auxiliary function $\mathcal{F}(\Xi)$ has no poles, then all model functions of the resulting theory will be well-defined for all $r\in(0,+\infty)$.
The asymptotic behavior of $B^2(r)$ is of the form 
\begin{equation}
    B^2(r) \to 1 + \frac{\eta^2 q^4 \lambda^4}{4r^4} + \mathcal{O}(r^{-6})\,,
\end{equation}
which respects the asymptotic condition \eqref{eq:B-asym-con}, while at the other extreme, i.e. as $r\to 0$, $B^2(r)$ takes the value $(1+\sqrt{1+\eta^2q^4})/2$.
Given the expression for $Z(X)$, obtaining a specific beyond-Horndeski theory together with the corresponding compact-object solutions requires specifying the functional form of $\mathcal{F}(\Xi)$. 
In the following subsections, we explore several possible choices for $\mathcal{F}(\Xi)$.

%==========================================
%Hypergeometric solution
%==========================================

\subsubsection{A family of black-hole solutions from power-law model functions}
\label{power_law_model}

As a first example, let us consider a power-law ansatz for $\mathcal{F}$, namely
\begin{equation}
    \mathcal{F}(\Xi) = -g_2 (\lambda^2 \Xi)^p\,,
    \label{eq:F-power}
\end{equation}
where $p\in \mathbb{R}$, and $\lambda$ is the same coupling constant that was introduced in Eq.~\eqref{eq:G2-G4}.
The constant $\lambda$ is used to render the quantity inside parentheses dimensionless, and thus the dimensions of $\mathcal{F}$ are carried by the coupling constant $g_2$.
Using Eqs. \eqref{eq:Z-X}, \eqref{eq:F-power} in \eqref{eq:G2-F}, \eqref{eq:G2-G4}, and \eqref{eq:def-Z}, it is straightforward to determine the model functions of the theory: 
\begin{gather}
    G_2(X)=-g_2\, \frac{\big(\lambda^2 X\big)^p}{ \big(1-\eta^2 X^2\big)^{p-1/2}}\,, \qquad 
    G_4(X) = \sqrt{1-\eta^2 X^2} - \frac{g_2\lambda^2}{2}\, \frac{\big(\lambda^2 X\big)^p}{ \big(1-\eta^2 X^2\big)^{p-1/2}}\,,
    \label{eq:model-power-G2G4}\\[5mm]
    F_4(X) = \frac{\eta^2}{4\sqrt{1-\eta^2 X^2}} + \frac{(2p-1)g_2 \lambda^{6}}{8} \frac{\big(1+\eta^2 X^2\big) \big(\lambda^2 X\big)^{p-2}}{(1-\eta^2 X^2)^{p+1/2}}\,.
    \label{eq:model-power-F4}
\end{gather}
Furthermore, Eq.~\eqref{eq:A-sol} gives us the metric function $A(r)$, which is evaluated to be
\begin{equation}
    A(r) = 1 -\frac{C_0}{r} + \frac{1}{2}\bigg(1+\frac{r^2}{\lambda^2}\bigg)[\Delta(r)-1] - \frac{g_2}{3}\,r^2\,\frac{(2p-1)(\lambda^2 q^2)^{p}}{2^p}\, \,_2F_1\bigg(\frac{3}{2},p\,;\frac{5}{2};-\frac{r^2}{\lambda^2}\bigg)\,.
    \label{eq:A-sol-power}
\end{equation}
According to Eq.~\eqref{eq:Fcal-asym-con} and Appendix \ref{App:Asym-III}, for $p=1/2$ and $ p>3/2$, the asymptotic behavior of $A(r)$ is flat and generates a Schwarzschild tail; therefore, an ADM mass can always be defined for the resulting compact object.
Additionally, for $p>3/2$, the hypergeometric function can be recast in terms of a real argument lying between zero and unity, making its expansion into a convergent power series obvious:
\begin{equation}
    _2F_1\bigg(\frac{3}{2},p\,;\frac{5}{2};-\frac{r^2}{\lambda^2}\bigg)=\frac{\,_2F_1\bigg(1,p\,;\frac{5}{2};\frac{r^2/\lambda^2}{1+r^2/\lambda^2}\bigg)}{\big(1+r^2/\lambda^2\big)^p}
    = \frac{3\sqrt{\pi}}{4\Gamma(p)} \sum_{n=0}^{\infty} \frac{\Gamma(p+n)}{\Gamma(5/2+n)}\frac{\big(r^2/\lambda^2\big)^n}{\big(1+r^2/\lambda^2\big)^{n+p}}\,.
\end{equation}
Notice that once we set $q=0$, the kinetic term vanishes as well.
In this case, the solution reduces to its GR limit, and the spacetime is described by the Schwarzschild geometry, as expected.
Finally, for $p=2$ or $p=\frac{2k+1}{2}\, \forall k\geq 2,\ k\in \mathbb{N}$, the above hypergeometric function reduces to elementary analytic functions.
We will exploit this property to study exact black-hole solutions in this family of theories.

%==========================================
%p=1/2 - Simplest case
%==========================================

\subsubsection{The simplest case: $p=1/2$}
\label{Sec:p-1/2}

The simplest member of this family corresponds to $p=1/2$, for which the prefactor $(2p-1)$ multiplying the hypergeometric term in Eq.~\eqref{eq:A-sol-power} vanishes identically, so that $A(r)$ collapses to the closed form
\begin{equation}
    A(r) = 1 - \frac{C_0}{r} -\frac{1}{2}\bigg(1+\frac{r^2}{\lambda^2}\bigg)\Bigg[1-\sqrt{1+\frac{\eta^2 q^4}{(1+r^2/\lambda^2)^2}}\ \Bigg]\,.
    \label{eq:A-sol-p12}
\end{equation}
Notice that we used Eq.~\eqref{eq:X-r} in \eqref{eq:A-sol-power}.
Expanding now the above function at large $r$ gives
\begin{equation}
    A(r) = 1 - \frac{C_0}{r} + \frac{\eta^2 q^4}{4}\frac{\lambda^2}{r^2} + \mathcal{O}(r^{-4})\,, \qquad r\to+\infty\,,
    \label{eq:A-asym-p12}
\end{equation}
confirming that the solution is asymptotically flat with a Schwarzschild
tail. In this case, $C_0=2M$, where $M$ is the ADM mass, while the
subleading $r^{-2}$ term encodes the contribution of the primary
scalar hair $q$.
The model functions of the theory can be easily determined by setting $p=1/2$ in Eqs.~\eqref{eq:model-power-G2G4}, \eqref{eq:model-power-F4}.

Let us now turn to the reciprocal frame obtained from this solution using the disformal transformation \eqref{eq:dis}.
Using the analysis performed in Sec.~\ref{Sec:Disformal-Frame}, it is straightforward to 
calculate the form of the model functions in the reciprocal frame, namely
\begin{equation}
    \widetilde{G}_2(\widetilde{X}) = -g_2 \big(\lambda^2 \widetilde{X}\big)^{1/2}\,,\qquad
    \widetilde{G}_4(\widetilde{X}) = 1 - \frac{g_2 \lambda^2}{2}\big(\lambda^2 \widetilde{X}\big)^{1/2}\,,\qquad 
    \widetilde{F}_4(\widetilde{X})=0\,.
\end{equation}
Notice that the above theory belongs to the Horndeski family since the beyond-Horndeski contribution $\widetilde{F}_4$ is identically zero.
The metric function in the reciprocal frame is evaluated through Eq.~\eqref{eq:metric-rec} and describes the Schwarzschild geometry.
Note that this reciprocal stealth Schwarzschild solution was first realized in \cite{Bakopoulos:2023fmv}.

%==========================================
%p=5/2 - Nonreciprocal extension of Bardeen solution
%==========================================

\subsubsection{Nonreciprocal extension of Bardeen solution}
\label{Sec:Nonrec-Bardeen}

The solution we will consider in this section is the one obtained for $p=5/2$, for which the metric function $A(r)$ can be evaluated using \eqref{eq:A-sol-power}, and \eqref{eq:X-r}.
By doing so, one finds
\begin{equation}
    A(r) = 1 - \frac{2M}{r} + \frac{g_2 \lambda^8 q^5}{3\sqrt{2}} \frac{1}{r} \bigg(1- \frac{r^3}{(r^2 +\lambda^2)^{3/2}}\bigg)-\frac{1}{2}\bigg(1+\frac{r^2}{\lambda^2}\bigg)\Bigg[1-\sqrt{1+\frac{\eta^2 q^4}{(1+r^2/\lambda^2)^2}}\ \Bigg]\,.
    \label{eq:A-sol-5/2}
\end{equation}
In the above expression, $M$ denotes the ADM mass of the solution.
According to the preceding discussion, it is clear that the only two independent parameters here are the ADM mass $M$ and the scalar hair $q$, while all remaining constants are coupling constants of the theory. 
The spacetime geometry is thus characterized by the metric ansatz \eqref{eq:metric} with the metric functions $A(r)$ and $B^2(r)$ given by Eqs.~\eqref{eq:A-sol-5/2} and \eqref{eq:B-r}, respectively. 
It is also apparent that, for $q=0$, we recover the Schwarzschild geometry, and thus the solution possesses a natural GR limit.

Asymptotically, the metric function $A(r)$ behaves as

\begin{equation}
    A(r)\xrightarrow{r\to +\infty} 1 - \frac{2M}{r} + \frac{q^4\eta^2\lambda^2}{4r^2}
    +\frac{g_2\, q^5\lambda^{10}}{2\sqrt{2}\,r^3}+\mathcal{O}\bigg(\frac{1}{r^4}\bigg)\,.
    \label{eq:A-asym-power}
\end{equation}
This structure resembles the asymptotic form of the Reissner-Nordstr\"{o}m solution. 
However, unlike the standard Reissner-Nordstr\"{o}m case, where the $1/r^2$ term is sourced by an electric charge, here it is generated by the primary scalar hair $q$.

On the opposite limit, the expansion at $r\to 0$ takes the form
\begin{equation}
    A(r)\xrightarrow{r\to 0}
    \frac{-12M + \sqrt{2}\, g_2\, q^5 \lambda^{8}}{6}\frac{1}{r}
    + \frac{1}{2}\Bigl(1+\sqrt{1+\eta^2 q^4}\Bigr)
    + \mathcal{O}(r^2)\, .
\end{equation}
We observe that, by fixing the free parameters $M$ and $q$ such that
\begin{equation}
    -12M + \sqrt{2}\, g_2\, q^5 \lambda^{8}=0\, ,
    \label{eq:reg-condition-power}
\end{equation}
the divergent contribution $1/r$ vanishes and the metric function $A(r)$ becomes regular at $r=0$.
Under this condition, one further notices that
\begin{equation}
    A(0)=B^2(0)
    =\frac{1+\sqrt{1+\eta^2 q^4}}{2}\, ,
    \label{eq:reg-condition}
\end{equation}
which precisely corresponds to the regularity condition required for all curvature invariants to remain finite (see Appendix \ref{App:Reg-Cond}), thereby removing entirely the spacetime singularity at the origin.
Since, in this case, the scalar hair parameter $q$ is not independent but instead determined by the mass parameter $M$, through \eqref{eq:reg-condition-power}, the resulting solution is characterized by secondary scalar hair.
Using \eqref{eq:reg-condition-power} in \eqref{eq:A-sol-5/2}, we can bring $A(r)$ in the form
\begin{equation}
    A(r)= 1 - \frac{2Mr^2}{(r^2+\lambda^2)^{3/2}}
    - \frac{1}{2}\bigg(1+\frac{r^2}{\lambda^2}\bigg)
    \Bigg[1-\sqrt{1+\frac{\eta^2 q^4}{(1+r^2/\lambda^2)^2}}\
    \Bigg]_{\text{\fontsize{10}{10}\selectfont $q=\sqrt[5]{\frac{6\sqrt{2}\, M}{g_2 \lambda^8}}$}}\,.
    \label{eq:A-sol-5/2-reg}
\end{equation}
The first two terms in the above expression are identical to those of the Bardeen solution \cite{Bardeen1968}, while the last term constitutes a correction to this geometry. 
Hence, the spacetime geometry described by the line element \eqref{eq:metric}, with metric functions \eqref{eq:B-r} and \eqref{eq:A-sol-5/2-reg}, represents a \emph{nonreciprocal extension of the Bardeen solution}. 
The connection to the exact Bardeen solution will become apparent when we move to the reciprocal frame.

\begin{figure}[t]
    \centering

    \begin{subfigure}[b]{0.35\textwidth}
        \centering
        \includegraphics[width=\textwidth]{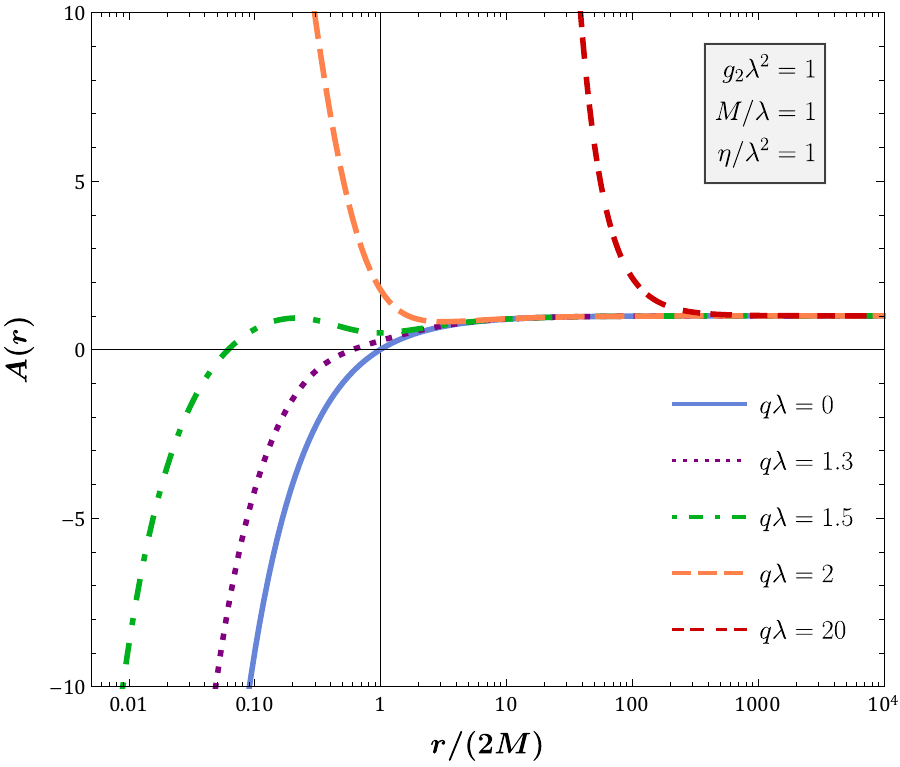}
        \caption{\hspace*{-1.5em}}
        \label{subf:sbh-5/2}
    \end{subfigure}
    \hfill
    \begin{subfigure}[b]{0.31\textwidth}
        \centering
        \includegraphics[width=\textwidth]{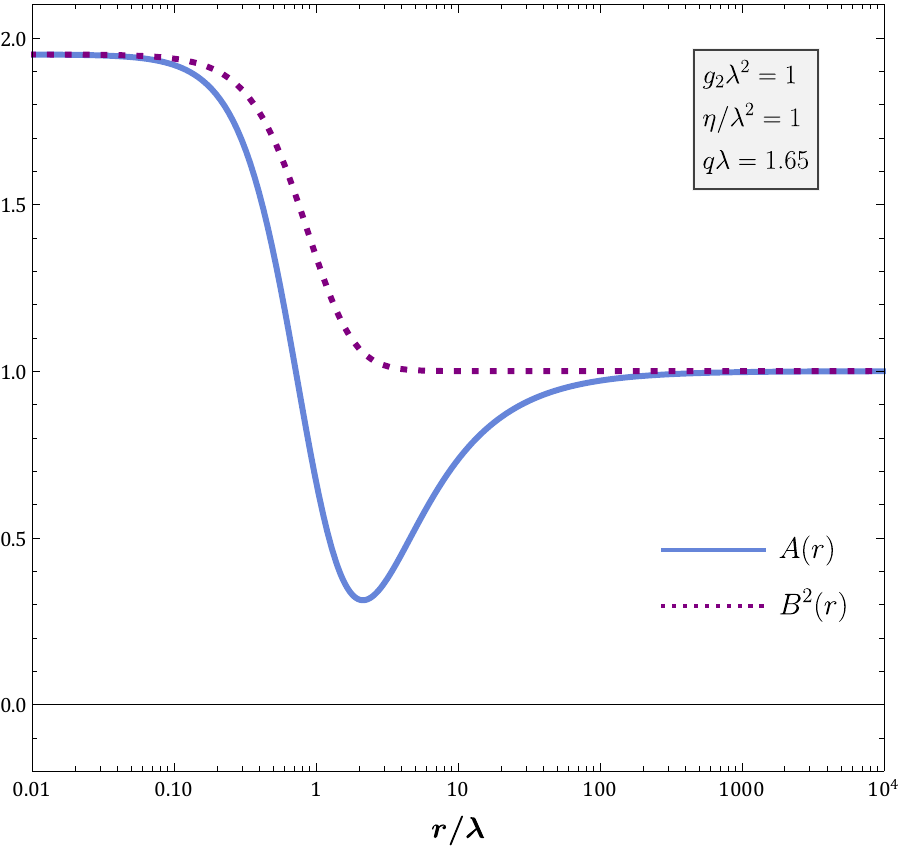}
        \caption{\hspace*{-0.7em}}
        \label{subf:soliton-5/2}
    \end{subfigure}
    \hfill
    \begin{subfigure}[b]{0.31\textwidth}
        \centering
        \includegraphics[width=\textwidth]{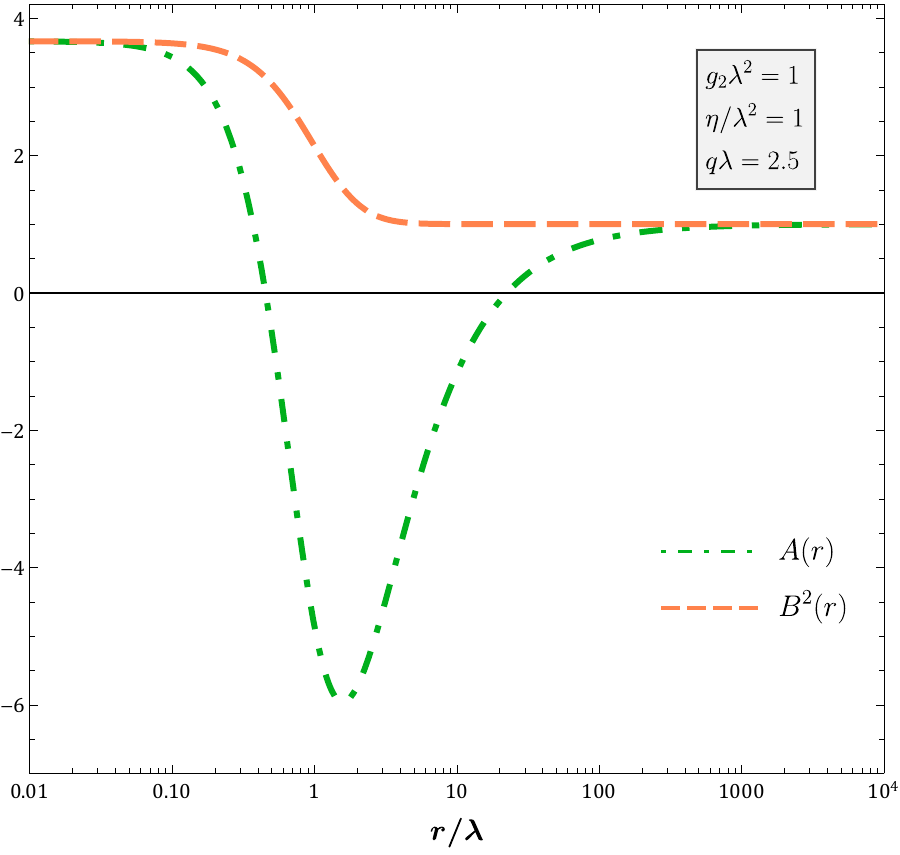}
        \caption{\hspace*{-0.7em}}
        \label{subf:rbh-5/2}
    \end{subfigure}

    \vspace*{-0.5em}

    \caption{The solutions depicted here originate from
    \eqref{eq:A-sol-5/2}. (a) For \(q\lambda=1.3,\,1.5\), the solutions
    correspond to single-horizon black holes with primary scalar hair.
    For comparison, the Schwarzschild geometry is depicted for \(q=0\).
    For \(q\lambda=2,\,20\), the resulting solutions describe naked
    singularities. After imposing the regularity condition
    \eqref{eq:reg-condition-power}, we also obtain: (b) solitonic
    solutions and (c) regular black holes. The horizontal axis is
    logarithmic in all panels.}
    \label{fig:Dis-Bardeen}
\end{figure}
%%%%%%%%%%%%%%%%%%%%%%

In Fig.~\ref{subf:sbh-5/2}, we illustrate the different types of compact objects that arise from the solution under consideration. 
Notice that all model and solution parameters have been rendered dimensionless by rescaling them with respect to the coupling constant $\lambda$. 
In this way, the description remains as general as possible and can be readily rescaled to different physical regimes.
First, using Eq.~\eqref{eq:A-sol-5/2}, we observe from Fig.~\ref{subf:sbh-5/2} that all black-hole solutions with primary scalar hair exhibit a single event horizon. 
As the scalar hair parameter $q$ approaches zero, the solution converges toward the Schwarzschild geometry, which is expected since in the limit $q\to0$ the solution reduces to the
Schwarzschild branch. 
Conversely, increasing the scalar hair leads to a progressive reduction of the event-horizon radius, yielding black holes that are more compact than the Schwarzschild solution for the same mass.
However, this trend does not persist indefinitely. 
Beyond a critical value of the scalar hair, the event horizon disappears, leaving behind a naked singularity.
In addition to singular configurations, the solution also admits regular geometries once the regularity condition \eqref{eq:reg-condition-power} is imposed and the mass parameter is expressed in terms of the scalar-hair parameter $q$.
In particular, Fig.~\ref{subf:soliton-5/2} depicts a solitonic compact object, which is massive, horizonless, and free of spacetime singularities. 
Meanwhile, Fig.~\ref{subf:rbh-5/2} illustrates a regular black hole, namely a massive and non-singular compact object possessing two distinct horizons.

Finally, it is important to mention that a disformal transformation of the form \eqref{eq:dis} recasts the above theory into the one discussed in \cite{Bakopoulos:2023sdm}, with the corresponding solutions mapped to those of the disformed DHOST theory, originating from Eqs.~\eqref{eq:G2G4F4new-Xnew} and \eqref{eq:F-power}.
A particularly interesting result of \cite{Bakopoulos:2023sdm} is that for $p=5/2$, the resulting reciprocal solution is of the form 
\begin{equation}
    \dd s^2 = -\widetilde{A}(r) \dd t^2 + \frac{\dd r^2}{\widetilde{A}(r)} + r^2 \dd \Omega^2\,, \qquad \widetilde{A}(r) = 1 - \frac{2M}{r} + \frac{g_2 \lambda^8 q^5}{3\sqrt{2}} \frac{1}{r} \bigg(1- \frac{r^3}{(r^2 +\lambda^2)^{3/2}}\bigg)\,,
\end{equation}
and describes a singular black-hole solution with primary scalar hair since both $M$ and $q$ remain free parameters of the solution.
The regularity condition \eqref{eq:reg-condition-power} remains the same even in this case, and once imposed, it yields the Bardeen solution, namely
\begin{equation}
    \widetilde{A}(r)=1-\frac{2Mr^2}{(r^2+\lambda^2)^{3/2}}\,.
\end{equation}
This makes explicit that the regularity condition is preserved under the disformal map, with the regular non-reciprocal solution being mapped exactly onto the Bardeen geometry.

%==========================================
%Adding solutions 
%==========================================

\subsubsection{Superposition of power-law contributions}
\label{sec:Superposition}

In the previous section, we demonstrated that the power-law ansatz \eqref{eq:F-power}, for the auxiliary function $\mathcal{F}$, gives rise, upon direct integration of Eq.~\eqref{eq:A-sol}, to the family of solutions described by Eq.~\eqref{eq:A-sol-power}.
In this section, we will show that it is straightforward to generalize this prescription and incorporate additional power-law terms in the expression for $\mathcal F$, thereby obtaining new black-hole solutions in a systematic and straightforward way.
To this end, we consider
\begin{equation}
    \mathcal{F}(\Xi) = -g_2 \sum_{n=1}^N \, \xi_n \big(\lambda^2\Xi\big)^{p_n} = -g_2\big[\xi_1\big(\lambda^2\Xi\big)^{p_1} + \xi_2\big(\lambda^2\Xi\big)^{p_2} + \ldots + \xi_N\big(\lambda^2\Xi\big)^{p_N} \big ]\,,
\end{equation}
for all dimensionless $\xi_i\in\mathbb{R}$ and $p_i\in\{1/2\} \cup (3/2,+\infty)$ (to obtain asymptotically flat solutions) and $p_i \neq p_j$ for $i \neq j$. 
Notice now that
\begin{equation}
    2\Xi\,\mathcal{F}_\Xi-\mathcal{F}(\Xi) = -g_2 \sum_{n=1}^N\,\xi_n (2p_n - 1) \big(\lambda^2 \Xi\big)^{p_n}\,.
\end{equation}
Substituting the above expression into Eq.~\eqref{eq:A-sol} and performing the integral, we obtain the general solution
{\fontsize{11}{11}\selectfont
\begin{equation}
    A(r) = 1 -\frac{C_0}{r}
    + \frac{1}{2}\bigg(1+\frac{r^2}{\lambda^2}\bigg)[\Delta(r)-1] 
    -\frac{g_2}{3}\, r^2
    \sum_{n=1}^N
    \frac{\xi_n(2p_n-1)(\lambda^2 q^2)^{p_n}}{2^{p_n}}\,
    {}_2F_1\bigg(
        \frac{3}{2},p_n;\frac{5}{2};
        -\frac{r^2}{\lambda^2}
    \bigg)\,.
    \label{eq:A-sol-power-add}
\end{equation}
}%

\noindent It is also worth noting that, even after applying the disformal transformation \eqref{eq:dis} to obtain the reciprocal frame, the above analysis continues to hold in the resulting DHOST theory. 
Similar observations for reciprocal solutions were presented in \cite{Bakopoulos:2023sdm, Charmousis:2025xug}, whereas here we extend the argument to include non-reciprocal solutions as well.

To illustrate the above result, we now turn to a specific example in which we add an extra power-law term, $p_2=7/2$, to the function $\mathcal{F}$ of the preceding example, thus, we have
\begin{equation}
    \mathcal{F}(\Xi)=-g_2\big[(\lambda^2\Xi)^{5/2}+\xi (\lambda^2\Xi)^{7/2}]\,,
    \label{CalF_SuperpositionP1P2}
\end{equation}
where $\xi$ is an additional dimensionless coupling constant.
From Eqs.~\eqref{eq:Z-X}, \eqref{eq:G2-F}, \eqref{eq:G2-G4}, and \eqref{eq:def-Z}, one can readily determine the model functions $G_2(X)$, $G_4(X)$, and $F_4(X)$ for the quadratic DHOST theory resulting from the above assumption.
To obtain the corresponding solution, one has to use \eqref{eq:A-sol-power-add} for $\xi_1=1$, $p_1=5/2$, $\xi_2=\xi$, and $p_2=7/2$.
By doing so, we obtain
\begin{align}
    A(r) = 1 &- \frac{2M}{r} -\frac{1}{2}\bigg(1+\frac{r^2}{\lambda^2}\bigg)\Bigg[1-\sqrt{1+\frac{\eta^2 q^4}{(1+r^2/\lambda^2)^2}}\ \Bigg] +\frac{g_2\, q^5\lambda^8}{30\sqrt{2}\, r}  
    \bigg\{(10 + 3\xi q^2\lambda^2)\nonumber\\[2mm]
    & - \frac{r^3}{(r^2 +\lambda^2)^{5/2}} 
    \bigg[10(r^2+\lambda^2) + \frac{3}{2}\xi q^2\lambda^2(2r^2+5\lambda^2) \bigg]\bigg\}\,,
\end{align}
where $M$ denotes the ADM mass of the solution.
In the asymptotic regime, the solution matches Eq.~\eqref{eq:A-asym-power} up to fourth order in the expansion.
However, in the limit $r\to 0$, the solution reads
\begin{equation}
    A(r)\xrightarrow{r\to 0}
    \bigg[-2M+\frac{g_2\, q^5\lambda^8(10 + 3\xi q^2\lambda^2)}{30\sqrt{2}}\bigg]\frac{1}{r}
    + \frac{1}{2}\Bigl(1+\sqrt{1+\eta^2 q^4}\Bigr)
    + \mathcal{O}(r^2)\, .
\end{equation}
Hence, the regularity condition in this case is
\begin{equation}
    -2M+\frac{g_2\, q^5\lambda^8(10 + 3\xi q^2\lambda^2)}{30\sqrt{2}}=0\,,
    \label{eq:reg-condition-power-add}
\end{equation}
under which Eq.~\eqref{eq:reg-condition} is satisfied and as a consequence the spacetime becomes regular everywhere.

%%%%%%%%%%%%%%%%%%%%%%
\begin{figure}[t]
    \centering

    \begin{subfigure}[b]{0.48\textwidth}
        \centering
        \includegraphics[width=\textwidth]{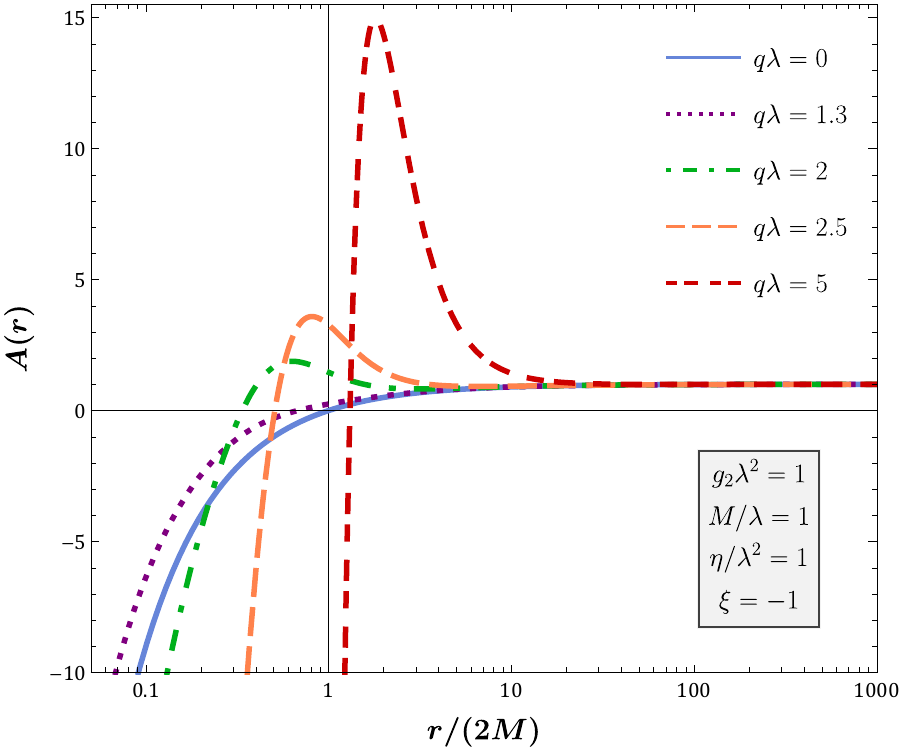}
        \caption{\hspace*{-1.5em}}
        \label{subf:sbh-5/2-7/2}
    \end{subfigure}
    \hfill
    \begin{subfigure}[b]{0.445\textwidth}
        \centering
        \includegraphics[width=\textwidth]{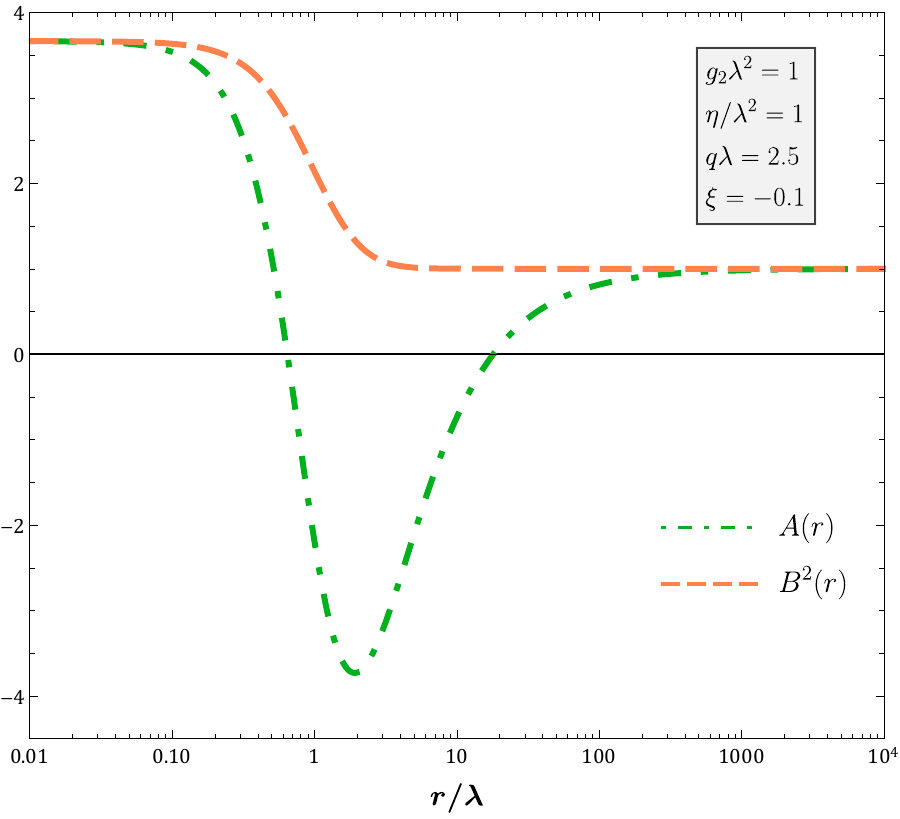}
        \caption{\hspace*{-0.5em}}
        \label{subf:rbh-5/2-7/2}
    \end{subfigure}

    \vspace*{-0.5em}

   %\captionsetup{justification=raggedright,singlelinecheck=false}
    \caption{The solutions depicted here originate from
    \eqref{eq:A-sol-power-add} with $p_1=5/2\,,p_2=7/2$. (a) Single-horizon black holes with
    primary scalar hair. For comparison, the Schwarzschild geometry is
    depicted for $q=0$. (b) Regular black-hole solution. The horizontal
    axis is logarithmic in both panels.}
    \label{fig:Add-5/2-7/2}
\end{figure}
%%%%%%%%%%%%%%%%%%%%%%

Figure~\ref{fig:Add-5/2-7/2} illustrates representative singular and regular black-hole solutions of the superposition model. 
Rather than repeating the features already discussed for the single power-law case, we focus here on the new behavior introduced by the additional coupling $\xi$.
For negative values of $\xi$, the dependence of the event-horizon radius on the scalar hair becomes non-monotonic, as shown in Fig.~\ref{subf:sbh-5/2-7/2}. 
Starting from the Schwarzschild limit at $q=0$, the horizon radius initially decreases as $q$ increases, reaches a minimum, and subsequently grows again. 
Consequently, within the corresponding range of parameters, two distinct black-hole solutions can have the same mass and the same horizon radius while carrying different values of the scalar hair.
A similar phenomenon has previously been observed in nonlinear electrodynamics~\cite{Bakopoulos:2024hah}.\,\footnote{In Ref.~\cite{Bakopoulos:2024hah}, such configurations were referred to as doppelg\"anger black holes.}

Regarding the regular compact objects described by Eq.~\eqref{eq:A-sol-power-add}, Fig.~\ref{subf:rbh-5/2-7/2} shows that the present framework naturally admits regular black-hole solutions. 
Moreover, since the scalar-hair parameter $q$ and the mass are connected through the regularity condition \eqref{eq:reg-condition-power-add}, decreasing either of them eventually causes the event horizon to disappear, giving rise to solitonic compact objects with a profile qualitatively similar to that shown in Fig.~\ref{subf:soliton-5/2}.

%==========================================
%Pade-type solutions
%==========================================

\subsubsection{Pad\'e-type ans\"atze for the model functions}

Having explored black-hole solutions generated by power-law ans\"atze for the auxiliary function $\mathcal{F}$, it is natural to consider other possibilities for its functional form. 
A particularly interesting class is provided by Pad\'e-type ans\"atze, where $\mathcal{F}$ is modeled by a rational function. 
Two relatively simple such scenarios are the following:
\begin{equation}
    \mathcal{F}_1(\Xi)=-g_2 \frac{\big(\lambda^2\Xi\big)^2}{1+\lambda^2\Xi}\,,\qquad \mathcal{F}_2(\Xi)=-g_2 \frac{\big(\lambda^2\Xi\big)^2}{\big(1+\lambda^2\Xi\big)^2}\,.
    \label{eq:F-pade1}
\end{equation}
By following the same procedure as before, we start from \eqref{eq:F-pade1} and make use of Eqs.~\eqref{eq:Z-X}, \eqref{eq:G2-F}, \eqref{eq:G2-G4}, and \eqref{eq:def-Z} to determine the model functions of the theories. 
In complete analogy, substituting $\mathcal{F}$ into Eq.~\eqref{eq:A-sol} and using \eqref{eq:B-r}, one directly finds the radial expression for the metric functions $A_1(r)$ and $A_2(r)$ of the corresponding models.
By doing so, one can verify that by defining
\begin{equation}
    \mathcal{U} \equiv \sqrt{2+q^2\lambda^2}\,,
\end{equation}
one obtains
\begin{equation}
\begin{aligned}
A_1(r)={}&1-\frac{2M}{r}
-\frac{1}{2}\left(1+\frac{r^2}{\lambda^2}\right)
\left[1-\sqrt{1+\frac{\eta^2q^4}
{\left(1+r^2/\lambda^2\right)^2}}\right]
+\frac{g_2\pi q^2\lambda^5}{4r\mathcal{U}}
\left[\sqrt{2}\left(1+q^2\lambda^2\right)-\mathcal{U}\right]
\\
&+\frac{g_2q^4\lambda^8}{2r}
\left\{
\frac{r}{2r^2+\lambda^2\mathcal{U}^2}
+\frac{1}{q^2\lambda^3}
\left[
\arctan\left(\frac{r}{\lambda}\right)
-\frac{\sqrt{2}\left(1+q^2\lambda^2\right)}{\mathcal{U}}
\arctan\left(\frac{\sqrt{2}\,r}{\lambda\mathcal{U}}\right)
\right]
\right\}.
\end{aligned}
\label{eq:A-pade-1}
\end{equation}
and
\begin{equation}
\begin{aligned}
A_2(r)={}&1-\frac{2M}{r}
-\frac{1}{2}\left(1+\frac{r^2}{\lambda^2}\right)
\left[1-\sqrt{1+\frac{\eta^2q^4}
{\left(1+r^2/\lambda^2\right)^2}}\right]
+\frac{g_2\pi q^4\lambda^7
\left(3+q^2\lambda^2\right)}
{4\sqrt{2}\,r\mathcal{U}^3}
\\
&+\frac{g_2q^4\lambda^8}{2r}
\left[
\frac{
r^3(6+4q^2\lambda^2)
+r\lambda^2(6+5q^2\lambda^2+q^4\lambda^4)}
{\mathcal{U}^2\left(2r^2+\lambda^2\mathcal{U}^2\right)^2}
-\frac{3+q^2\lambda^2}
{\sqrt{2}\lambda\mathcal{U}^3}
\arctan\left(\frac{\sqrt{2}\,r}{\lambda\mathcal{U}}\right)
\right].
\end{aligned}
\label{eq:A-pade-2}
\end{equation}
In both solutions, $M$ is the ADM mass, as can be observed by their asymptotic expansion at infinity, which up to third order is identical, namely
\begin{equation}
    A_1(r) = A_2(r) \xrightarrow{r\to\infty} 1 - \frac{2M}{r} + \frac{q^{4}\lambda^{2}\left(\eta^{2} + 3 g_{2}\lambda^{6}\right)}{4 r^{2}} + \mathcal{O}\bigg(\frac{1}{r^4}\bigg)\,.
\end{equation}
Both solutions can describe singular black holes with primary scalar hair and regular black holes with secondary scalar hair, while solitonic solutions can also arise for specific values of the parameters.
The regularity condition for case 1 is
\begin{equation}
    M = \frac{g_{2}\pi q^{2}\lambda^{5}\big[\sqrt{2}(1 + \,q^{2}\lambda^{2}) - \sqrt{2+q^{2}\lambda^{2}}\big]}{8\sqrt{2+q^{2}\lambda^{2}}}\,,
    \label{reg_condition_PadeI}
\end{equation}
which essentially corresponds to eliminating the second and fourth terms in \eqref{eq:A-pade-1}.
Similarly, for case 2, the regularity condition reads 
\begin{equation}
    M = \frac{g_2\pi q^4\lambda^7(3+q^2\lambda^2)}{8\sqrt{2}\,(2+q^2\lambda^2)^{3/2}}\,.
\end{equation}
Finally, it is important to stress that even after recasting the above solutions into their reciprocal frame using the disformal transformation \eqref{eq:dis}, the resulting solutions will still describe both singular and regular black holes.
To our knowledge, these solutions are new to the literature in both frames.\\

The Pad\'e families are qualitatively different from the finite power-law examples discussed above. 
Indeed, they resum an infinite tower of power-law contributions in the small-\(\Xi\) expansion, which translates to large $r$ values through \eqref{eq:Xi}.
One can verify that
\begin{equation}
    \mathcal{F}_1(\Xi) = -g_2 \left[ \left(\lambda^2\Xi\right)^2
        - \left(\lambda^2\Xi\right)^3
        + \left(\lambda^2\Xi\right)^4
        - \ldots \right],
\end{equation}
and
\begin{equation}
    \mathcal{F}_2(\Xi) = -g_2\left[ \left(\lambda^2\Xi\right)^2
        - 2\left(\lambda^2\Xi\right)^3 
        + 3\left(\lambda^2\Xi\right)^4
        - \ldots \right].
\end{equation}
Therefore, both solutions share the same leading weak-field behavior as the quadratic power-law model, but they differ in the strong-field region through an infinite sequence of higher-order scalar-hair corrections. 
This explains why the two Pad\'e geometries have the same asymptotic expansion up to the order displayed above, while their near-core structures are different.

This difference can be seen directly from the combination that sources the
radial integral in Eq.~\eqref{eq:A-sol}. For the two Pad\'e choices, one finds
\begin{equation}
    2\Xi\mathcal{F}_{1\Xi}-\mathcal{F}_1 = 
    -g_2\,\frac{\left(\lambda^2\Xi\right)^2 \left(3+\lambda^2\Xi\right)}{\left(1+\lambda^2\Xi\right)^2},
\end{equation}
whereas
\begin{equation}
    2\Xi\mathcal{F}_{2\Xi}-\mathcal{F}_2  =  -g_2\,
    \frac{\left(\lambda^2\Xi\right)^2\left(3-\lambda^2\Xi\right)}{\left(1+\lambda^2\Xi\right)^3}.
\end{equation}
Thus, the first model gives a source term with a fixed sign, while the second one can change sign inside the compact object. Since
\begin{equation}
    \lambda^2\Xi(r)=
    \frac{q^2\lambda^2}{2\left(1+r^2/\lambda^2\right)},
\end{equation}
the sign change in the second Pad\'e model occurs whenever \(q^2\lambda^2>6\), at the radius
\begin{equation}
    r_\star^2 = \lambda^2 \left(\frac{q^2\lambda^2}{6}-1\right).
\end{equation}
This provides a simple mechanism for generating a layered core, namely the scalar sector can contribute with one sign near the center and with the opposite sign at larger radii. 
This feature is absent in the single power-law examples.

%==========================================
%Different Z(X(r)) ansatz
%==========================================

\subsection{Different $Z(X)$ profiles and connection between solutions}
\label{sec:IVB}

In the preceding subsection, we based our analysis on a specific choice for the auxiliary function $Z(X)$, namely \eqref{eq:Z-X}. 
It is important to stress, however, that within Branch III the relation between the invariant combination $\Xi\equiv Z^2X$ and the radial coordinate, i.e. Eq. \eqref{eq:Xi}, follows entirely from the branch condition $\mathcal G_2/\mathcal G_4=2/\lambda^2$, Eq. \eqref{eq:G2-G4} together with the field equation \eqref{eq2-new}, and does not depend on which functional form is chosen for $Z(X)$ itself. 
The choice of $Z(X)$ only enters the algebraic step of solving Eq.~\eqref{eq:Xi} for $X(r)$, and hence affects the metric function $B(r)=Z(X(r))$.
Consequently, different $Z(X)$ ans\"atze generate genuinely different non-reciprocal geometries, with different $X(r)$, $B(r)$, and $A(r)$, within different DHOST models.
However, as we showed in Sect.~\ref{Sec:Disformal-Frame}, for a given function $\mathcal{F}(\Xi)$, regardless of the choice of $Z(X)$, the disformal map \eqref{eq:dis} will always lead to the same theory in the reciprocal frame and the same solution \eqref{eq:A-rec-frame}. 
In other
words: any two theories built from the same $\mathcal F(\Xi)$
but different $Z(X)$ profiles are disformally mapped, through their own
($Z$-dependent) transformation~\eqref{eq:dis}, onto one and the same reciprocal
DHOST theory and the same reciprocal solution, even though their
non-reciprocal geometries $A(r)$, $B(r)$ are genuinely different.

It is also important to notice that since $\Xi(r)$ is fixed once and for all by the branch condition \eqref{eq:G2-G4}, for the class of theories characterized by Eq.~\eqref{eq:G2-F}, the choice of $Z(X)$ does not affect in any way the term in $A(r)$ which is sourced from $\mathcal{F}(\Xi)$, i.e. the last term of Eq.~\eqref{eq:A-sol}.
Two immediate consequences follow from this fact.
First, the near-origin regularity condition that removes the $1/r$ divergence of $A(r)$, which fixes the scalar hair $q$ in terms of the ADM mass $M$ and leads to regular black-hole solutions, is determined solely by the $1/r$ behavior of $C_0/r$-term and the integral term near $r=0$, since $B^2(r)$ is regular there for any admissible $Z(X)$.
The resulting relation between $M$ and $q$, which is necessary for the existence of regular black holes, is therefore the same for every $Z(X)$ profile sharing the same $\mathcal F(\Xi)$. Second, the value $A(0)=B^2(0)$ singled out by the regularity condition (see Appendix \ref{App:Reg-Cond}) does depend on $Z(X)$, through $B^2(0)$, even though $M(q)$ does not.

Let us now showcase the above observations by considering an alternative ansatz for the auxiliary function $Z(X)$.
To this end, we assume that
\begin{equation}
    Z(X)=\sqrt{\frac{2}{1+\sqrt{1-\eta^2X^2}}}\,.
    \label{eq:Z-2}
\end{equation}
Using Eq.~\eqref{eq:Xi}, the radial expression of $X(r)$ can be immediately determined and is given by
\begin{equation}
    X(r) = \frac{8q^2\big(1+r^2/\lambda^2\big)}{16\big(1+r^2/\lambda^2\big)^2 +\eta^2q^4}\,.
    \label{eq:X-2}
\end{equation}

Notice that $X(r)$ is positive-definite and finite for all $r\in[0,+\infty)$.
Additionally, $X(r)\to 0$ smoothly as $q\to 0$, recovering the GR limit.
Using also the fact that $B^2(r)=Z^2(X(r))$, we obtain the closed form 
\begin{equation}
    B^2(r) = \frac{2(1+\varkappa^2)}{4\varkappa^2}\big(1+\varkappa^2-|1-\varkappa^2|\big)\,,\qquad \varkappa^2 \equiv \frac{\eta^2 q^4}{16(1+r^2/\lambda^2)^2}\,,
\end{equation}
which for $\eta^2q^4\leq 16$ it reduces to the remarkably simple form
\begin{equation}
    B^2(r) = 1 + \frac{\eta^2q^4}{16\big(1+r^2/\lambda^2\big)^2}\,.  \label{eq:B-2}
\end{equation}
It is clear that $B^2(r)$ is finite and positive over the whole domain $r\in[0,+\infty)$.
Moreover, when $\eta^2q^4\leq 16$, it is also monotonically decreasing, attaining its maximum $B^2(0)=1+\eta^2 q^4/16$ at $r=0$ and approaching unity as $r\to +\infty$, with
\begin{equation}
    B^2(r)\to 1+\frac{\eta^2q^4}{16} \frac{\lambda^4}{r^4}+\mathcal O(r^{-6})\,,
    \qquad r\to+\infty\,,
\end{equation}
which is consistent with the asymptotic-flatness requirement discussed in Appendix \ref{App:Asym-III}.

To proceed further with the analysis, we need to assume a specific form for $\mathcal{F}(\Xi)$ and study the resulting solution.
Hence, we choose the same power-law ansatz that we have already used in Sec.~\ref{Sec:Nonrec-Bardeen} to construct the nonreciprocal extension of the Bardeen solution, namely $\mathcal F(\Xi)=-g_2(\lambda^2\Xi)^{5/2}$.
Making use of Eq.~\eqref{eq:A-sol}, it is straightforward to derive the form of the metric function $A(r)$, namely
\begin{equation}
    A(r) = 1 - \frac{2M}{r}  + \frac{g_2 \lambda^8 q^5}{3\sqrt{2}} \frac{1}{r} \bigg(1- \frac{r^3}{(r^2 +\lambda^2)^{3/2}}\bigg) + \frac{\eta^2q^4}{16\big(1+r^2/\lambda^2\big)}\,,
    \label{eq:A-2}
\end{equation}
where $M$ denotes the ADM mass of the black hole. 
The preceding form of $A(r)$ should be compared term by term with Eq.~\eqref{eq:A-sol-5/2}.
The two solutions share exactly the same $1/r$ and $g_2$-terms, and differ only through the last, $\eta$-term, which is the sole trace of the underlying $Z(X)$ ansatz.
This is also manifest at the level of the large-$r$ expansion,
\begin{equation}
    A(r)\xrightarrow{r\to+\infty}1-\frac{2M}{r}+\frac{\eta^2q^4\lambda^2}{16\,r^2}
    +\frac{g_2q^5\lambda^{10}}{2\sqrt2\,r^3}+\mathcal O(r^{-4})\,.
    \label{eq:A-2-asym}
\end{equation}
Comparing the expansions \eqref{eq:A-asym-power} and \eqref{eq:A-2-asym}, one notices that the $g_2$-terms at order $r^{-3}$ are identical in both solutions, while the $r^{-2}$ terms differ by a factor of $1/4$.
The two profiles thus give rise to compact objects of the same mass $M$ but with a quantitatively different subleading structure, originating from the difference in the $B(r)$-metric function.

On the other hand, expanding around the origin, we find
\begin{equation}
    A(r)\xrightarrow{r\to0}\frac{-12M+\sqrt2\,g_2q^5\lambda^8}{6}\,\frac1r
    +\Big(1+\frac{\eta^2q^4}{16}\Big)+\mathcal O(r^2)\,.
    \label{eq:A-2-reg}
\end{equation}
The coefficient of the $1/r$ divergence is identical to that of Eq.~\eqref{eq:reg-condition-power}, confirming the general argument above: the regularity condition that removes the central singularity is insensitive to the choice of $Z(X)$. 
In contrast, the value of $A(0)=B^2(0)$, which in this case is equal to $1+\eta^2q^4/16$, differs from that of \eqref{eq:reg-condition}; therefore, it is clear that this quantity depends on the choice of $Z(X)$. 
Hence, upon fixing the scalar hair $q$ in terms of mass $M$ through \eqref{eq:reg-condition-power}, the metric function $A(r)$ becomes
\begin{equation}
    A(r)=1-\frac{2Mr^2}{(r^2+\lambda^2)^{3/2}}+\frac{\eta^2q^4\lambda^2}
    {16\big(r^2+\lambda^2\big)}\bigg|_{q=\left(\frac{6\sqrt2\,M}{g_2\lambda^8}\right)^{\frac{1}{5}}}\,.
\label{eq:Anew_reg}
\end{equation}
Similar to the solution \eqref{eq:A-sol-5/2-reg}, the above metric function can describe everything from regular black holes to solitonic solutions.
We also observe that both metric functions share the same Bardeen-like mass term $-2Mr^2/(r^2+\lambda^2)^{3/2}$ and differ only in the scalar-hair correction, which is now a particularly simple rational function of $r$ rather than the square-root structure of Eq.~\eqref{eq:A-sol-5/2-reg}.

\begin{figure}[t]
    \centering

    \begin{subfigure}[b]{0.48\textwidth}
        \centering
        \includegraphics[width=\textwidth]{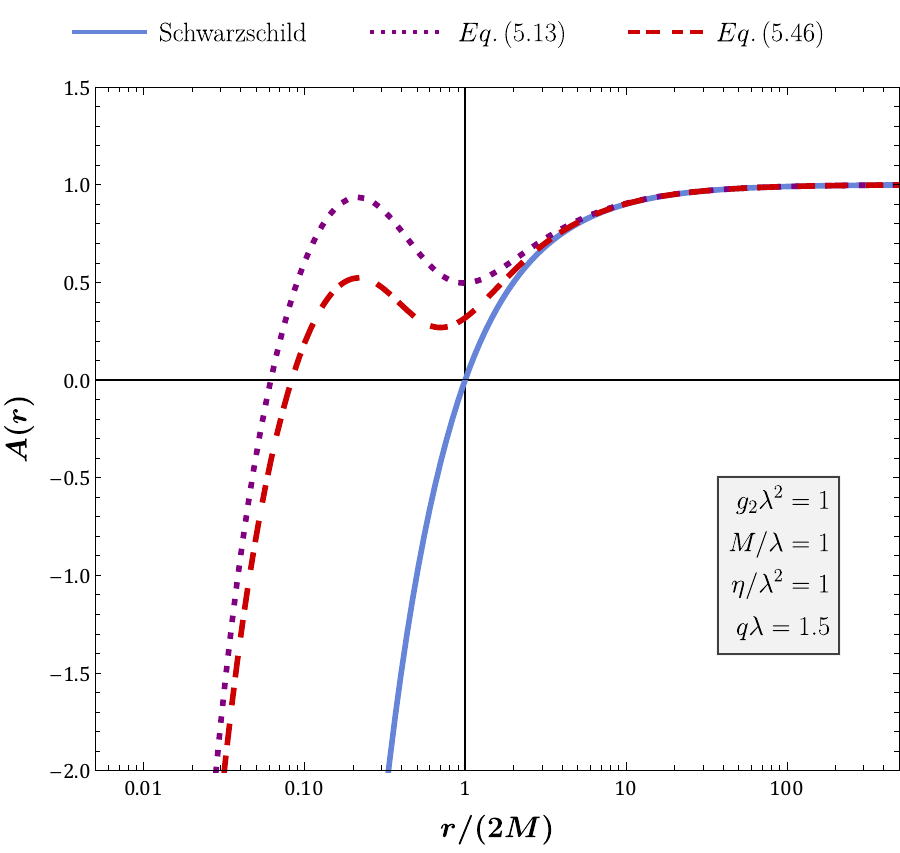}
        \caption{\hspace*{-2em}}
        \label{subf:sbh-2-A-1}
    \end{subfigure}
    \hfill
    \begin{subfigure}[b]{0.47\textwidth}
        \centering
        \includegraphics[width=\textwidth]{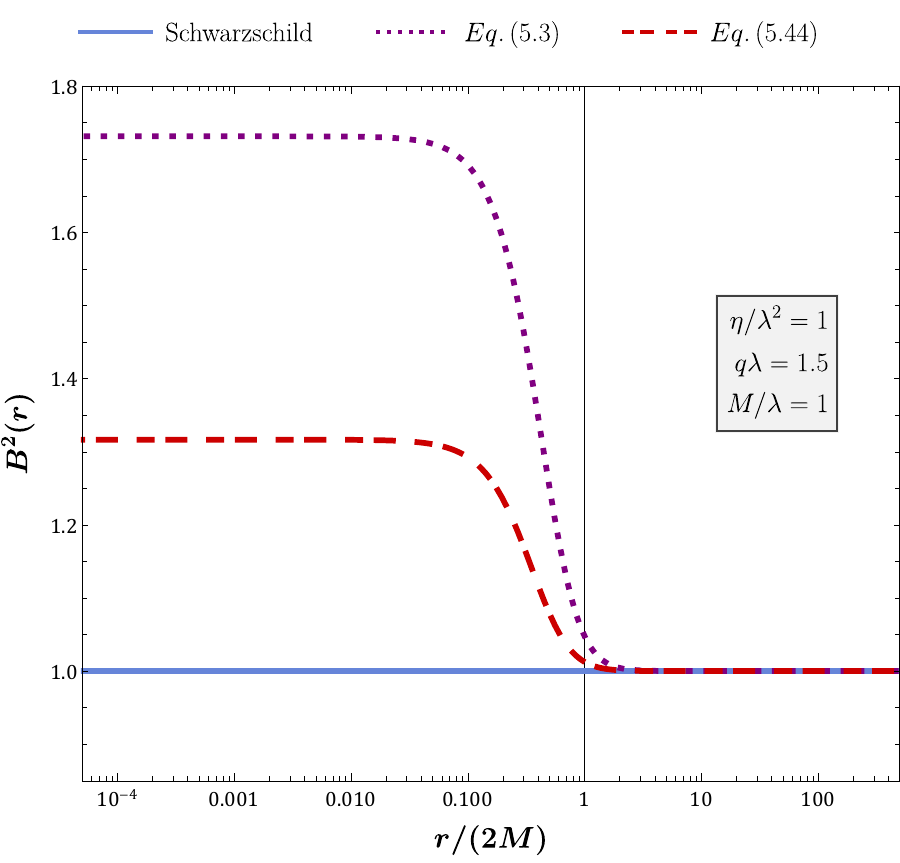}
        \caption{\hspace*{-2em}}
        \label{subf:sbh-2-B-1}
    \end{subfigure}

    \vspace*{-0.5em}

    %\captionsetup{justification=raggedright,singlelinecheck=false}
    \caption{Singular black-hole solutions for the parameters indicated in the panels.
    (a) Comparison of the metric function $A(r)$ for the two non-reciprocal solutions and the Schwarzschild geometry.
   \newline (b) Corresponding metric function $B^2(r)$ for the two non-reciprocal solutions.}
    \label{fig:sbh-2-1}
\end{figure}

\begin{figure}[t]
    \centering

    \begin{subfigure}[b]{0.475\textwidth}
        \centering
        \includegraphics[width=\textwidth]{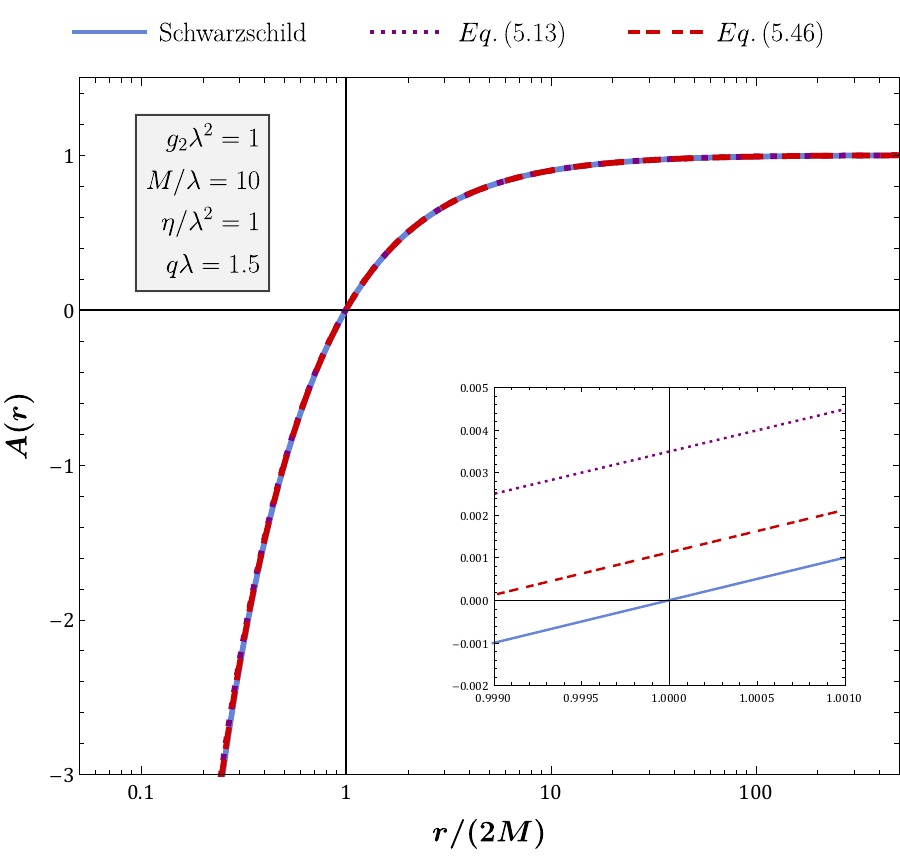}
        \caption{\hspace*{-2em}}
        \label{subf:sbh-2-A-2}
    \end{subfigure}
    \hfill
    \begin{subfigure}[b]{0.47\textwidth}
        \centering
        \includegraphics[width=\textwidth]{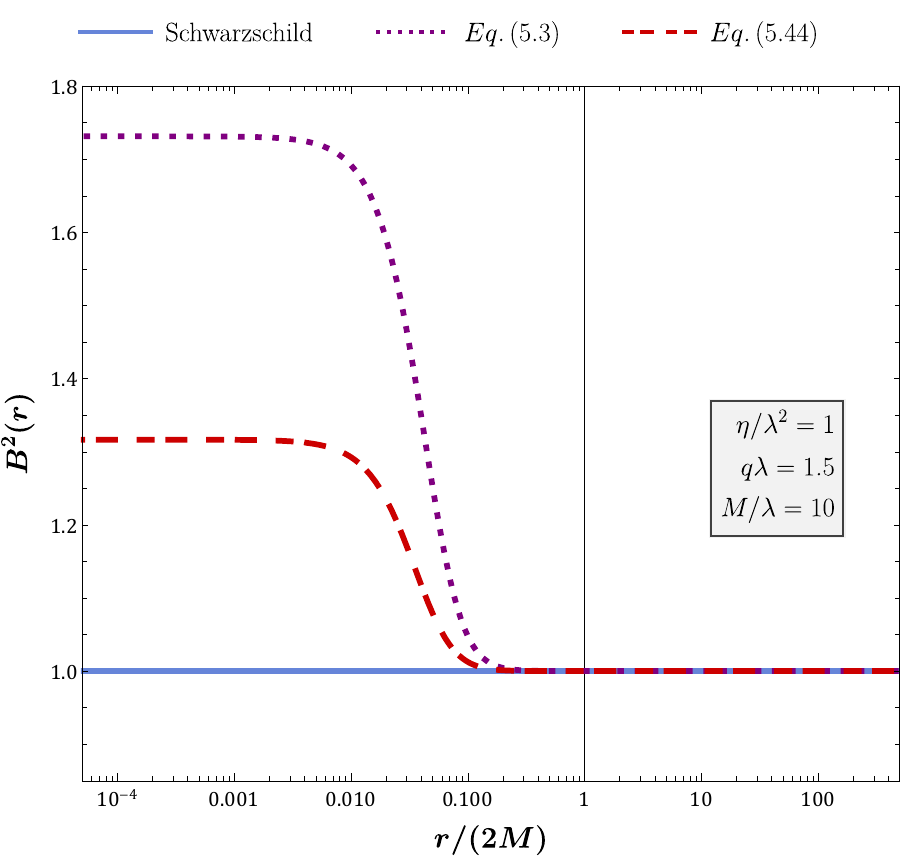}
        \caption{\hspace*{-2em}}
        \label{subf:sbh-2-B-2}
    \end{subfigure}

    \vspace*{-0.5em}

    %\captionsetup{justification=raggedright,singlelinecheck=false}
    \caption{Singular black-hole solutions for the parameters indicated in the panels.
  (a) Comparison of the metric function $A(r)$ for the two non-reciprocal solutions and the Schwarzschild geometry.\newline
    (b) Corresponding metric function $B^2(r)$ for the two non-reciprocal solutions.}
    \label{fig:sbh-2-2}
\end{figure}

Let us now gauge the departure from General Relativity of the two solutions that originate from the same $\mathcal{F}(\Xi)=-g_2(\lambda^2\Xi)^{5/2}$ but different $Z(X)$ profiles, namely \eqref{eq:Z-X} and \eqref{eq:Z-2}.
To this end, we compare the metric functions $A(r)$ and $B^2(r)$ built from each profile directly to the Schwarzschild geometry ($A_{\rm Schw}(r)=1-2M/r$ and $B^2_{\rm Schw}(r)=1$), for three representative choices of parameters.
For each profile, the choices of parameters describe two equally hairy singular black holes, one light and one heavy (compared to the length scale $\lambda$ introduced by the theory), corresponding to Figs.~\ref{fig:sbh-2-1} and ~\ref{fig:sbh-2-2}, respectively, and a regular black hole obtained once the mass and the scalar hair are tied together via the regularity condition \eqref{eq:reg-condition-power}.
In each scenario, whether singular or regular, the parameters are kept the same to allow for an easy comparison between the solutions. 
In all three cases, the first profile refers to the $Z(X)$ of Eq.~\eqref{eq:Z-X}, with $B^2(r)$ and $A(r)$ given by \eqref{eq:B-r} and \eqref{eq:A-sol-5/2}, respectively, while the second one refers to the $Z(X)$ of Eq.~\eqref{eq:Z-2}, with $B^2(r)$ and $A(r)$ given by Eqs.~\eqref{eq:B-2} and \eqref{eq:A-2}.
Specifically for regular black holes, we use the metric functions obtained by substituting the regularity condition \eqref{eq:reg-condition-power} into the general solutions.
In all figures, the radial coordinate is normalized by the black hole mass, while every other parameter of the theory, or the solution, is normalized in terms of the coupling constant $\lambda$.

%%%%%%%%%%%%%%%%%%%%%%
%%%%%%%%%%%%%%%%%%%%%%

%%%%%%%%%%%%%%%%%%%%%%
%%%%%%%%%%%%%%%%%%%%%%

Fig.~\ref{fig:sbh-2-1} depicts singular black holes with parameters $q\lambda=1.5$, $M/\lambda=1$, $\eta/\lambda^2=1$, and $g_2\lambda^2=1$, for some fixed scale $\lambda$.
The mass is comparable to the coupling constant $\lambda$, while the scalar hair has the largest possible effect relative to the horizon size.
We therefore refer to these configurations as light hairy black holes.
The panel \ref{subf:sbh-2-A-1} compares the metric function $A(r)$ of the hairy solutions with the Schwarzschild one.
All three solutions possess a single horizon, but the two hairy
solutions cross zero at a smaller value of $r/(2M)$ than Schwarzschild,
which means that these are therefore more compact black holes than their Schwarzschild counterpart of equal mass. 
Outside the horizon, both hairy solutions overshoot the Schwarzschild curve, reaching a local maximum and then dipping to a positive local minimum around $r/(2M)\sim1$.
After that, they merge smoothly with the Schwarzschild tail at large $r$.
This non-monotonic ``bump'' is a direct signature of the primary scalar hair and disappears, as expected, together with $B^2(r)-1$, panel \ref{subf:sbh-2-B-1}, once $r\gg 2M$.

Figure~\ref{fig:sbh-2-2} repeats the comparison for the same parameters, but this time for black holes ten times more massive, $M/\lambda=10$, than before.
Since, in both solutions under consideration, $B^2(r)$ depends on $q$, $\eta$, and $\lambda$ alone and not on $M$, panel \ref{subf:sbh-2-B-2} shows
exactly the same behavior for $B^2(r)$ as in Fig.~\ref{subf:sbh-2-B-1}.
The difference here is that the horizontal axis is measured in units of $r/(2M)$, with $M$ being ten times greater than before.
Hence, the transition region where $B^2(r)$ departs from unity is now confined to
$r/(2M)\lesssim 0.1$, an order of magnitude closer to the origin than for
$M/\lambda=1$. 
In other words, increasing the mass at fixed $\lambda$ squeezes the ``hairy core'' of size $\sim\lambda$ into an ever smaller fraction of the gravitational radius.
This is also apparent from panel \ref{subf:sbh-2-A-2}, which shows the dramatic consequence for the metric function $A(r)$.
Both hairy solutions are essentially indistinguishable from the Schwarzschild geometry over the entire plotted range $r/(2M)\in[0.1,300]$. 
The inset, zooming in on $r/(2M)\in[0.999,1.001]$, reveals that the three curves are nonetheless distinct at the $10^{-3}$ level.
This illustrates explicitly the decoupling of the primary scalar hair in the $M\gg\lambda$ limit relevant for astrophysical black holes.
An external observer effectively sees the Schwarzschild geometry, with the imprint of the scalar hair confined to a small near-horizon correction that becomes increasingly
difficult to resolve as $M/\lambda$ grows.

%%%%%%%%%%%%%%%%%%%%%%
%%%%%%%%%%%%%%%%%%%%%%
\begin{figure}[t]
    \centering

    \begin{subfigure}[b]{0.48\textwidth}
        \centering
        \includegraphics[width=\textwidth]{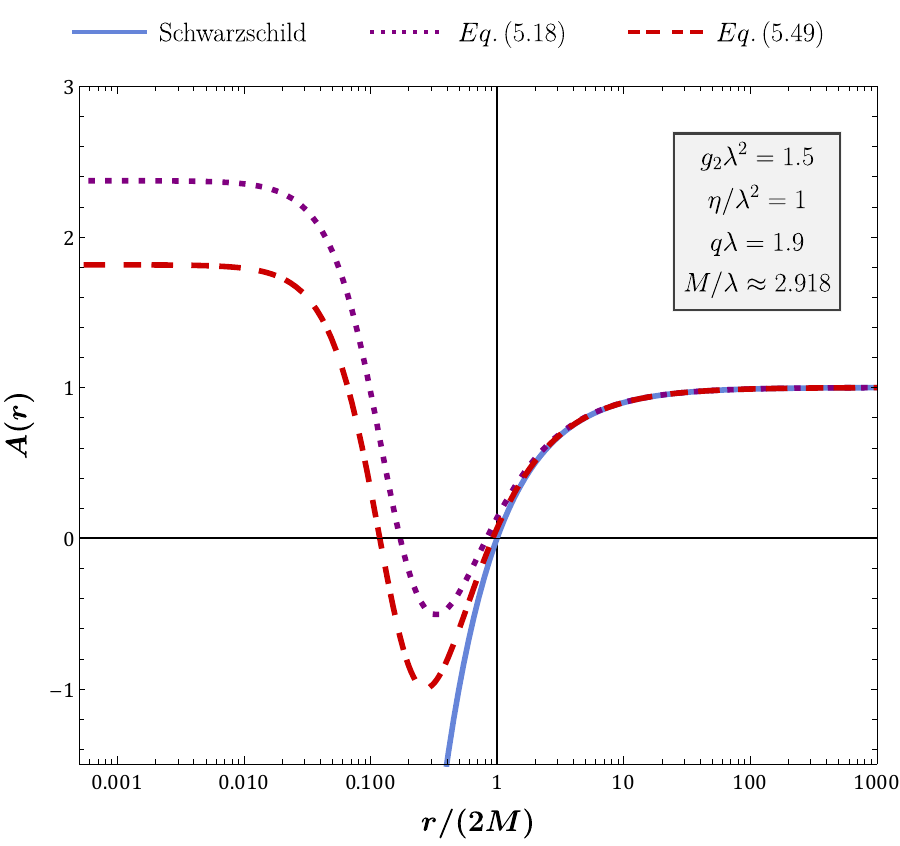}
        \caption{\hspace*{-1.5em}}
        \label{subf:rbh-2-A}
    \end{subfigure}
    \hfill
    \begin{subfigure}[b]{0.47\textwidth}
        \centering
        \includegraphics[width=\textwidth]{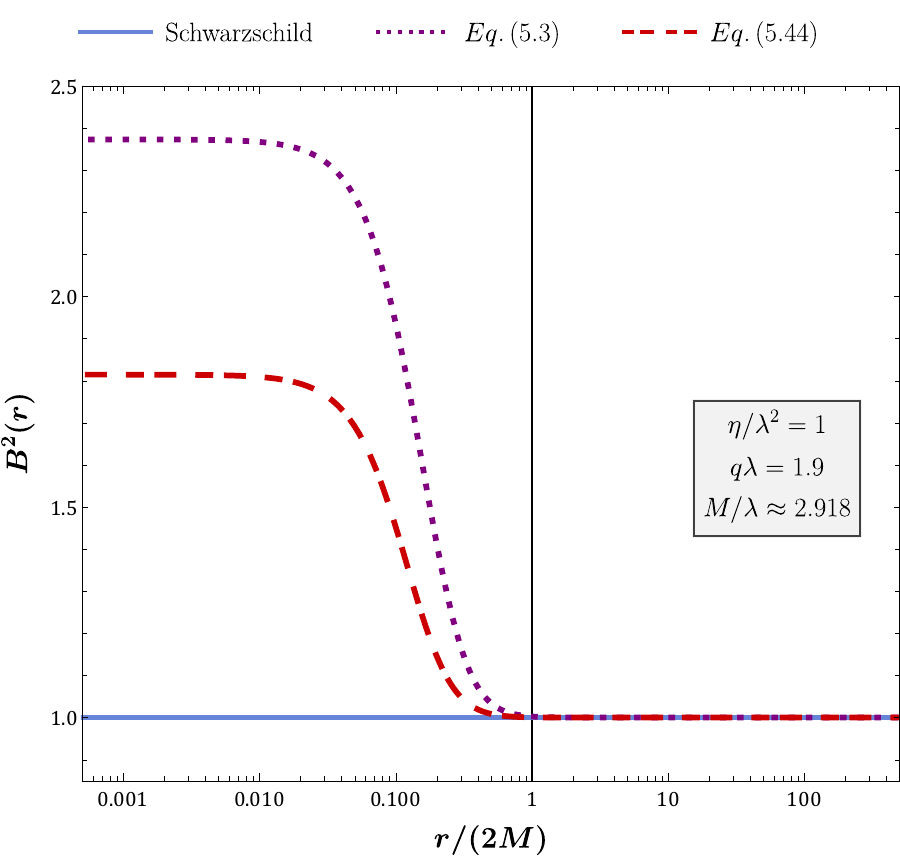}
        \caption{\hspace*{-2em}}
        \label{subf:rbh-2-B}
    \end{subfigure}

    \vspace*{-0.5em}

    \caption{Regular black-hole solutions for the parameters indicated in the panels.
    (a) Comparison of the metric function $A(r)$ for the two non-reciprocal solutions and the Schwarzschild geometry.
\newline    (b) Corresponding metric function $B^2(r)$ for the two non-reciprocal solutions.}
    \label{fig:rbh-2}
\end{figure}
%%%%%%%%%%%%%%%%%%%%%%

Finally, Fig.~\ref{fig:rbh-2} compares the two regular solutions, obtained once $M$ and $q$ are linked by the regularity condition \eqref{eq:reg-condition-power}, for $q\lambda=1.9$, $g_2\lambda^2 =1$, and $\eta/\lambda^2=1$, corresponding to $M/\lambda\approx 2.918$. 
Panel~\ref{subf:rbh-2-A} shows the metric function $A(r)$, together with $A_{\rm Schw}(r)$.
In sharp contrast with the singular cases above, and with the Schwarzschild solution itself, both regular solutions remain finite at the origin, with $A(0)=B^2(0)$ as required by the regularity condition.
Moving outward, $A(r)$ decreases from this finite value, crosses zero at an inner horizon, reaches a deep negative minimum, and rises back through zero at an outer (event) horizon, before joining the common Schwarzschild profile outside the event horizon. 
The resulting causal structure (an inner and an outer horizon surrounding a smooth core) is qualitatively the same for both $Z(X)$ profiles.
In both cases, the deviations from Schwarzschild decrease at sufficiently large radii, with the most significant differences concentrated in the inner and near-horizon regions. Although the regularization mainly affects the interior geometry, subleading corrections remain present outside the event horizon, where both solutions approach the Schwarzschild geometry asymptotically. As for the metric function $B^2(r)$, panel \ref{subf:rbh-2-B} shows a qualitative behavior similar to the singular cases. At the outer horizon of the regular black holes, $B^2(r)$ is effectively unity, confirming once more that in the causal region of spacetime neither regular solution can be easily distinguished from the Schwarzschild solution. 

Taken together, Figs.~\ref{fig:sbh-2-1}--\ref{fig:rbh-2} confirm that the two $Z(X)$ profiles produce quantitatively different, but qualitatively identical, departures from Schwarzschild, confined in every case to distances smaller or equal to a few $\lambda$ from the black-hole center.
These departures become parametrically small whenever the black hole mass is large compared to $\lambda$, while, once the regularity condition is imposed, they are replaced by a smooth, singularity-free core bounded by an inner and an outer horizon, leaving the Schwarzschild exterior undisturbed.

%==========================================
% Linear stability
%==========================================

\section{Linear stability}
\label{Sec:Stability}
We now investigate the linear stability of the black-hole solutions constructed above.
To this end, we consider both axial (odd-parity) perturbations and the monopole
$\ell=0$ sector of polar (even-parity) perturbations.
We begin by introducing the general perturbative setup common to both sectors.

We perturb both the metric and the scalar field around the static, spherically
symmetric backgrounds discussed above. To this end, we write
\begin{equation}
    g_{\mu\nu} = \bar g_{\mu\nu} + h_{\mu\nu}\,,\qquad
    \phi = \bar\phi + \delta\phi\,,
    \label{eq:pert-ansatz}
\end{equation}
where $\bar g_{\mu\nu}$ and $\bar\phi(t,r)$ characterize the background solution,
while $h_{\mu\nu}$ and $\delta\phi$ are treated as infinitesimal perturbations.
Since the background is spherically symmetric, it is natural to expand
$h_{\mu\nu}$ and $\delta\phi$ in terms of spherical harmonics
$Y_{\ell m}(\vartheta,\varphi)$ and their vector and tensor extensions.
The staticity of the background further allows a Fourier decomposition in time,
so that each $(\ell,m,\omega)$ mode evolves independently.

A crucial simplification follows from parity.
Under the spatial inversion
$(\vartheta,\varphi)\to(\pi-\vartheta,\varphi+\pi)$, the spherical harmonics
transform as $Y_{\ell m}\to(-1)^\ell Y_{\ell m}$, and the perturbations split
into two decoupled sectors: polar (even-parity) perturbations, with parity
$(-1)^\ell$, and axial (odd-parity) perturbations, with parity
$(-1)^{\ell+1}$.
Since $\delta\phi$ is a scalar perturbation, it belongs exclusively to the
even-parity sector. Thus, the axial sector contains only metric perturbations,
whereas the scalar degree of freedom enters the polar sector.

As shown in \cite{Charmousis:2026bhe,Langlois:2022ulw}, both monopole and axial perturbations are invariant under invertible disformal transformation. 
Following that remark, for the monopole sector we will work in the reciprocal frame where the computations are easier although all the computations can be carried over in any frame and give the same result.
As for the axial sector, the analysis will start in the original non-reciprocal frame and then a general disformal transformation will be used to recast the analysis in an effective GR frame.
For instance, the following analysis could be applied to the power-law model described in Sec.~\ref{power_law_model}, however, since it is disformally related to the solutions found in~\cite{Bakopoulos:2023sdm}, its analysis is already carried over in~\cite{Charmousis:2025xug,Charmousis:2026bhe}. We thus focus on the new classes of theory, characterized by $\mathcal F$, introduced in this article.

%==========================================
% Axial perturbations
%==========================================

\subsection{Axial perturbations}
\label{Sec:Axial}

Axial perturbations involve only the metric, since $\delta\phi$ has even parity, and constitute the simplest part of the linear problem.
We follow the effective-metric formulation of Refs.~\cite{Langlois:2022ulw,Charmousis:2025xug} (see also Ref. \cite{Kobayashi:2012kh}), in which the odd-parity sector of shift-symmetric quadratic DHOST theories reduces to the odd-parity sector of GR on a different geometry.
We collect below only the ingredients needed to follow the analysis and refer to those works for the derivations.

On a spherically symmetric background, different $(\ell,m)$ modes decouple and the resulting equations depend on $\ell$ only, so it suffices to consider a single mode.
In Regge--Wheeler gauge \cite{Regge:1957td}, the odd-parity metric perturbation of a given $(\ell,m)$ mode is encoded in two functions $h_0$ and $h_1$ of $(t,r)$, namely
\begin{gather}
    h_{t\vartheta} = \frac{h_0}{\sin\vartheta}\,\partial_\varphi Y_{\ell m}\,,\qquad
    h_{t\varphi} = -h_0\sin\vartheta\,\partial_\vartheta Y_{\ell m}\,,\nonumber\\
    h_{r\vartheta} = \frac{h_1}{\sin\vartheta}\,\partial_\varphi Y_{\ell m}\,,\qquad
    h_{r\varphi} = -h_1\sin\vartheta\,\partial_\vartheta Y_{\ell m}\,,
    \label{eq:axial-metric}
\end{gather}
with all other components vanishing.
The dipole $\ell=1$ is non-radiative and describes an infinitesimal shift of the angular momentum of the object.
The radiative spectrum therefore starts at $\ell\geq2$, as in GR.

Because $\bar\phi=qt+\psi(r)$ depends on time, the linearized equations mix $t$ and $r$.
This mixing is removed by the shifted time
\begin{equation}
    t_*=t-\int\!\mathrm{d}r\,\Psi(r)\,,\qquad
    \Psi=\frac{q\,\psi' A_1}{\Upsilon}\,,\qquad
    \Upsilon\equiv A\,Z-q^2A_1\,,\qquad
    A_1=\frac{Z-G_4}{2X}\,,
    \label{eq:t-star}
\end{equation}
where $A_1(X)$ is the quadratic-DHOST function of Eq.~\eqref{eq:quad-dhost2}, rewritten with the help of Eq.~\eqref{eq:def-Z}.
Since the background is static, the functions $h_0$ and $h_1$ can be decomposed into Fourier modes
\begin{equation}
h_{0,1}(t_*,r)=\int \dd \omega\, e^{-i\omega t_*}h_{0,1}(\omega,r)\,.
\label{eq:fourier}
\end{equation}
From this point on, we work exclusively in the frequency domain.
The two independent axial equations then form a first-order system for $\mathbf Y=(Y_1,Y_2)^{\rm T}$, with $Y_1=h_0$ and $\omega Y_2=h_1+\Psi h_0$,
\begin{equation}
    \frac{\mathrm{d}\mathbf Y}{\mathrm{d}r}=
    \begin{pmatrix}
        2/r & -i\omega^2+2i\lambda_\ell\Phi/r^2\\[1mm]
        -i\Gamma & \Delta
    \end{pmatrix}\mathbf Y\,,
    \qquad \lambda_\ell=\frac{\ell(\ell+1)}{2}-1\,,
    \label{eq:axial-system}
\end{equation}
whose coefficients depend on the background only through
\begin{equation}
    \Phi=\frac{\Upsilon}{Z}\,,\qquad
    \Gamma=\frac{B^2G_4Z}{\Upsilon^2}\,,\qquad
    \Delta=\frac{B'}{B}-\frac{\Upsilon'}{\Upsilon}\,.
    \label{eq:axial-coeffs}
\end{equation}

Combining the two first-order equations for $Y_1$ and $Y_2$ we can bring the system to a single Schr\"odinger-like equation for a master variable $\mathcal Y$,
\begin{equation}
    \frac{\mathrm{d}^2 \mathcal{Y}}{\mathrm{d} r_*^2}
    + \left[\frac{\omega^2}{c_*^2(r)} - \mathcal{V}(r)\right]\mathcal{Y} = 0 \,,
 \label{eq:schro-axial}
\end{equation}
where the tortoise coordinate $r_*$ is defined by
\begin{equation}
\frac{\mathrm{d} r}{\mathrm{d} r_*}=n(r)\,,
\end{equation}
and the master variable follows from the rescaling $Y_2=N(r)\mathcal{Y}(r)$, with 
\begin{equation}
    N^2(r) \propto n\, r^2 \Gamma \, \exp\bigg(\int \Delta \,\dd r\bigg) = \frac{n\, r^2 \Gamma B}{\Upsilon}\,.
\end{equation}
The propagation speed of the odd-parity mode is given by
\begin{equation}
    c_*^2(r)=\frac{1}{n^2 \Gamma}\,,
    \label{eq:axial-speed}
\end{equation}
while the effective potential reads
\begin{equation}
    \mathcal{V}(r) = \frac{(\ell-1)(\ell+2)}{c_*^2}\frac{\Phi}{r^2}+\frac{1}{\mathcal{P}}\frac{\dd^2 \mathcal{P}}{\dd r_*^2}\,,
    \qquad \mathcal{P} = \bigg(\frac{c_*^2}{r^4 G_4 Z}\bigg)^{1/4}\,.
    \label{eq:Veff-bg}
\end{equation}
In the forthcoming discussion on the effective metric of axial perturbations, it will become clear that $\mathcal{P}$ is related to the areal radius of an effective metric $\hat{g}_{\mu\nu}$, for which $c_*$ is identically unity.
For the Schwarzschild solution in GR, $Z=G_4=B=1$, $A_1=0$, $\Psi=0$, $\Phi=A$, $\Gamma=1/A^2$, $n=A=1-2M/r$, and $n^2\Gamma=1$, so that $c_*=1$ and Eq.~\eqref{eq:schro-axial} matches precisely the Regge--Wheeler equation.
In DHOST theories, on the contrary, $c_*$ generally differs from unity.
For the background metric, radial null rays satisfy $\dd t/\dd r =\pm (B/A)$.
By choosing $n=A/B$, then $\dd r_*/\dd t =\pm 1$, thus $r_*$ becomes the radial coordinate of the background metric $g_{\mu\nu}$ in which light propagates radially with unit speed. 
However, speed $c_*$ is the radial speed of axial gravitons relative to light.
For $c_*=1$ gravitons travel on the light cone of $g_{\mu\nu}$, whereas for $c_*\neq1$ their cone is wider ($c_*>1$) or narrower ($c_*<1$).
For the present family of theories and for $n=A/B$, one finds that
\begin{equation}
    c_*^2=\frac{\Upsilon^2}{A^2\,G_4Z}\,.
    \label{eq:cstar}
\end{equation}
This speed vanishes at the points where $\Upsilon=0$ and diverges at the event horizon, where $A=0$ but $\Upsilon=-q^2A_1\neq0$.
Axial gravitons therefore do not propagate on the light cone of $g_{\mu\nu}$.
Moreover, the sign of $c_*^2$ carries physical information: $c_*^2<0$, i.e.\ $G_4Z<0$, signals a gradient instability \cite{Tomikawa:2021pca, Takahashi:2021bml}.
The value of $c_*$ relative to unity, by contrast, depends on the light cone used as a reference, so $c_*^2>1$ is not a pathology by itself \cite{Langlois:2022ulw}.
The appropriate geometrical description is the effective metric introduced next, whose light cone is the one followed by axial gravitons.

\subsubsection{Effective metric}

The central result of Refs.~\cite{Langlois:2022ulw} (see also Ref.~\cite{Charmousis:2025xug}) is that Eqs.~\eqref{eq:axial-system} and \eqref{eq:axial-coeffs} are exactly the odd-parity equations of GR for the static metric
\begin{equation}
    \mathrm{d}\hat s^2=\sqrt{G_4Z} \Big[-\Phi\,\mathrm{d}t_*^2+\frac{G_4 Z}{\Phi}\,\mathrm{d}r^2+r^2\mathrm{d}\Omega^2\Big]\,.
    \label{eq:eff-metric}
\end{equation}
In the above, we have also used the fact that $B^2=Z^2$.
Note also that the condition $G_4 Z>0$, which is natural in this case, agrees with the no-ghost condition discussed in \cite{Tomikawa:2021pca, Takahashi:2021bml}.
Axial gravitons therefore follow the null geodesics of $\hat g_{\mu\nu}$, whereas light and matter follow those of the background metric $g_{\mu\nu}$.
The two metrics have different light cones and, in general, different horizons: the horizon $A(r_l)=0$ bounds the region from which light cannot escape, and the horizon $\Phi(r_g)=0$, i.e.\ $\Upsilon(r_g)=0$, plays the same role for axial gravitons.
For $q\neq0$ and $A_1\neq0$ at the light horizon, one has $\Upsilon|_{A=0}=-q^2A_1\neq0$, so the two horizons do not coincide. Only the region outside $r_g$ is relevant for the axial dynamics seen by an external observer. As shown in Appendix~\ref{App:Axial-horizons}, for the Pad\'e and power-law models satisfying $\mathcal F\leq0$ and $\mathcal F=o(\Xi)$ as $\Xi\to0$, the outermost light horizon of $g_{\mu\nu}$ lies inside the graviton horizon of $\hat g_{\mu\nu}$. Consequently, the divergence of $c_*$ at $A=0$ does not affect the causal exterior of the effective metric $\hat g_{\mu\nu}$.

One can also confirm that the above metric $\hat{g}_{\mu\nu}$ results from the background metric $g_{\mu\nu}$ through the disformal transformation
\begin{equation}
    \hat g_{\mu\nu}=C(X)\,g_{\mu\nu}+D(X)\,\partial_\mu\phi\, \partial_\nu\phi\,,
    \qquad C=\sqrt{G_4Z}\,,
    \qquad D=\frac{G_4(Z-G_4)}{2X\sqrt{G_4Z}}\,.
    \label{eq:disformal}
\end{equation}
Substituting $\phi=qt+\psi(r)$ and passing to $t_*$, one verifies that $\hat g_{t_*r}=0$ and that Eq.~\eqref{eq:eff-metric} follows.
The conformal function $C(X)$ rescales all null cones uniformly and does not change the causal structure; it nevertheless changes the effective areal radius $\hat R^2=C(X)\,r^2 = r^2 \sqrt{G_4 Z}$.
The disformal function $D(X)$ acts along the scalar gradient, i.e.\ orthogonally to the hypersurfaces of constant $\phi$.
The map is invertible for $C\neq 0$ and $C-2XD>0$.
For the choice \eqref{eq:disformal} one finds $C-2XD=G_4^2/\sqrt{G_4 Z}$, so the condition $G_4 Z>0$ already guarantees that $C$ is real and that the transformation is invertible.

Since the odd-parity perturbation equations coincide with the corresponding
GR odd-parity equations written on the effective geometry
$\hat g_{\mu\nu}$, they can be reduced to a Schr\"odinger equation with unit
propagation speed in the associated tortoise coordinate,
\begin{equation}
    \bigg[-\frac{\mathrm{d}^2}{\mathrm{d} \hat{r}_*^2} + \hat{\mathcal{V}}(r)\bigg]\mathcal{Y} = \omega^2 \mathcal{Y} \,,\qquad
    \frac{\mathrm{d}\hat r_*}{\mathrm{d}r}=\frac{\sqrt{G_4 Z}}{\Phi}\,,
    \label{eq:Schro-eff}
\end{equation}
The potential $\hat{\mathcal{V}}(r)$ has the following Regge--Wheeler form
\begin{equation}
    \hat{\mathcal{V}}(r) = (\ell-1)(\ell+2)\frac{\Phi}{r^2}+\hat R\,\frac{\mathrm{d}^2}{\mathrm{d}\hat r_*^2}\!\left(\frac1{\hat R}\right)  \,,
    \qquad \hat{R}^2=r^2\sqrt{G_4 Z}\,.
    \label{eq:Veff-eff}
\end{equation}
Comparing Eqs.~\eqref{eq:Veff-bg} and \eqref{eq:Veff-eff} makes the relation between $\mathcal{V}$ and $\hat{\mathcal V}$ explicit.
Essentially, the only difference between them is the value of $c_*$.
First, both potentials consist of the same angular term proportional to $(\ell-1)(\ell+2)$, except for the factor $c_*^2$ in $\mathcal{V}$.
Additionally, in the effective frame $\mathcal P=1/\hat R$, while in the background frame $\mathcal P=\sqrt{c_*}/\hat R$. 
The areal radius of the effective metric is therefore already present in $\mathcal V$, even though that potential is written in terms of the background metric. 
The additional factor $\sqrt{c_*}$ is the imprint of the propagation speed. 
For $c_*=1$, the two tortoise coordinates coincide and $\hat{\mathcal V}=\mathcal V$. 
Working with unit speed is more convenient, because the stability of the axial sector then follows from the standard theory of Schr\"odinger operators.
To prove the stability of the axial perturbations, we first examine the self-adjointness of the Schr\"{o}dinger operator associated with \eqref{eq:Schro-eff} and then study its positivity.

\subsubsection{Self-adjointness and positivity of the Schr\"{o}dinger operator}

A Schr\"odinger-like operator $H=-\dd^2/\dd x^2+V(x)$ on an interval $x\in(x_L,x_R)$ defines a unique, energy-conserving evolution only if it is self-adjoint. 
Only then is its spectrum real, so that the statement $\omega^2>0$ is meaningful. 
For two perturbations $\mathcal{Y}_1$ and $\mathcal{Y}_2$, integration by parts gives
\begin{equation}
    \langle\mathcal{Y}_2,H\mathcal{Y}_1\rangle-\langle H\mathcal{Y}_2,\mathcal{Y}_1\rangle
    =\bigg[\mathcal{Y}_1\frac{\dd \mathcal{Y}_2^*}{\dd x} - \mathcal{Y}_2^*\frac{\dd \mathcal{Y}_1}{\dd x}\bigg]_{x_L}^{x_R}\,,
    \label{eq:SA-boundary}
\end{equation}
and self-adjointness requires this boundary term to vanish for every finite-energy perturbation. 
Whether this happens without further input depends on the two independent solutions of $H\mathcal{Y}_1=\omega^2\mathcal{Y}_1$ near each endpoint~\cite{Reed:1975mmp2, Kral_Zettl}.
If only one of them is square integrable there (\emph{limit-point} case), normalizability alone fixes the behavior, and no boundary condition is needed. 
If both are square integrable (\emph{limit-circle} case), we must impose a boundary condition, i.e.\ choose a combination of the two. 
Different choices define different self-adjoint operators with different spectra, and the choice is additional physical input. 
The Friedrichs extension selects the less singular (principal) solution and always preserves the lower bound of $H$ on smooth, localized perturbations; other choices may or may not do so~\cite{Reed:1975mmp2}.

The operator governing the axial perturbations is $H=-\dd^2/\dd\hat r_*^2+\hat{\mathcal V}$. 
Since $\dd\hat r_*/\dd r=\sqrt{G_4Z}/\Phi$ and $\Phi$ vanishes at $r_g$, the tortoise coordinate diverges there, logarithmically for a simple zero as in the Schwarzschild case
(and faster for a degenerate one).
Hence, the horizon lies at $\hat r_*\to-\infty$, while the spatial infinity is at $\hat r_*\to+\infty$. 
Moreover, every term of $\hat{\mathcal V}$ in Eq.~\eqref{eq:Veff-eff} carries a factor $\Phi$, because each derivative with respect to $\hat r_*$ brings a factor $\Phi/\sqrt{G_4Z}$. 
The potential therefore vanishes at the horizon ($\hat r_*\to-\infty$), while $\hat{\mathcal V}\simeq\ell(\ell+1)/r^2$ at large $r$.
Since $\hat{\mathcal V}$ is also continuous in between, because the background is smooth and $G_4Z>0$ for $r>r_g$, it is bounded on the whole line.
Near either end, $\hat{\mathcal V}\to0$, and for a negative value of the spectral parameter, $\omega^2=-k^2$ with $k>0$ (i.e.\ a candidate growing mode $e^{-i\omega t_*}=e^{kt_*}$), the two solutions behave as $e^{\pm k\hat r_*}$. 
Only $e^{-k\hat r_*}$ is normalizable as $\hat r_*\to+\infty$, and only $e^{+k\hat r_*}$ as $\hat r_*\to-\infty$, so both endpoints are of limit-point type. 
Consequently, $H$ is self-adjoint with $\mathcal{Y}$, $\dd\mathcal{Y}/\dd\hat r_*$, and $\dd^2\mathcal{Y}/\dd\hat r_*^2$ square integrable, exactly as for a quantum particle scattered by a localized potential on the real line. 
In particular, the ingoing and outgoing conditions of quasinormal-mode calculations are not boundary conditions of this problem: they select non-normalizable mode solutions and play no role in the evolution of finite-energy perturbations. 
The spectrum of $H$ is thus real and its positivity can be established with the $S$-deformation method, which we now describe.

The $S$-deformation method~\cite{Kodama:2003jz, Ishibashi:2003ap, Kodama:2003kk, Kimura:2018eiv, Kimura:2018whv}, summarized in Appendix~\ref{App:S-deformation}, shows that $H$ has a non-negative spectrum even when $\hat{\mathcal V}$ is not positive everywhere. 
For any continuous function $S(\hat r_*)$, integration by parts rewrites the energy of a perturbation in terms of the deformed potential
\begin{equation}
    \hat{\mathcal V}_S\equiv\hat{\mathcal V}+\frac{\dd S}{\dd\hat r_*}-S^2\,,
\end{equation}
up to a boundary term. 
If this boundary term vanishes and $\hat{\mathcal V}_S\geq0$ on $\hat r_*\in(-\infty,+\infty)$, no mode with $\omega^2<0$ can exist. 
The task, therefore, reduces to finding a suitable $S$. For the potential~\eqref{eq:Veff-eff},the natural choice is built from $1/\hat R$, which is a nodeless solution of the
$\ell=1$ master equation at zero frequency, namely we have
\begin{equation}
    S\equiv\frac{\dd\ln\hat R}{\dd\hat r_*} = \frac{\Phi}{\sqrt{G_4Z}}\bigg(\frac{1}{r}+\frac{(G_4Z)'}{4G_4Z}\bigg)
    \qquad\Longrightarrow\qquad
    \hat{\mathcal V}_S=(\ell-1)(\ell+2)\,\frac{\Phi}{r^2}\,.
    \label{eq:Veff-eff-S}
\end{equation}
The function $S$ is bounded and vanishes at both ends of the real line.
It vanishes as $\Phi$ when $r\to r_g$ and as $1/r$ when $r\to+\infty$, thus, the boundary term vanishes as well (see Appendix \ref{App:S-deformation}). 
The deformed potential is strictly positive for $\ell\geq2$ on the whole domain, since $\Phi>0$ for $r>r_g$. 
Hence, no axial mode with $\omega^2<0$ exists for $\ell\geq2$, and axial
mode stability follows analytically without requiring
$\hat{\mathcal V}$ itself to be positive.

\subsubsection{Horizonless configurations}

When $\Upsilon>0$ for all $r>0$, as for the solitons, the effective metric has no horizon and the axial problem lives on $r>0$. 
At a regular center, $\Phi$ and $G_4Z$ are smooth, with $(G_4Z)(0)>0$ and $\Phi(0)>0$.
The tortoise coordinate is then finite there, $\hat r_*\in(0,+\infty)$ with $\hat r_*\propto r$, and the center becomes an endpoint at finite distance. 
Since at $\hat{r}_*\to 0$, $\hat R\propto\hat{r}_*[1+\mathcal{O}(\hat{r}_*)]$, the function $S$ of Eq.~\eqref{eq:Veff-eff-S} behaves as $S\simeq1/\hat r_*+\mathcal{O}(1)$, and the potential takes the form,
\begin{equation}
    \hat{\mathcal V}\simeq\frac{\hat{\mathcal V}_0}{\hat r_*^2}\,,\qquad
    \hat{\mathcal V}_0=2+(\ell-1)(\ell+2)\,\frac{G_4Z(0)}{\Phi(0)}\geq2\,.
\end{equation}
Consequently, since $\hat{\mathcal V}_0>3/4$, for $\ell\geq 2$, the center is of limit-point type and does not require a boundary condition (see Theorem X.10 in \cite{Reed:1975mmp2}).
Normalizable perturbations vanish there as $\hat r_*^{\,s_+}$ with $s_+\geq2$, so the boundary term of the $S$-deformation vanishes, while the behavior at infinity is
unchanged. 
The argument of Eq.~\eqref{eq:Veff-eff-S} thus carries over unchanged, and the
horizonless configurations are axially stable whenever $G_4Z>0$ for all $r\geq0$.

\subsubsection{Branch II solutions}

In Branch II, $G_4Z=g_4$ is constant, and so is the conformal factor $C=\sqrt{g_4}$ in the effective metric \eqref{eq:eff-metric}. 
For $g_2=0$, one can show that $\Phi = g_4 + C_0/r$, where $C_0$ is the integration constant of the expression \eqref{Branch2MetricA}.
For a vanishing integration constant, $\Phi=g_4$ and the effective metric is flat up to a constant factor, i.e.
\begin{equation}
    \dd\hat s^2=\sqrt{g_4}\,\big[-\dd\tau^2+\dd r^2+r^2\dd\Omega^2\big]\,,\qquad
    \tau=\sqrt{g_4}\,t_*\,.
\end{equation}
Correspondingly, $\hat r_*=r/\sqrt{g_4}$ and
\begin{equation}
    \hat{\mathcal V}=\frac{\ell(\ell+1)}{\hat r_*^2}\,.
\end{equation}
The axial problem is that of flat space, thus, the center is of limit-point type
($\hat{\mathcal{V}}_0=\ell(\ell+1)\geq6$), the spectrum is purely continuous and non-negative, and there is neither a bound state nor quasinormal ringing. 
All the hair, carried by $q$, $X(r)$ and $G_4(X)$, drops out of the axial dynamics.

For the regular solutions with $g_4=1$, the effective metric is exactly Minkowski, while the original metric $g_{\mu\nu}$ has horizons, $A=0$, inside the domain.
Axial gravitons therefore cross the horizon of $g_{\mu\nu}$ smoothly: matter is trapped, gravitons are not.

For $g_2=0$ and a nonzero integration constant $C_0$, the effective metric is Schwarzschild up to a constant factor, and axial stability reduces to the classic Schwarzschild result~\cite{Wald:1979lth}. 
In every case, the only requirement is $g_4>0$, which ensures $G_4Z>0$ and a real conformal factor.

%==========================================
%Polar Perturbations and the monopole sector
%==========================================

\subsection{Monopole Sector}
\label{Sec:Polar}

In this section, we study the polar $\ell=0$ (monopole) perturbations.
As in the axial sector, the monopole sector of generic scalar-tensor
theories contains a single propagating degree of freedom, in contrast
to GR, where no propagating monopole mode exists as a consequence of
Birkhoff's theorem.
Throughout this section, unless stated otherwise, $A$, $X$, $G_i$
and $F_4$ denote reciprocal-frame quantities; we omit tildes to simplify
the notation.

We closely follow the analysis and notation of
Ref.~\cite{Charmousis:2026bhe}, to which we refer the interested reader
for further details. Here we recall only the ingredients needed for
our analysis. We parametrize the polar $\ell=0$ perturbations as follows:
\begin{align}
&h_{t t}  =A(r)  H_0(t, r) \, ,\quad
h_{t r} = H_1(t, r) \, , \quad 
h_{r r}  =\frac{1}{A(r)}  H_2(t, r) \, , \\
&h_{a b}  = K(t, r)\,  \bar{g}_{a b}
\quad (a,b \in \{\vartheta,\varphi \})
\,, \qquad
\delta \phi= \delta \phi(t, r) \, .
\end{align}
and we adopt the gauge conditions 
 \begin{equation}
 \label{gauge}
     K=0,\qquad \delta \phi=0 \, ,
 \end{equation}
which require $q\neq0$. Then, one can solve the perturbed EOMs for one adequately defined master variable, and, provided that the determinant $\Det$ introduced in \cite{Charmousis:2026bhe} is non-vanishing, obtain a second-order equation. This equation can then be cast into Schr\"odinger form,
although the coordinates differ from those used in the axial sector.
\begin{equation}
\label{Schrodinger_equation_generic}
\Schroop\sigma=-\frac{\mathrm{d}^2 \sigma}{\mathrm{d}r_*^2}+ \Veff \, \sigma \; = \; \omega^2 \, \sigma \, ,
\end{equation}
where $\Schroop$ is the analog of $H$ and $\sigma$ the analog of $\mathcal{Y} $ in the monopole case and where the coordinates are now defined via 
\begin{equation}
     r_*=\int \frac{\mathrm{d}r}{n(r)}\, ,\qquad  t_*=t +\int W(r)\mathrm{d}r \, .
     \label{Schro_coordinates}
\end{equation}
As for the axial perturbation, we study the stability of the monopole perturbation, proceeding in two steps, first examining the self-adjointness of the Schr\"odinger operator $\Schroop$ associated with \eqref{Schrodinger_equation_generic} before analyzing its positivity.

Before turning to Branch~III, let us briefly comment on Branch~II.
As established in Sec.~\ref{Sec:Rec-BranchII}, provided that the corresponding
conformal--disformal transformation is regular and invertible (i.e. that $0<g_4<\infty$), Branch~II
propagates only the two tensor degrees of freedom and contains no independent
propagating scalar mode. Accordingly, there is no independent scalar-monopole
stability problem to analyse in this branch. Consistently with this result, the
determinant $\Det$ entering the generic monopole reduction of
Ref.~\cite{Charmousis:2026bhe} vanishes identically for Branch~II. However, for completeness and to highlight that this conclusion might not be so trivial to obtain at the level of linear perturbations, we give more details on the monopole equations of Branch II in Appendix~\ref{App:BranchII}. 

In contrast to the khronometric-hair Branch~II, Branch~III supports an
independent propagating scalar monopole degree of freedom, consistently
with its interpretation as the scalar-hair branch. For the non-degenerate Branch-III models considered below, the monopole
perturbation equations can be reduced to the Schr\"odinger form introduced
above, provided that the determinant of the generic reduction does not vanish.

\subsubsection{Schr\"odinger formulation.}
We first determine the coordinates entering Eq.~\eqref{Schro_coordinates}.
For the non-degenerate Branch-III theories considered here, the time coordinate $t_*$ is directly
proportional to the background scalar field:
\begin{equation}
\label{tstar_p=2}
t_*=t+\frac1q\psi=\frac1q\bar\phi\,.
\end{equation}
Unlike the original static coordinates $(t,r)$, which become singular
at the metric horizon, the scalar-adapted time coordinate
$t_*=\bar\phi/q$ remains regular across the future horizon.
Moreover, since $X=\frac{q^2}{2}\frac{1}{1+r^2/\lambda^2}>0$, the gradient of $t_*$ is timelike wherever
the background is regular. This coordinate can therefore be extended
into the black-hole interior, up to the central singularity for
singular solutions, as discussed in Ref.~\cite{Charmousis:2026bhe}.

The corresponding radial coordinate is given by
\begin{equation}
   r_*=\int \frac{\mathrm{d}r}{n(r)}\,,
   \qquad
   n(r)=\frac{q^2 }{r}\sqrt{\lambda ^2\frac{A(r)^2 \psi '(r)^2}{8 X^3}\frac{  G_{4X}}{ \left(2 X G_{4XX}+G_{4X}\right)}}\,.
\end{equation}
From the expression above, in order for the radial coordinate to be real and non-degenerate
(with $X>0$ and a real scalar profile), we consider only the domain of $r$ where
\begin{equation}
\frac{G_{4X}}
     {G_{4X}+2XG_{4XX}}>0,
\qquad
G_{4X}+2XG_{4XX}\neq0,
\end{equation}
We restrict our analysis to the parameter ranges for which $n(r)$
is real and non-degenerate throughout the radial domain $0<r<+\infty$.
In this case, the monopole radial coordinate $r_*$ extends beyond
the exterior region of the black hole and covers its interior.
The endpoint $r=0$ corresponds to a curvature singularity for
singular black holes and to a regular center for regular configurations.
We choose the integration constant such that $r_*(0)=0$.

Moreover, as we shall see, at spatial infinity
\begin{equation}
    n(r)\propto r^2
\end{equation}
so that the integral for $r_*$ converges both as $r\to0$ and as $r\to+\infty$, so that $r_*$ is defined on a compact interval of the form $[0,\rmax]$.

Finally, the effective potential of equation ~\eqref{Schrodinger_equation_generic} takes the compact form:
\begin{align}
\Veff =  S^2 - \frac{\dd S}{\dd r_*}\,,\qquad S \equiv - \frac{\dd}{\dd r_*} \ln\bigg[X \sqrt{\frac{n}{A(r)^2 \psi '(r)^2G_{4X}}}\,\bigg]
\label{eq:Veff}
\end{align}

and we notice that in the monopole case, the potential $V_S$ introduced in \eqref{eq:Def_VS} vanishes.

\subsubsection{Asymptotic Analysis}
In order to assess the self-adjointness of the Schr\"odinger operator, one needs to analyze the asymptotic behavior of the potential \eqref{eq:Veff} in the new radial coordinate $r_*$.
When $r\to 0$, two cases arise depending on whether we consider a singular or a regular BH.
When the BH is singular, one can show that
\begin{equation}
\label{V_limit_0}
    \Veff(r_*) \; = \; 
    \frac{11 }{100\ r_*^2} (1+ {\cal O}(r_*^{4/5}))\,\quad {\rm when} \quad r_*\to 0 \, .
\end{equation}
while when it is regular, due to behavior described by Eq.~\eqref{eq:regular-AB}, 
\begin{equation}
    \Veff(r_*)=\frac{2}{r_*^2}+\mathcal O(1)\quad {\rm when} \quad r_*\to 0 \, .
    \label{Vreg_limit_0}
\end{equation}
In the opposite limit, when $r\to\infty$, $r_*$ tends to a finite limit, which we will call $\rmax$, and we introduce the notation:
\begin{equation}
\label{ustar}
          u_*\, \equiv \, \rmax - r_*  \,.
\end{equation}
\noindent Notice that for all the theories at stake, one can easily check that
\begin{equation}
    G_4(0)=1, \qquad G_{4X}(0)=0\, ,
\end{equation}
and two cases arise depending on the value of $G_{4XX}(0)$.
Either, $G_{4XX}(0)\neq0$, like for the Pad\'{e} theories, and the effective potential tends to a constant given by:
\begin{equation}
    \Veff(r_*) \; = \frac{3 G_{4XX}(0)-q^2G_{4XXX}(0) }{9 \lambda ^2 G_{4XX}(0)}(1+ {\cal O}(u_*))
\end{equation}
When instead $G_{4XX}(0)= 0$, like for the superposition theory when $\pmin\equiv\min(p_1,p_2)>2$, which we assume in the following, then:
\begin{equation}
    \Veff\; \propto \, \frac{1}{u_*^2} \left( 1 + {\cal O}(u_*)\right)\quad {\rm when} \quad r_*\to \rmax
   \, .
\end{equation}
Now that the asymptotic behaviors of the potentials are known, one can proceed to define a self-adjoint Schr\"odinger operator. 

\subsubsection{Self-adjointness of the Schr\"odinger Operator}
The  construction of a self-adjoint extension of the Schr\"odinger operator $\SchroopE$ (see \cite{Kral_Zettl} for instance)  depends  on the asymptotic behaviors of the solutions at both extremities of the interval $[0,\rmax]$, which can be determined explicitly.
Let's first consider the superposition theory such that $\pmin>2$. In this case, the two end points of the interval $[0,\rmax]$ are singular~\cite{Reed:1975mmp2}.
Starting with the limit $r_*\to \rmax$, we find that the two asymptotic solutions of the Schr\"odinger equation are given by
\begin{equation}
\label{asymptr02}
\sigma=\hat\sigma_+ + \hat\sigma_- \, , \qquad 
   \hat\sigma_\pm(r_*) = b_\pm \, u_*^{s_\pm} \left[ 1 +{\cal O}(u_*) \right] \, , \quad u_*=\rmax - r_* \, ,
\end{equation}
with  $s_+=p_{\rm min}-1$ and $s_-=2-p_{\rm min}$. The solution $\hat\sigma_+$ is square integrable for all
$p_{\rm min}>2$, whereas
\begin{equation}
\int^{\epsilon}\dd u_*\,|\hat\sigma_-|^2
\propto
\int^{\epsilon}\dd u_*\,u_*^{\,4-2p_{\rm min}}
\end{equation}
converges only for $p_{\rm min}<5/2$. Therefore, for
$2<p_{\rm min}<5/2$ the asymptotic endpoint is limit-circle, whereas for
$p_{\rm min}\geq5/2$ it is limit-point. In the explicit superposition model
considered here, $p_{\rm min}=5/2$, and hence the endpoint is limit-point.
We notice that this result is exactly the one given in  \cite{Charmousis:2026bhe}. The reason is that since $X\to 0$ at spatial infinity, the asymptotic analysis of $G_4$ is dominated by the lowest power of $X$, $\pmin$ and the superposition model is asymptotically approximated by the power law model described in\cite{Charmousis:2026bhe}.

Turning now to the limit $r_*\to 0$.
Starting with the case of a regular BH, the Schr\"odinger equation described by \eqref{Vreg_limit_0} admits the two following asymptotics solutions:
\begin{equation}
\label{asymptu0Mu0}
\sigma=\sigma_+ + \sigma_- \, , \qquad 
\sigma_\pm\left(r_*\right) = a_\pm \, r_*^{s_\pm}
\left[1+ {\cal O}(r_*^2)\right]\,,
\end{equation}
with $s_{+}=2,s_{-}=-1$. Although the background geometry is regular at $r=0$, the endpoint $r_*=0$ is singular for the Schr\"odinger operator. Since only $\sigma_+$ is square integrable, this endpoint is of limit-point type, and no additional boundary condition is required there to obtain a self-adjoint extension. In the case of a singular BH, using \eqref{V_limit_0}, the  solutions of the Schr\"odinger equation  behave as
\begin{equation}
\label{asymptr0}
\sigma_\pm\left(r_*\right) = a_\pm \, r_*^{s_\pm}
\left[1+ {\cal O}(r_*^{4/5})\right]\,,
\end{equation}
with $s_+=11/10$ and $s_-=-1/10$.
Both solutions $\sigma_\pm$ are square-integrable, which means that the endpoint $r_*=0$ is in the so-called limit-circle case. 
There is therefore a one-parameter family of boundary conditions at the point $r_*=0$ that defines a self-adjoint extension of the operator. These boundary conditions, that lead to a symmetric operator, can be expressed in terms of the asymptotics \eqref{asymptr0} as
\begin{eqnarray}
\label{boundcondalpha}
a_+  \cos \Theta  +  a_-  \sin\Theta  =  0 \ ,
\end{eqnarray}
where $\Theta  $ is, at this stage, an arbitrary angle.  If one further requires the solution to be regular, one obtains the so-called Friedrichs extension~\cite{Reed:1975mmp2}, which corresponds to imposing $a_-=0$, or equivalently to choosing $\Theta =\pi/2$ in the above parametrisation.

We now consider the Pad\'{e} theories.
Since the behaviors \eqref{V_limit_0} and \eqref{Vreg_limit_0} are common to
the superposition and Pad\'e theories, the same analysis applies at $r\to0$. More explicitly, for the regular solution, Eq.~\eqref{Vreg_limit_0}  applies and the endpoint is limit-point, so no additional boundary condition is needed. For the singular solutions, Eq.~\eqref{V_limit_0} applies and the endpoint is limit-circle; choosing the Friedrichs extension then amounts to setting the non-regular coefficient to zero. Numerically these choices may lead to the same vanishing behaviour at the origin, but conceptually the two cases should not be conflated.

Regarding now the limit $r \to \infty$, since the potential tends to a constant, $r_*=\rmax$ is a regular, and therefore limit-circle, endpoint.
In other words, at this end too, the two solutions are square integrable and one needs to fix boundary conditions.
For simplicity, we pick the restrictive condition  that $\sigma$ satisfies the same boundary condition as in the superposition theories discussed above: we select the Dirichlet self-adjoint extension $\sigma(\rmax)=0$. 
The positivity analysis below therefore refers to this choice of domain.
To conclude, for the models considered here, the admissible perturbations vanish at both endpoints, either as a consequence of the limit-point behavior or through the boundary conditions selected at limit-circle endpoints. We denote by $\SchroopE$ the corresponding self-adjoint realization of the Schr\"odinger operator.

\subsubsection{Positivity of the spectrum}
We now briefly discuss the boundary terms entering the $S$-deformation argument described in App.~\ref{App:S-deformation}. Since the analysis closely follows that performed previously for the power-law theories, we only report here the main steps. Since, for the monopole sector, $V_S=0$, it is sufficient to show that the boundary terms appearing in \eqref{eq:S-identity} vanish.

Let us first consider the superposition theories. At spatial infinity, the asymptotic behavior is controlled by the smallest power $\pmin$ entering $G_4$. For the explicit superposition model considered here,
$p_{\rm min}=5/2$, and the only square-integrable asymptotic solution is
\begin{equation}
    \sigma(r_*)
    =
    b_+\,u_*^{\pmin-1}
    \left[1+\mathcal O(u_*)\right].
\end{equation}
The corresponding asymptotic behavior of the $S$ function is
\begin{equation}
    S(r_*)
    =
    \frac{2-\pmin}{u_*}
    \left[1+\mathcal O(u_*)\right],
\end{equation}
and hence
\begin{equation}
   \sigma^*
   \left(
       \frac{\dd\sigma}{\dd r_*}+S\sigma
   \right)
    =
    (3-2\pmin)|b_+|^2
    u_*^{2\pmin-3}
    \left[1+\mathcal O(u_*)\right].
\end{equation}
Since $p_{\rm min}=5/2$ for the explicit superposition model, this quantity vanishes as $r_*\rightarrow\rmax$. We therefore obtain
\begin{equation}
    {\Bterm}_{\rm max}=0
    \qquad\text{for the superposition theories.}
\end{equation}

For the Pad\'e theories, the endpoint $r_*=\rmax$ is regular. Contrary to the superposition theories, the $S$ function vanishes linearly as $r_*\to\rmax$,
\begin{equation}
    S_{\rm I}(r_*)
    =
    \frac{1+q^2\lambda^2}{3\lambda^2}\,u_*
    +\mathcal O(u_*^2),
    \qquad
    S_{\rm II}(r_*)
    =
    \frac{1+2q^2\lambda^2}{3\lambda^2}\,u_*
    +\mathcal O(u_*^2).
\end{equation}
As specified in the previous subsection, we select the Dirichlet self-adjoint extension, $\sigma(\rmax)=0$. Since the endpoint is regular, the corresponding solution behaves as
\begin{equation}
    \sigma(r_*)
    =
    b_+\,u_*
    \left[1+\mathcal O(u_*)\right].
\end{equation}
It then follows that the boundary term in \eqref{eq:S-identity} vanishes,
\begin{equation}
   \sigma^*
   \left(
       \frac{\dd\sigma}{\dd r_*}+S\sigma
   \right)
    =
    -|b_+|^2u_*
    +\mathcal O(u_*^2)
    \longrightarrow0,
\end{equation}
and consequently
\begin{equation}
    {\Bterm}_{\rm max}=0
\end{equation}
for the Pad\'e theories as well. Contrary to the superposition case, this result follows from the choice of the Dirichlet self-adjoint extension at the regular endpoint.

Let us now turn to the boundary at $r_*=0$. The analysis is common to the superposition and Pad\'e theories, since they share the same asymptotic behavior at the origin. For the regular solutions, using the asymptotic behavior given in Eq.~\eqref{eq:regular-AB}, one finds
\begin{equation}
    S(r_*)
    =
    \frac{1}{r_*}
    +\mathcal O(1).
\end{equation}
Together with the square-integrable behavior selected by the self-adjoint domain, this gives
\begin{equation}
    \sigma^*
    \left(
        \frac{\dd\sigma}{\dd r_*}+S\sigma
    \right)
    =
    3|a_+|^2r_*^3
    \left[1+o(1)\right]
    \longrightarrow0,
\end{equation}
and hence
\begin{equation}
    {\Bterm}_0=0.
\end{equation}

For the singular black-hole solutions, using instead the asymptotic behavior given in Eq.~\eqref{asymptr0}, the $S$ function behaves as
\begin{equation}
    S(r_*)
    =
    \frac{1}{10r_*}
    \left[1+\mathcal O(r_*^{4/5})\right].
\end{equation}
For the Friedrichs extension selected in the previous subsection, $a_-=0$, and therefore
\begin{equation}
    \sigma^*
    \left(
        \frac{\dd\sigma}{\dd r_*}+S\sigma
    \right)
    =
    \frac{6}{5}|a_+|^2r_*^{6/5}
    \left[1+o(1)\right]
    \longrightarrow0.
\end{equation}
Thus,
\begin{equation}
    {\Bterm}_0=0.
\end{equation}

Combining this result with the analysis at $r_*=\rmax$, we obtain
\begin{equation}
    {\Bterm}_0={\Bterm}_{\rm max}=0
\end{equation}
for both the superposition and Pad\'e theories. Thus, although $V_{\rm eff}$ is not pointwise positive, the corresponding self-adjoint Schr\"odinger operator is non-negative on the chosen domain. Its spectrum therefore contains no negative eigenvalues, excluding exponentially growing normal modes with $\omega^2<0$.

Before concluding , let us comment on the
validity of the scalar perturbative description.
For the Branch-III backgrounds analysed in this section, $X\to0$
and $A\to1$ as $r\to+\infty$, so that $\psi'/q\to\pm1$, and the scalar gradient becomes null.
The radial characteristics of the time-dependent equation associated
with Eq.~\eqref{Schrodinger_equation_generic} satisfy
$\dd t_*=\pm\dd r_*$. Using Eqs.~\eqref{tstar_p=2}
and \eqref{Schro_coordinates}, one obtains
\begin{equation}
    \frac{\dd t}{\dd r}
    =-\frac{\psi'}{q}\pm\frac{1}{n(r)}
   \xrightarrow{r\to+\infty}\mp1.
    \label{eq:monopole-asymptotic-characteristics}
\end{equation}
Since $n(r)\propto r^2$, the two radial characteristics merge in the
asymptotically Minkowski limit. This degeneration raises a possible
strong-coupling problem for scalar
fluctuations~\cite{Candan:2025fbl}.

At finite radius, perturbative control depends on the background
cutoff and on the local frequency and momentum scales of the
perturbations. An asymptotic loss of control may therefore become
relevant already at a finite radius. Determining whether this occurs
requires an analysis of the nonlinear
interactions~\cite{deRham:2019gha,Takahashi:2021bml} and cannot be
inferred from the positivity of the linear operator. The analysis above concerns the spectral properties of the reduced
monopole equation for the self-adjoint domains specified here.
It does not resolve the asymptotic strong-coupling issue.
This distinction is relevant to physical predictions involving the
scalar sector, including scalar or coupled polar quasinormal modes,
whose boundary conditions must be imposed or matched within a
controlled description. The nonlinear analysis of the asymptotic
regime and its consequences for such predictions are left for
future work. 

%==========================================
%Conclusions
%==========================================

\section{Conclusions}
\label{Sec:Conclusions}

In this work, we have developed a systematic and fully analytic method for constructing non-reciprocal, exact, static, spherically symmetric, and asymptotically flat black-hole solutions with primary hair in shift- and $Z_2$-symmetric beyond Horndeski (quadratic DHOST) theories. Unlike previous constructions, which impose the restrictive reciprocal condition $g_{tt}g_{rr}=-1$, our approach retains two independent metric functions, $A(r)$ and $B(r)$, and thereby accesses a substantially larger space of exact solutions.

We showed in Sec.~\ref{Sec:Disformal-Frame} that every solution obtained in this way is nevertheless disformally equivalent, through a \emph{unique} regular and invertible purely disformal transformation, to a solution of a (generally different) DHOST theory satisfying the reciprocal condition. The field equations organize the solution space into two main branches: Branch~II, which carries khronometric hair, for which the Noether current identically vanishes and no independent scalar degree of freedom propagates; and Branch~III, which carries genuine scalar hair, with a nonvanishing Noether current and an independent propagating scalar degree of freedom. For Branch~II, we constructed an explicit exact asymptotically flat regular
black-hole family with primary khronometric hair, for which both the metric
and the scalar field admit smooth extensions through the center. Focusing on Branch~III, in Sec.~\ref{sec:branch3} we constructed four distinct families of exact non-reciprocal black-hole solutions and showed, in Sec.~\ref{sec:IVB}, that the construction is not tied to a single choice of the auxiliary function $Z(X)$. Dressing the same $\mathcal F(\Xi)$ with a different, independently chosen $Z(X)$ produces a genuinely different, inequivalent geometry in the original frame, while both remain disformally connected to one and the same reciprocal-frame solution. In every case, the solutions are generically singular at the origin and carry primary scalar hair, but we showed that a single relation between the black-hole mass and the scalar charge can remove the central curvature singularity, yielding either a regular two-horizon black hole or a horizonless soliton, with the scalar hair becoming secondary in the process.

We further examined the linear stability of every family constructed above. For the axial (odd-parity) sector (Sec.~\ref{Sec:Axial}), the perturbation equations reduce to the corresponding GR odd-parity problem on an effective geometry. The associated Schr\"odinger operator is non-negative whenever the effective metric is regular, i.e. $G_4Z>0$ throughout the
relevant domain, and the boundary conditions entering the $S$-deformation are satisfied. Under these conditions, exponentially growing axial modes are excluded analytically. For the polar monopole ($\ell=0$) sector (Sec.~\ref{Sec:Polar}), Branch~II contains no independent propagating scalar mode, whereas Branch~III generically supports one. For the non-degenerate superposition and Pad\'e models analyzed explicitly, we show that the corresponding Schr\"odinger operators are non-negative on the specified self-adjoint domains. This follows from the $S$-deformation method, for which the deformed potential vanishes and the boundary terms vanish under the selected boundary conditions. Consequently, exponentially growing monopole normal modes are excluded within the quadratic analysis.

Taken together, these results show that non-reciprocal black holes with khronometric or scalar hair, together with the regularized secondary-scalar-hair configurations of Branch~III, constitute a broad class of exact solutions of beyond Horndeski gravity, extending well beyond the reciprocal ansatz previously considered in the literature. For the representative configurations studied perturbatively, our results further provide evidence for linear stability under the sectors considered.
As discussed in Sec.~\ref{Sec:Polar}, the scalar monopole sector of
Branch~III becomes degenerate in the asymptotic Minkowski limit.
Indeed, as $X\to0$, the two radial monopole characteristics merge,
signalling strong coupling of the scalar degree of freedom in this limit.
Although our analysis establishes the non-negativity of the reduced
Schr\"odinger operator, this does not guarantee perturbative control
of the scalar sector. In particular, determining how far this
strong-coupling regime extends requires studying the nonlinear
interactions and the corresponding cutoff. An interesting possibility
would be to consider non-asymptotically flat configurations with
$X\to X_\infty\neq0$, which could avoid this particular asymptotic
degeneracy.
To this end, one could consider families of solutions that respect a different relation between $\mathcal{G}_2$ and $\mathcal{G}_4$, not necessarily the one used in Branch-III, i.e. \eqref{eq:G2-G4}.

Several directions naturally follow from this work. 
On the model-building side, it would be interesting to classify more systematically the space of admissible pairs $(Z(X),\mathcal F(\Xi))$, and to explore whether other branches of the general quadratic DHOST action~\eqref{eq:Quad-DHOST} admit similarly rich, non-reciprocal solution spaces. 
On the perturbative side, a natural continuation is the analysis of the full
polar $\ell\geq2$ sector, where the scalar and metric perturbations mix
nontrivially.
Finally, extending the present construction to slowly and rapidly rotating configurations, along the lines already developed for reciprocal DHOST black holes~\cite{BenAchour:2020fgy, Charmousis:2019fre, Candan:2025fbl}, and to compact-star analogues in the spirit of~\cite{Boumaza:2026dix}, would open the way to confronting these solutions with astrophysical and gravitational-wave observations.

%==========================================
%Acknowledgements
%==========================================

\begin{center}
    \textbf{Acknowledgements}
\end{center}
The work of S.B. and T.N. is supported by the Institute for Basic Science (IBS-R018-D3). We are particularly grateful to Karim Noui for his invaluable feedback and discussions, and we also thank Pavel Petrov and Mohammad Ali Gorji for helpful discussions.

\appendix

%==========================================
%Appendix-Asymptotic Behavior
%==========================================

\section{Asymptotic behavior of solutions in Branch III}
\label{App:Asym-III}

In this Appendix, we discuss general characteristics of the asymptotic behavior of the metric function $A(r)$ as it is given in the Branch III, i.e. Eq.~\eqref{eq:A-sol}. 
Since we are interested in asymptotically flat solutions describing black holes, the asymptotic expansion of $A(r)$, as $r\to+\infty$, should be of the form
\begin{equation}
A(r) = 1 - \frac{2M}{r}+ o(r^{-1})\,,
\end{equation}
where $M$ is identified as the ADM mass of the compact object.
From \eqref{eq:A-sol}, we notice that the first two terms already capture the desired asymptotic behavior.
Consequently, we deduce that the remaining two terms must decay at least as fast $1/r$ as $r\to +\infty$.
Hence, we readily infer that the following relation must be satisfied
\begin{equation}
    \bigg(1+\frac{r^2}{\lambda^2}\bigg)[B^2(r)-1]=\mathcal{O}(r^{-1})\,,\qquad r\to+\infty\,.
\end{equation}
The above condition implies that
\begin{equation}
    B^2(r) \to 1 + \mathcal{O}(r^{-3})\,,\qquad \text{as $r\to+\infty$}.
    \label{eq:B-asym-cond}
\end{equation}
Additionally, the last term in the expression of $A(r)$ must respect the condition
\begin{equation}
    \lim_{R\to+\infty}
    \int_{r_0}^{R}\dd r\,r^2
    \big[2\,\Xi\,\mathcal{F}_\Xi-\mathcal{F}(\Xi)\big]
    \in\mathbb{R}\,,
    \label{eq:F-asym-condition}
\end{equation}
where $\Xi$ is given by Eq.~\eqref{eq:Xi} and
$r_0$ is an arbitrary fixed radius.
To analyze the asymptotic behavior of the above integral, we first assume a specific form for $\mathcal{F}(\Xi)$. 
In particular, we consider
\begin{equation}
\mathcal{F}(\Xi)\propto\Xi^s,
\end{equation}
where $s\in \mathbb{R}$.
We then determine the values of $s$ for which the last term exhibits the desired asymptotic behavior. 
Once these values are established, the analysis can be extended to functions $\mathcal{F}(\Xi)$ admitting an asymptotic expansion in integer or fractional powers of $\Xi$, by considering the leading nonvanishing contribution to $2\Xi\mathcal{F}_\Xi-\mathcal{F}$.

\textbf{Proposition.} For $\mathcal{F}(\Xi)\propto \Xi^s$ and $\Xi$ given by \eqref{eq:Xi}, the integral $\int \dd r\, r^2 \big[2\,\Xi\, \mathcal{F}_\Xi -\mathcal{F}(\Xi)\big]$ takes the form
\begin{equation}
    \mathcal{I}(r) \equiv \int \dd r\, r^2 \big[2\,\Xi\, \mathcal{F}_\Xi -\mathcal{F}(\Xi)\big] \propto \frac{2s-1}{3}\,\left(\frac{q^{2}}{2}\right)^s r^3 \ {}_{2}F_{1}\left(\frac{3}{2},s;\frac{5}{2};-\frac{r^{2}}{\lambda^2}\right)\,.
\end{equation}
and its asymptotic behavior is of the form
\begin{equation}
\lim_{r\to\infty} \mathcal{I}(r) =
\begin{cases}
0, & s=1/2\,,\\[2mm]
\pm\infty, & \text{$s \le 3/2$\ \ \&\ \ $s\neq 1/2$}\,,\\[2mm]
\text{$C_1$ (finite \& nonzero)}, & s > 3/2\,.
\end{cases}\\
\label{eq:proposition}
\end{equation}

\textit{Proof.}\ \ We start by using the Euler's integral representation for the hypergeometric function, which reads
\begin{equation}
\begin{split}
{}_{2}F_{1}(a,b;c;z) = \frac{\Gamma(c)}{\Gamma(a)\Gamma(c-a)}\int_{0}^{1} t^{a-1}(1-t)^{c-a-1}(1-zt)^{-b}\ \dd t
\end{split}
\label{eq:euler-general}
\end{equation}
valid for $c>a>0$ and $z\notin[1,+\infty)$. 
The formula is symmetric under $a\leftrightarrow b$ since the hypergeometric function is.
Taking $a=3/2$, $b=s$, and $c=5/2$, the hypothesis $c>a>0$ reads $5/2>3/2>0$, which is true unconditionally, with \emph{no constraint on $s$}.
Since $c-a=1$, the factor $(1-t)^{c-a-1}$ is unity, and with $z=-r^{2}/\lambda^2\le 0$ (so $1-zt=1+r^{2}t/\lambda^2>0$ on $[0,1]$) Eq.~\eqref{eq:euler-general} gives, for every real $s$ and every $r\ge 0$,
\begin{align}
\mathcal{I}(r)& \propto (2s-1) r^3\int_{0}^{1} t^{1/2}\bigg(1+\frac{r^{2}}{\lambda^2}\,t\bigg)^{-s}\dd t,
\label{eq:euler-applied}
\end{align}
where we have neglected non-divergent constant factors.
Substituting now $t=u\lambda^2/r^2$, the above integral takes the form
\begin{equation}
    \mathcal{I}(r) \propto (2s-1)\int_{0}^{r^2/\lambda^2} u^{1/2}(1+u)^{-s}\,\dd u.
\end{equation}
\begin{itemize}
    \item \textbf{Case 1:} For $s=3/2$ we have
        \begin{equation}
            \mathcal{I}_{\frac{3}{2}}(r) \propto \text{arcsinh}(r/|\lambda|)-\sqrt{\frac{r^2}{r^2+\lambda^2}}\,,
        \end{equation}
        which asymptotically leads to
        \begin{equation}
            \lim_{r\to+\infty} \mathcal{I}_{\frac{3}{2}}(r) = \lim_{r\to+\infty}[ 2\ln(r/|\lambda) + (2\ln 2 -2) + \mathcal{O}(r^{-2})] = +\infty\,.\\
        \end{equation}
        
    \item \textbf{Case 2:} For $s\neq 3/2$ one finds that as $r\to +\infty$
        \begin{equation}
            \mathcal{I}(r) \propto  (2s-1)\left\{\frac{\sqrt{\pi}}{2}\,\frac{\Gamma(s-3/2)}{\Gamma(s)} + (r^2)^{\frac{3}{2}-s}\bigg[\alpha_0 + \frac{\alpha_2}{r^2} + \mathcal{O}(r^{-4}) \bigg]\right\}\,.
        \end{equation}
        From the above, it is straightforward to deduce that for $s=1/2$ the above limit becomes identically zero, for $s>3/2$ the above limit converges to a constant, while for $s<3/2\ \& \ s\neq 1/2$ the limit diverges to $\pm \infty$. 
\end{itemize}
The preceding analysis proves Eq.~\eqref{eq:proposition}. \hfill $\blacksquare$\\

Now, using the previous result, we conclude that for functions $\mathcal{F}(\Xi)$ admitting an asymptotic expansion in integer or fractional powers of $\Xi$, the integral in Eq.~\eqref{eq:F-asym-condition} converges if the leading nonvanishing term in $2\Xi\mathcal{F}_\Xi-\mathcal{F}$ scales as $\Xi^s$ with $s>3/2$. The special case $\mathcal{F}(\Xi)\propto\Xi^{1/2}$ also satisfies this condition, since its contribution to the integral vanishes identically. This, together with Eq. \eqref{eq:B-asym-cond}, guarantees asymptotic flatness of the spacetime \eqref{eq:metric}.

%==========================================
%Appendix-Regularity conditions
%==========================================

\section{Regularity conditions for static and spherically symmetric compact objects}
\label{App:Reg-Cond}

In this Appendix, we examine the conditions under which the line element \eqref{eq:metric} describes a regular compact object, such as a regular black hole, or a gravitational soliton.
In this context, regularity means that the spacetime is free of curvature singularities, i.e., that the components of the Riemann tensor $R^{\rho}{}_{\lambda\mu\nu}$ as well as the curvature invariants $R\equiv R_{\mu\nu} g^{\mu\nu}$, $\mathcal{R}\equiv R_{\mu\nu}R^{\mu\nu}$, $\mathcal{K}\equiv R_{\mu\nu\rho\sigma}R^{\mu\nu\rho\sigma}$, remain finite at $r=0$.
One can verify by brute force that, in terms of the metric functions \(A(r)\) and \(B(r)\), a regular center at \(r=0\) requires
%.
\begin{equation}
    A(r)=A_0 + \sum_{n=1}^{+\infty} A_{2n} r^{2n} ,
    \qquad
    B(r)=B_0 + \sum_{n=1}^{+\infty} B_{2n} r^{2n},
    \qquad
    A_0 = B_0^2 .
    \label{eq:regular-center-AB}
\end{equation}
In particular, the equality
\begin{equation}
    A(0)=B^2(0)
    \label{eq:regular-center-condition}
\end{equation}
is the leading condition required to remove the central curvature singularity.
Indeed, when the expansions \eqref{eq:regular-center-AB} hold, the curvature invariants at $r\to 0$ behave as
\begin{equation}
\begin{aligned}
R &= \frac{12}{B_0^2}(B_0 B_2-A_2) 
+ \sum_{n=1}^{+\infty} (C_R)_n\, r^{2n},
\\[2mm]
\mathcal{R} &= \frac{12}{B_0^4} \left( 3A_2^2-6B_0A_2 B_2+4B_0^2B_2^2 \right) 
+ \sum_{n=1}^{+\infty} (C_\mathcal{R})_n\, r^{2n},
\\[2mm]
\mathcal{K} &=
\frac{24}{B_0^4} \big(A_2^2 - 2B_0 A_2 B_2 + 2 B_0^2 B_2^2 \big)
+ \sum_{n=1}^{+\infty} (C_\mathcal{K})_n\, r^{2n},
\end{aligned}
\label{eq:curvature-invariants-regular-center}
\end{equation}
where for any given $n$, the constants $(C_x)_n$ are built from the expansion coefficients of $A(r)$ and $B(r)$ up to order $2(n+1)$, that is
\begin{equation}
    (C_x)_n=(C_x)_n(A_0,A_2,\ldots,A_{2(n+1)},B_0,B_2,\ldots,B_{2(n+1)})\,.
\end{equation}
Additionally, one can verify that for a spacetime \eqref{eq:metric} whose metric functions $A(r)$ and $B(r)$ admit an expansion of the form \eqref{eq:regular-center-AB} at $r=0$, the terms $\square^p R$, $\square^p \mathcal{R}$, and $\square^p \mathcal{K}$ remain finite for any positive integer $p$.
For an in-depth analysis of higher-order regularity conditions in local and nonlocal quantum gravity, the reader is referred to \cite{Burzilla:2020utr}.

%===============================================
%Appendix-Light and graviton horizons of axial
%===============================================

\section{Location of the light and graviton horizons for the axial sector}
\label{App:Axial-horizons}

\textbf{Proposition}. \emph{If $\mathcal F\leq0$, every zero of $A$ lies inside $r_g$. 
On $(r_g,+\infty)$ one has $A>0$, $\Upsilon>0$ and $\Phi>0$.}

\textit{Proof.} For both choices \eqref{eq:Z-X} and \eqref{eq:Z-2} of the auxiliary
function, $Z^2\geq1$. 
For $\mathcal F\leq0$, Eq.~\eqref{eq:G2-G4} gives $G_4Z=1+\lambda^2\mathcal F/2\leq1$. 
Also, since $\Xi=Z^2X>0$ by Eq.~\eqref{eq:Xi}, we deduce that $X>0$, and therefore
\begin{equation}
    A_1=\frac{Z-G_4}{2X}=\frac{Z^2-G_4Z}{2XZ}>0\,,
    \label{eq:A1-positivity}
\end{equation}
where equality is excluded unless $Z=1$ and $\mathcal F=0$, i.e.\ for the trivial stealth configuration. 
Consequently, at the light horizon $r_l$ for which $A(r_l)=0$, $\Upsilon(r_l)=-q^2A_1(r_l)<0$. 
Additionally, we have $\Upsilon \to 1$ as $r\to+\infty$. 
To see that $\Upsilon\to1$ at infinity, we make use of Eq.~\eqref{eq:G2-G4} and $X=\Xi/Z^2$, and we have
\begin{equation}
    A_1=\frac{Z^2-1}{2XZ}-\frac{\lambda^2 Z}{4}\,\frac{\mathcal F(\Xi)}{\Xi}\,.
\end{equation}
For both choices \eqref{eq:Z-X} and \eqref{eq:Z-2}, $Z^2-1=\mathcal O(X^2)$, so the first term vanishes as $X\to 0$. 
Since $A,Z\to1$, one obtains $\Upsilon\to1+\frac{q^2\lambda^2}{4}\lim_{\Xi\to0}\mathcal F(\Xi)/\Xi$, and hence $\Upsilon\to1$ whenever $\mathcal F(\Xi)=o(\Xi)$. 
This holds for the Pad\'e and for all power-law functions $\mathcal{F}\propto \Xi^p$ with $p>1$.
Since $\Upsilon(r_l) < 0$ and $\Upsilon\to1$ as $r\to+\infty$, $\Upsilon$ vanishes somewhere outside the light horizon of $A$, and $r_g$ exceeds the outermost zero $r_l$ of $A$. 
Therefore, on $(r_g,+\infty)$ we have $\Upsilon>0$. \hfill $\blacksquare$

In conclusion, for the models satisfying $\mathcal{F}\leq0$ and $\mathcal{F}=o(\Xi)$ as $\Xi\to0$, the outermost light horizon of $g_{\mu\nu}$ lies inside the graviton horizon of $\hat g_{\mu\nu}$. Consequently, the divergence of $c_*$ at $A=0$ does not affect the causal exterior of the effective metric $\hat g_{\mu\nu}$.

%===============================================
%Appendix-S-deformation
%===============================================

\section{The \texorpdfstring{$S$}{S}-deformation and the positivity of \texorpdfstring{$\omega^2$}{omega squared}}
\label{App:S-deformation}
 
Both the axial and the monopole sectors reduce to a Schr\"odinger equation
\begin{equation}
    H\Phi\equiv\left[-\frac{\dd^2}{\dd x^2}+V(x)\right]\Phi=\omega^2\,\Phi=E\Phi\,,
    \qquad x\in(x_L,x_R)\,,
    \label{eq:S-schro}
\end{equation}
where $\Phi$ here is a complex function, which should not be confused with $\Phi$ defined in Eq.~\eqref{eq:axial-coeffs}.
Additionally, $x$ is the relevant tortoise coordinate, either $\hat r_*$ or $r_*$, and the interval may be infinite.
For perturbations $\propto e^{-i\omega t}$, an instability is a normalizable solution with ${\rm Im}\,\omega>0$.
If $H$ is self-adjoint on a domain $\mathcal D\subset L^2(x_L,x_R)$, its spectrum $E$ is real, and such a mode requires $E=\omega^2<0$. 
Indeed, given a positive self-adjoint operator $H$, the time evolution of $H$ is given by:
\begin{equation}
 \Phi(t)= \cos(H^{1/2} t) \Phi_0 + H^{-1/2} \sin(H^{1/2} t) \dot{\Phi}_0 \,,
\end{equation}
where $\Phi_0$ and $\dot{\Phi}_0 $ are some smooth initial conditions with compact support~\cite{Wald:1979lth}. 
If $H$ is self-adjoint and positive then 
$\cos (H^{1/2} t)$ and $H^{-1/2} \sin(H^{1/2} t)$  are bounded operators. Therefore the time
evolution of $\Phi$ remains bounded at all times, which guarantees the stability of the linear perturbations. 
Linear stability therefore amounts to showing that $H$ has no negative spectrum, i.e.\ that $\langle\Phi,H\Phi\rangle\geq0$ for all $\Phi\in\mathcal D$.
When $V$ is positive, the previous condition is immediately satisfied upon appropriate boundary conditions, while the $S$-deformation method \cite{Kodama:2003jz, Ishibashi:2003ap, Kodama:2003kk, Kimura:2018eiv, Kimura:2018whv} extends the argument to potentials of indefinite sign.
 
Let $S(x)$ be any continuous function on $(x_L,x_R)$, it is straightforward to show that
\begin{equation}
    \langle\Phi,H\Phi\rangle
    =\int_{x_L}^{x_R}\,\dd x\left[\,\Big|\frac{\dd\Phi}{\dd x}+S\,\Phi\Big|^2+\bigg( V+\frac{\dd S}{\dd x}-S^2 \bigg)\,|\Phi|^2\right]
    -\Big[\Phi^*\Big(\frac{\dd\Phi}{\dd x}+S\,\Phi\Big)\Big]_{x_L}^{x_R}\,,
    \label{eq:S-identity}
\end{equation}
with $\Phi^*$ the complex conjugate of $\Phi$. 
It is convenient to introduce $\Bterm_x\equiv\left. 
\Phi^*\left(\frac{\dd\Phi}{\dd x}+S\,|\Phi|\right)
\right|_{x}$.

Therefore, for a non-negative deformed potential on  $(x_L,x_R)$, namely
\begin{equation}
    V_S \equiv V+\frac{\dd S}{\dd x}-S^2 \geq 0\,,
    \label{eq:Def_VS}
\end{equation}
a sufficient condition to prove positivity of the spectrum is to have positive boundary terms as well as positive $V_S$.
Then it holds that $\langle\Phi,H\Phi\rangle \geq 0$, which by its turn results in $E=\omega^2\geq 0$.

%===============================================
%Appendix-Absence of scalar degree of freedom 
%===============================================

\section{Absence of a propagating scalar degree of freedom for Branch II}
\label{App:BranchII}
As explained earlier, Branch~II does not propagate a scalar mode. It might nevertheless be useful to exhibit perturbation structure in some detail. Indeed, understanding how the absence of the scalar mode manifests itself at the level of the perturbation equations may provide useful insight into other, less straightforward degenerate cases, such as the $p=1/2$ model discussed in \cite{Charmousis:2026bhe}.
For this branch, the determinant introduced in \cite{Charmousis:2026bhe}, $\Det$ vanishes signaling that the system of perturbation equations is degenerate and cannot be inverted. Indeed, if one defines the variable
\begin{equation}
    \sigma= H_2-\frac{2 (g_4-1) q^3 A(r) \psi '(r) }{4 A(r)^2 X^2+(g_4-1) q^4}H_1+\frac{(g_4-1) q^2  A(r)^2 \psi '(r)^2}{4 A(r)^2 X^2+(g_4-1) q^4}H_0 \, ,
\end{equation}
then, one can show, using two of the three independent equations at hand, that $\sigma$ is a function depending only on $r$
\begin{equation}
   \dot{\sigma}=0
  \label{eq:br2-dot-sigma}
\end{equation}
\begin{equation}
\sigma'+F(X,A,g_2,g_4)\ \sigma=0 \, ,
\end{equation}
where $F$ is a function of the undetermined variable $X$.
These equations give a constraint on both $H_0,H_1$
\begin{equation}
    [H_1,H_0,\sigma,H_0',H_1',\overset{.}{H_0},\overset{.}{H_1}]=0
\end{equation}
where the notation $[\cdots]$ means a linear combination of the terms enclosed within the brackets with coefficients that depend on the background solution. 
The resulting perturbation system is therefore rather peculiar. As expected from the disformal relation to GR, the three equations do not describe an independent propagating monopole degree of freedom. This structure is reminiscent of the monopole sector of cuscuton theory studied in \cite{Saito_2023}, where a similarly degenerate system of perturbation equations was obtained. In that case, additional boundary conditions were imposed to show that the monopole perturbation vanishes identically. A related analysis of the monopole sector in degenerate scalar--tensor theories was also carried out in \cite{De_Felice_2018}.

%======================= 
\bibliography{Refs.bib}
%======================= 

\end{document}